\documentclass[pre,amsmath,aps,twocolumn,superscriptaddress,letterpaper]{revtex4}
\usepackage{graphicx,xspace,units,subfigure}
\usepackage{float}
\usepackage{amsmath}
\usepackage{amssymb}
\usepackage[normalem]{ulem}
\usepackage{appendix}
\input{epsf}

\newcommand{\be}{\begin{equation}}
\newcommand{\ee}{\end{equation}}

\begin{document}


\title{Collisional Statistics and Dynamics of 2D Hard-Disk Systems: From Fluid to Solid}

\author{Alessandro Taloni}
\affiliation{CNR-IENI, Via R. Cozzi 53, 20125 Milano, Italy}
\affiliation{Dipartimento di Fisica, University of Milan, Via Giovanni Celoria, 16, 20133 Milano, Italy}
\author{Yasmine Meroz}
\affiliation{School of Engineering and Applied Sciences, Harvard University, Cambridge, Massachusetts 02138, USA}
\author{Adri\'an Huerta}
\affiliation{
Facultad de F\'isica, Universidad Veracruzana,
Circuito Gonz\'alo Aguirre Beltr\'an s/n Zona Universitaria,
Xalapa, Veracruz 91000, M\'exico
}

\begin{abstract}  We perform extensive MD simulations of two-dimensional systems of hard disks, focusing on the \emph{on}-collision statistical properties. We analyze the distribution functions of velocity, free flight time and free path length for packing fractions ranging from the fluid to the solid phase. The behaviors of the mean free flight time and path length between subsequent collisions are found to drastically change in the coexistence  phase. We show that single particle dynamical properties behave analogously in collisional and continuous time representations, exhibiting apparent crossovers between the fluid and the solid phase. We find that, both in collisional and continuous time representation, the mean square displacement, velocity autocorrelation functions, intermediate scattering functions and self part of the van Hove function (propagator), closely reproduce the same behavior exhibited by  the corresponding quantities in granular media, colloids and supercooled liquids close to the glass or jamming transition.   
\end{abstract}
\maketitle

\section{Introduction}

A system of two-dimensional  (2D) hard disks is one of the simplest models of a classical fluid.
Despite the apparent simplicity, the transport properties and the nature of the phase transitions remain areas of active investigation ever since the  pioneering work of Alder and Wainwright~\cite{alder1967, alder1970} some $50$ years ago. The appearance of slow power-law decaying tails in the velocity autocorrelation function for moderately dense systems has deeply  
changed the classical understanding and the ensuing formulation of the kinetic theory (see for instance Refs.~\cite{pomeau1975, dorfman1977}). Moreover, controversy about the nature of the fluid-solid transition occurring at higher densities has persistently involved generations of scientists, debating on whether its best representation is the Kosterlitz-Thouless-Halperin-Nelson-Young scenario \cite{Kosterlitz1973,Halperin1978}, or a  first order transition \cite{alder1962,hoover1968}. This debate seems to have only recently arrived to a conclusion thanks to the initial massive use of event-chain Monte Carlo algorithm~\cite{bernard2011}, and by later adopting local Monte Carlo algorithm 
and event-driven molecular dynamics simulations~\cite{engel2013}.

A system may be followed via an {\it external} clock represented by continuous time $t$, or an {\it internal} clock associated with the number of collisions $n$ of each particle. The continuous time representation is said to be subordinated to the underlying collisional process~\cite{feller}. 
Though the system is clearly characterized by collision events, a systematic analysis of this system in collisional representation, spanning from the fluid to the solid phase, is missing. To the best of our knowledge the only analytic and numerical studies in this sense concern  the non-Poissonian nature  of the number of collisions within a certain time interval in the dilute limit \cite{ViscoJCP2008,Visco2008,Lue2005}.
In what follows we study statistical and dynamical properties of a 2D system of hard-disks in collisional representation, with packing fractions ranging from the fluid to the solid regimes. 

The first part of our analysis focuses on the statistics of \emph{on}-collision velocity, free flight time $\tau$ and the path undergone by a disk between subsequent collisions, the free path length $\langle\left|\xi\right|\rangle$. 
Within this context we study correlations between the $x$ and $y$ components of the velocity, and the dependence of free flight times on the \emph{on}-collision velocities in systems with various packing fractions spanning from the fluid to the solid phase. We also compare our findings to the kinetic approach and Enskog theory, reporting inconsistencies even in the fluid phase. 
Furthermore, our analysis of  $\langle\tau\rangle$ and  $\langle\left|\xi\right|\rangle$  as a function of packing fractions has unveiled plateau-like regions corresponding to the fluid-solid coexistence phase, which to the best of our knowledge have not been reported.

The second part focuses on dynamics, comparing single particle dynamical observables in continuous time  (as a function of $t$) and in collisional representation (as a function of the number of collisions, $n$). We find a full analogy, in spite of the fact that the process is not truly Poissonian~\cite{Visco2008}. We show that the velocity is never a Markovian process in either representations and in both fluid and solid phases ~\cite{Puglisi2006}.
Furthermore we have analyzed the mean squared displacement, intermediate scattering function and self-part of the van Hove function (or propagator) in both representations, showing remarkable similarities to glassy behavior. 

We consider a 2D system of size $L_{x} \times L_{y}$ where $L_{x}$ ($L_{y}$) corresponds to the length in the $x$ ($y$) direction, consisting of $N$ hard disks
with diameter $\sigma$. The {\emph packing fraction} $\eta$ of the system is defined 
as the ratio between the area occupied by the disks, $N\pi \frac{\sigma^2}{4}$, and the available area 
${L_{x} \times L_{y}}$, yielding $\eta = \frac{N}{L_{x} \times L_{y}}\frac{\pi \sigma^2}{4}$.
In collisional representation we follow a single particle counting the collisions it undergoes
with a collision index $n$. For a collision $n$ we now define the particle's position $\mathbf{x}_{n}$, 
the external (continuous) time $t_n$, and the \emph{on}-collision velocity $\mathbf{v}_n$.

We run event-driven molecular dynamics (MD)~\cite{allen1987} numerical 
simulations. Simulations start from a random configuration, according to the desired $\eta$, obtained from 
NVT Monte Carlo simulations.
The temperature is kept constant by scaling the magnitude of velocities of each hard disk 
such that the kinetic energy of the system agrees with the equipartition theorem.
The system is rectangular, where the ratio between $L_{x}$ and $L_{y}$ is kept at $\sqrt{3}$.
Periodic boundary conditions are applied in both directions and the system length is adjusted 
to keep constant the number of particles $N$, yielding the desired $\eta$. 
We set $k_BT=1$, $\sigma=1.$ and the lengths $L_x$ and $L_y$ are measured in units of $\sigma$.
Throughout this work we consider the following packing fractions: $\eta = 0.3, 0.56, 0.695, 0.713, 0.72$, plotted respectively in black, blue, red, green and magenta. We thus span the different phases: fluid phase for low $\eta$, the coexistence phase for
$0.69\leq \eta\leq 0.723$ \cite{truskette1998, mak2006}, and above that the solid phase
(the maximum possible packing fraction is given by $\eta=\pi/\sqrt{12}\approx0.907$~\cite{fejes1940}).

%
%

\section{Collisional representation: statistics}

The analysis of collisional statistics is based on three fundamental observables: 
(i) Free flight times $\tau_n =t_{n+1} - t_{n}$,  the times between consecutive collisions. 
The index $n$ represents the number of collisions suffered by a particle in the system. 
(ii)  \emph{On-}collision velocity $\mathbf{v}_n$, the $x$ and $y$ components of the 
particle velocity between collisions. (iii) Free path length, namely the vector connecting 
the particle's position on the $n$-th and  $(n+1)$-th collisions:  $\boldsymbol{\xi}_n= 
\mathbf{x}_{n+1} - \mathbf{x}_{n}$. We first focus on the statistics of these measurable 
observables by means of event-driven MD numerical simulations, identifying the stationary distributions and the corresponding average values of \emph{on}-collision velocities, free flight times and free path lengths.

\subsection{Instantaneous velocity distribution function}
\begin{figure}[]
\begin{centering}
\includegraphics[width=0.5\textwidth]{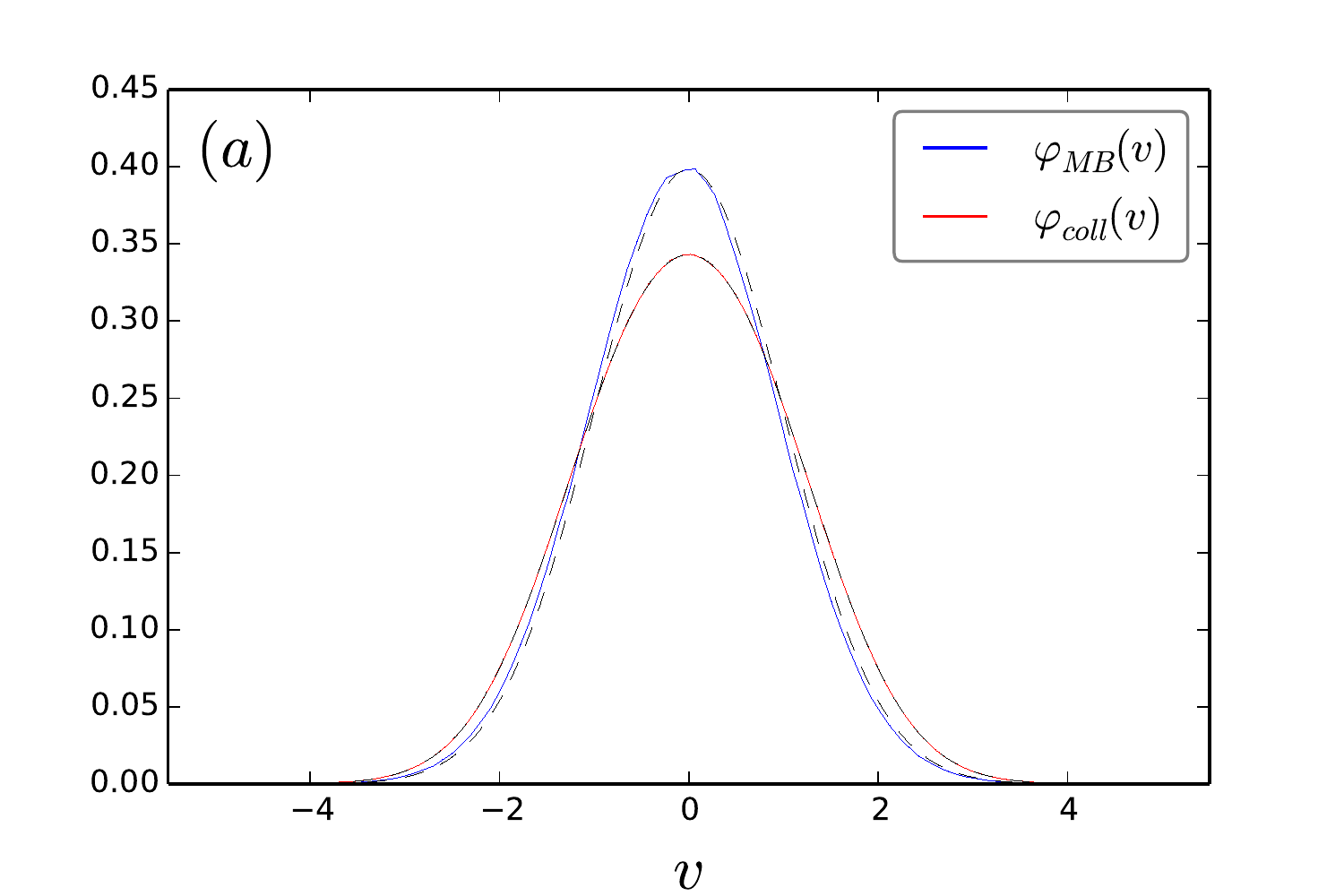}
\includegraphics[width=0.5\textwidth]{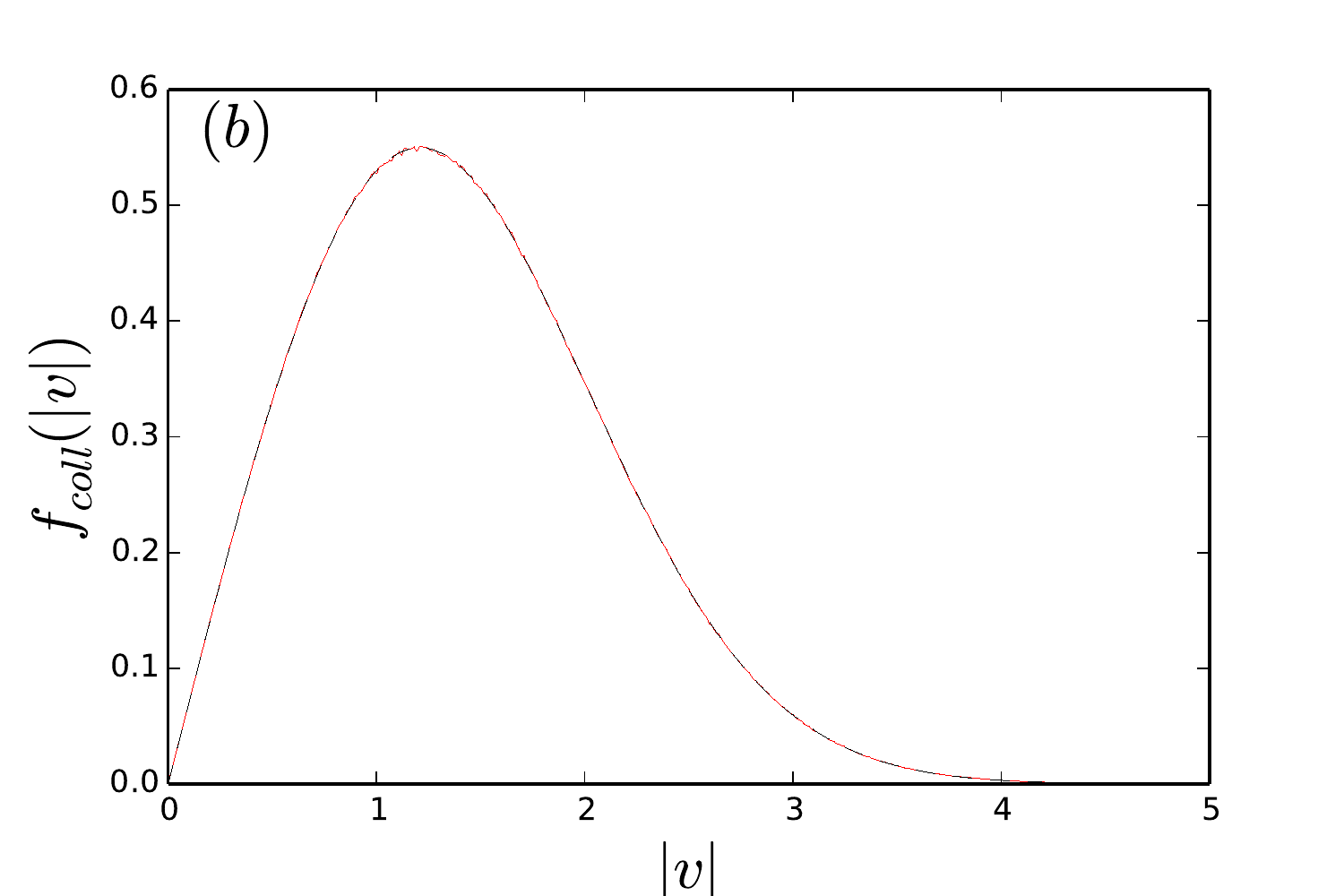}
\end{centering}
\caption{(Color online) Comparison of the velocity distribution in continuous time and collisional representation. (a) Maxwell-Boltzman distribution function of single component velocity for the continuous time representation, $\varphi_{MB}(v)$, plotted with a blue solid line,  and the collisional $\varphi_{coll}(v)$ in a red solid line. Black dashed lines represent, respectively, the usual Maxwell-Boltzmann expression, and the theoretical estimate obtained by numerical integration of Eq.(\ref{eq:phi_coll}) over the $y$-component. (b) Speed distribution function in collisional representation $f_{coll}\left(\left|\mathbf{v}\right|\right)$: numerics are shown as a solid red line, while the theoretical expression $f_{coll}\left(\left|\mathbf{v}\right|\right)$ is reported in dashed black line. $k_BT=1$, $\sigma=1$, $L_x$ and $L_y$ are measured in units of $\sigma$.}
\label{fig:figure_1}
\end{figure}

The distribution of instantaneous velocities of the disks in the
continuous time representation is given by the  
Maxwell-Boltzmann equilibrium distribution function:
\be\label{eq:MB}
\phi_{MB}(\mathbf{v})=\frac{1}{2\pi k_BT} e^{-\frac{\left|\mathbf{v}\right|^2}{2k_BT}},
\ee

where $T$ is the system temperature and $k_B$ is the Boltzmann constant. In collisional representation, however, it is known that the stationary 
distribution of \emph{on}-collision velocities $\phi_{coll}(\mathbf{v})$ takes a different  functional
form ~\cite{Puglisi2006, ViscoJCP2008, Visco2008}:

\be
\begin{array}{l}
\phi_{coll}(\mathbf{v})= \frac{1}{2^{3/2}\pi k_BT} e^{-\frac{\left|\mathbf{v}\right|^2}{2k_BT}}\\ \times \left[\left(\frac{\left|\mathbf{v}\right|^2}{2k_BT}+1\right) I_0\left(\frac{\left|\mathbf{v}\right|^2}{4k_BT}\right)+\frac{\left|\mathbf{v}\right|^2}{2k_BT} I_1\left(\frac{\left|\mathbf{v}\right|^2}{4k_BT}\right)\right],
\end{array}
\label{eq:phi_coll}
\ee
where $I_0$ and $I_1$ represent the cylindrical modified Bessel functions of $0$-th and $1$-st order respectively. As shown in Fig.~\ref{fig:figure_1}(a) the collisional distribution exhibits wider tails, i.e. a higher probability for higher velocities, as compared to its continuous time counterpart. This discrepancy can
be understood intuitively with the following argument: in 
continuous time representation, the velocity distribution is calculated across 
snapshots of the system at different intervals, therefore higher velocities will be counted 
less often since their respective free flight times are shorter on average, spanning across less 
consecutive snapshots. In contrast, in collisional representation
each \emph{on}-collision velocity is counted only once, upon collision.

The expression in Eq.~(\ref{eq:phi_coll}) highlights two important features of the velocity collisional process. 
The first is that the collisional trajectory is an isotropic process, since the velocity distribution depends only on its speed. The speed distribution $f_{coll}\left(\left|\mathbf{v}\right|\right)$ can indeed be derived 
from Eq.(\ref{eq:phi_coll}), yielding  $f_{coll}\left(\left|\mathbf{v}\right|\right) = 2\pi \phi_{coll}(\mathbf{v})=\frac{1}{\sqrt{2}k_BT} e^{-\frac{\left|\mathbf{v}\right|^2}{2k_BT}}\left|\mathbf{v}\right|\left[\left(\frac{\left|\mathbf{v}\right|^2}{2k_BT}+1\right)I_0\left(\frac{\left|\mathbf{v}\right|^2}{4k_BT}\right)+\frac{\left|\mathbf{v}\right|^2}{2k_BT}I_1\left(\frac{\left|\mathbf{v}\right|^2}{4k_BT}\right)\right]$.  $f_{coll}\left(\left|\mathbf{v}\right|\right)$ is plotted in Fig.~\ref{fig:figure_1}(b) against the numerical data: this extends  to $d=2$  the  analytic expression of the speed distribution in collisional representation furnished for  $d=3$ and $d=5$ in Ref.\cite{Lue2005}. 

The second property highlighted by Eq.~(\ref{eq:phi_coll})  is that the velocity components are correlated within the collisional process, i.e. $\phi_{coll}(\mathbf{v})\neq\varphi_{coll}(v^x)\varphi_{coll}(v^y)$ although $\varphi_{coll}(v^x)=\varphi_{coll}(v^y)$ (where $\varphi_{coll}(v)$ is the single component velocity 
distribution). The single component velocity distribution function
can be obtained by integrating Eq.(\ref{eq:phi_coll}) over the other component, i.e.   $\varphi_{coll}(v^x)=\int_{-\infty}^{+\infty}dv^y \phi_{coll}\left(\mathbf{v}\right)$. The numerical evaluation of the integral expression of $\varphi_{coll}(v^x)$ is shown in Fig.~\ref{fig:figure_1}(a), perfectly reproducing the numerical data. We note that Maxwell molecules, interacting through a pairwise potential such as
 $V(\left|\mathbf{r}\right|)\sim \left|\mathbf{r}\right|^{-2}$ \cite{Ernst1981}, exhibit identical behavior in both continuous time and collisional representations: they are both isotropic, and the $x$ and $y$ components are in fact uncorrelated, yielding the same Gaussian form of the velocity distribution functions as in Eq.~(\ref{eq:MB})~\cite{Maxwell1860}.
 
Noting that the \emph{on}-collision velocity is isotropic, and that the $x$ and $y$ components
are statistically identical (even though they are correlated), we can henceforth limit our 
analysis to the single component $x$ (or $y$ without loss of generality), i.e. $v_n\equiv v_n^x$ 
and $\xi_n =x_{n+1} - x_{n}$.

\begin{figure}[t]
\begin{centering}
\includegraphics[width=0.5\textwidth]{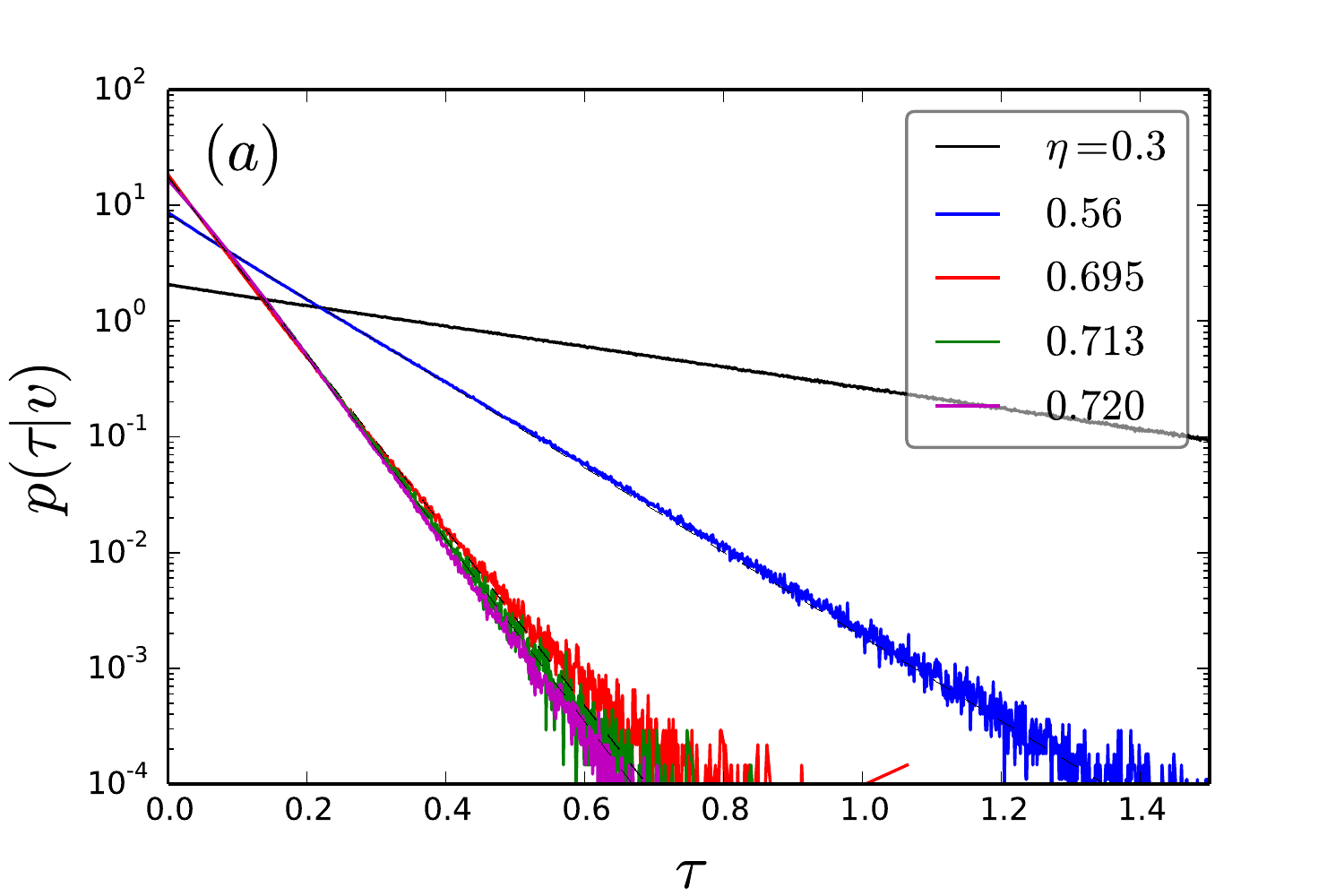}
\includegraphics[width=0.5\textwidth]{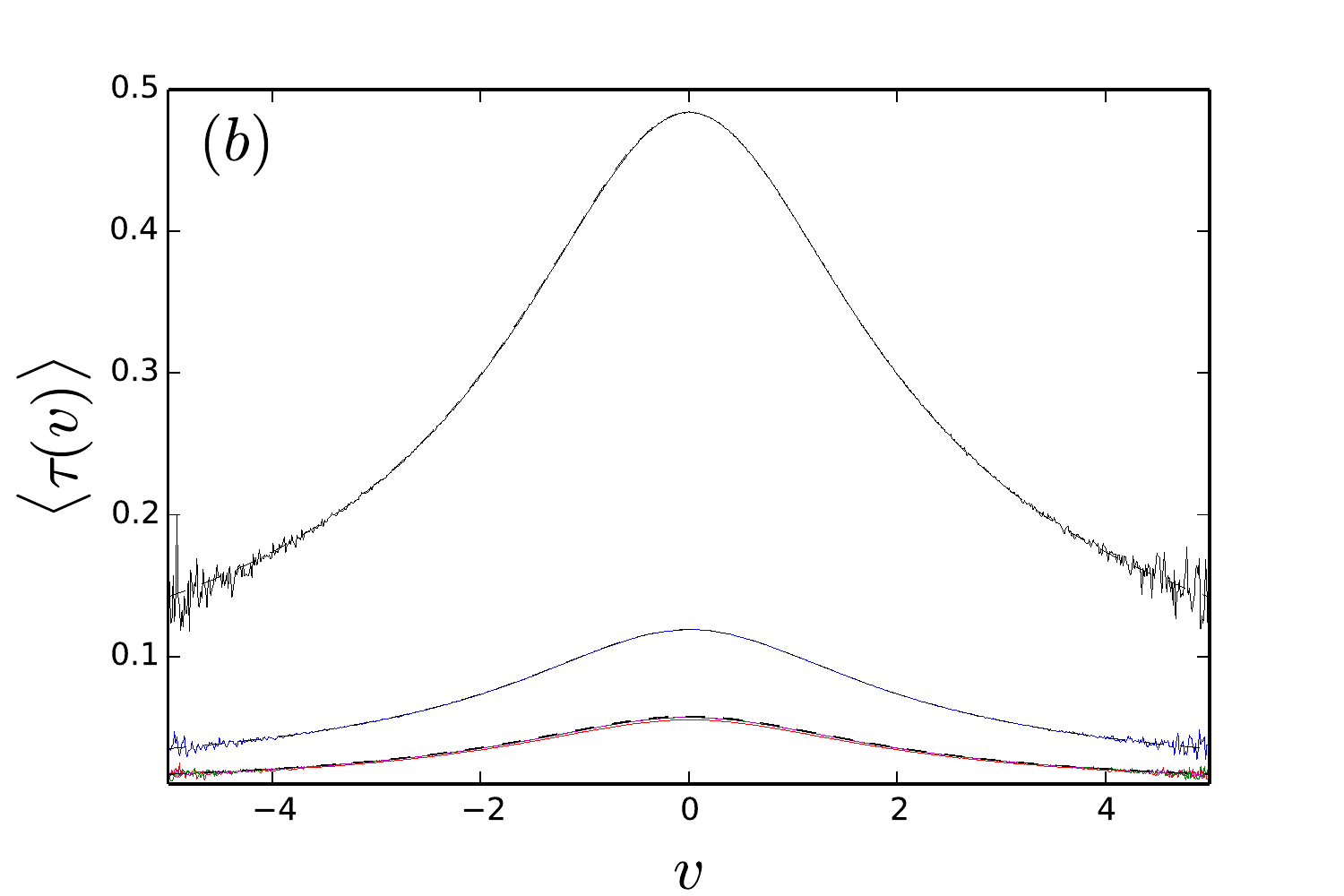}
\end{centering}
\caption{(Color online) Dependence of free flight times on velocity. (a) The probability distribution $p(\tau|v)$ for different $\eta$ with $v=0.0\pm 0.05$ (arbitrary units). The exponential expression $\frac{e^{-t/\langle\tau(v)\rangle}}{\langle\tau(v)\rangle}$ (dashed lines) successfully describes the numerical outcomes only for the lower value of $\eta$, i.e. $\eta=0.3, 0.56$.   (b) Average free flight time as a function of $v$, $\langle\tau(v)\rangle$,  for several $\eta$. The dashed lines are obtained from the numerical evaluation of Eq.~(\ref{eq:time_relation}).}
\label{fig:figure_2}
\end{figure}

\subsection{Free-flight time distribution function\label{sec:free_flight}}

\begin{figure}
\begin{center}
\includegraphics[width=0.5\textwidth]{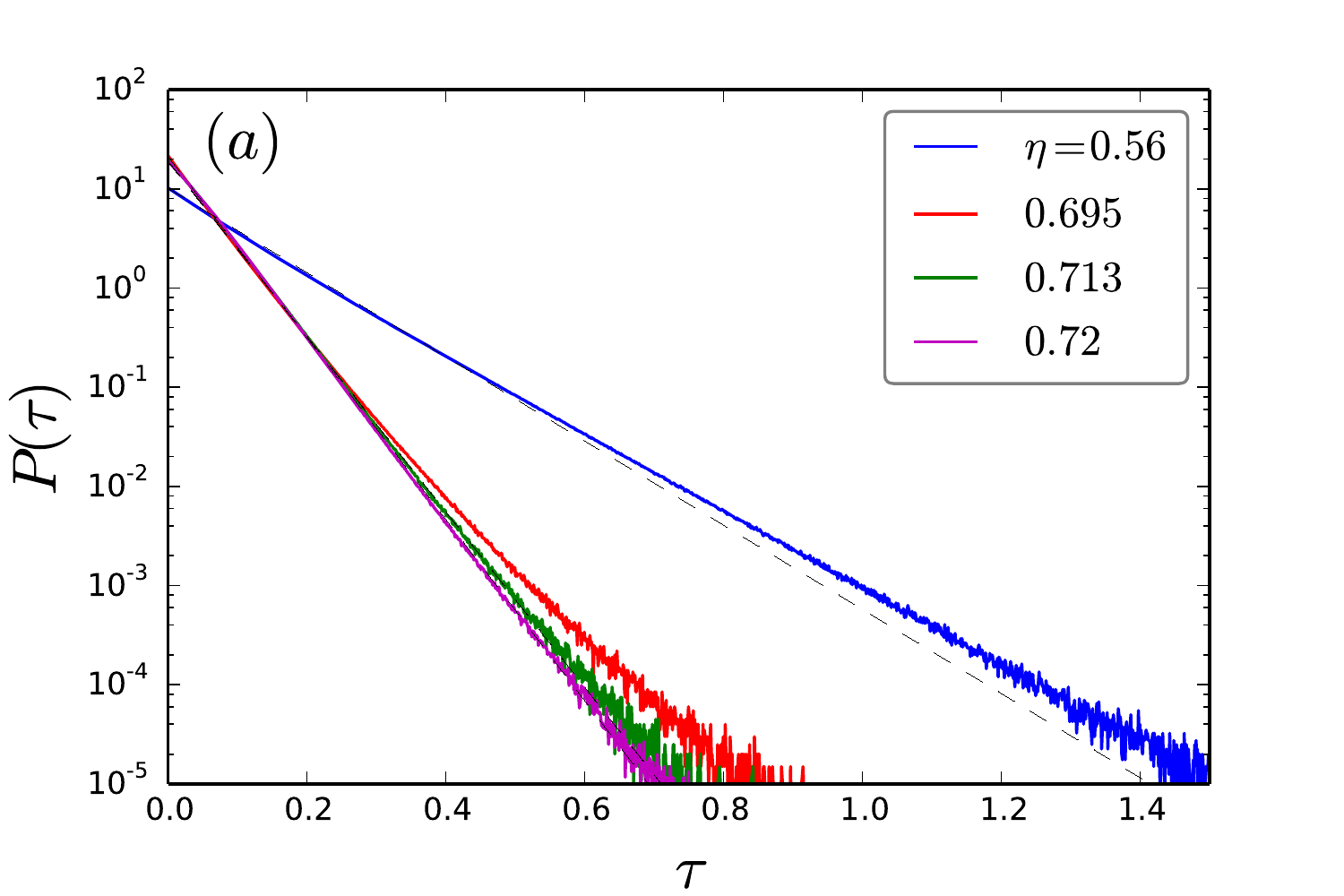}
\includegraphics[width=0.5\textwidth]{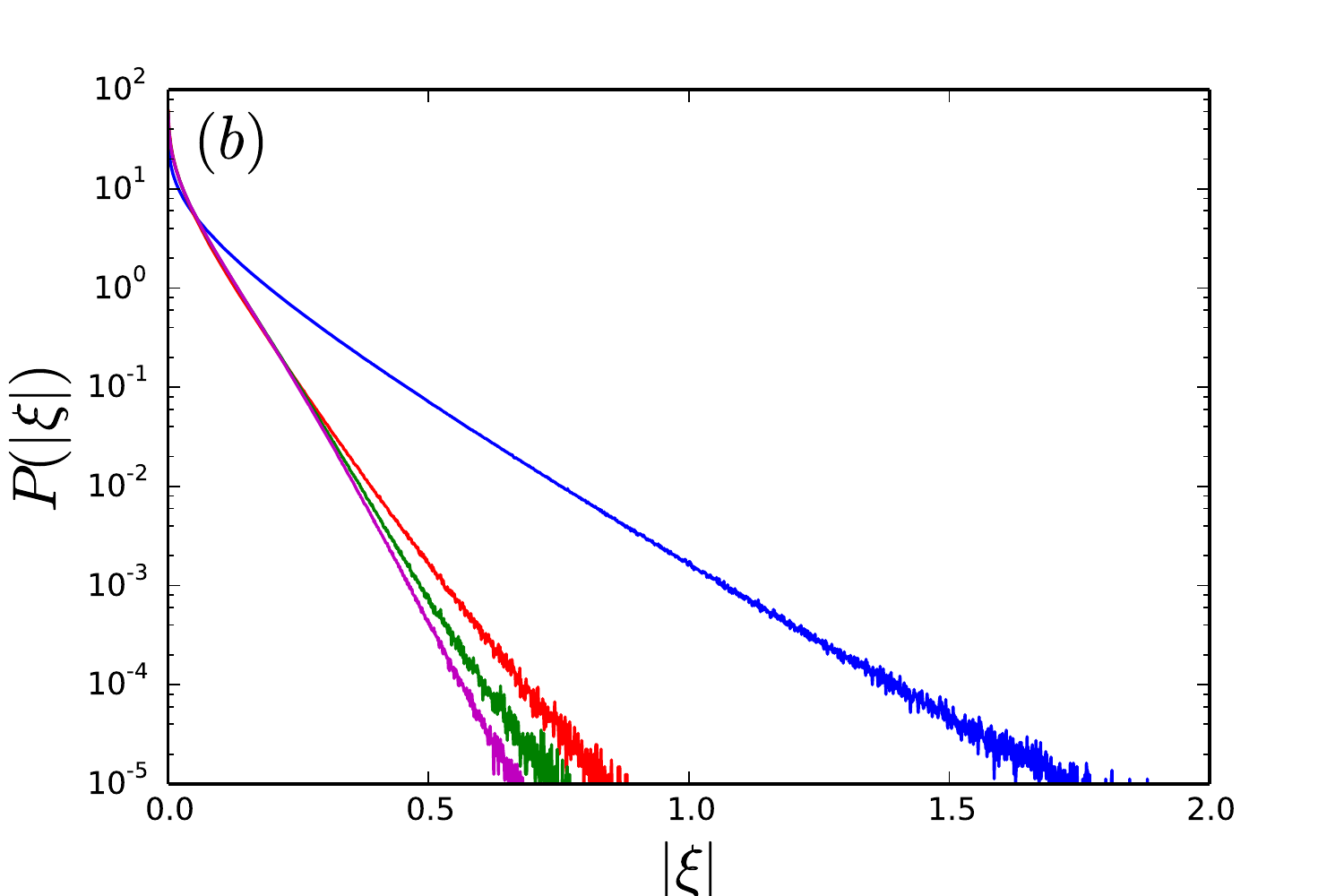}
\end{center}
\caption{(Color online) Free flight time  (a) and free path  (b) probability distribution functions for several values of $\eta$. The exponential expression $\frac{e^{-t/\langle\tau\rangle}}{\langle\tau\rangle}$ (dashed lines in panel (a)) do not capture the numerical outcomes in the low as well as in the high $\eta$ regime.}
\label{fig:figure_3}
\end{figure}

The probability distribution of free-flight times $P(\tau)$ was first analyzed in
Ref.s~\cite{Wiegel1976,Michels1979,Visco2008}, where it was shown that it
cannot be purely exponential, not even in the low density limit ~\cite{Wiegel1976}. 
This is rooted in the fact that the characteristic free-flight time (or its inverse 
the collision frequency) of particles with a given velocity, $\langle \tau (\mathbf{v}) \rangle$, depend on the speed $\left|\mathbf{v}\right|$ \cite{Ernst1981,Visco2008}. 
To explain the non-exponential shape of $P(\tau)$ analytically, we introduce
the conditional probability of having a free-flight time $\tau$ given that the 
particle has a velocity $\mathbf{v}$: $P\left(\tau|\mathbf{v}\right)$. One can then derive
the free-flight time distribution, following~\cite{ViscoJCP2008,Visco2008}:
\be
P(\tau)=\int_{-\infty}^{\infty}d\mathbf{v}\, P\left(\tau|\mathbf{v}\right)\phi_{coll}(\mathbf{v}).
\label{eq:free_flight_prob_def_vect}
\ee
The conditional probability $P\left(\tau|\mathbf{v}\right)$ is related to the viscosity
of the system~\cite{Wiegel1976}, and is therefore interesting to study.
By introducing the single component conditional probability  $p\left(\tau|v\right)$ (shown in  Fig.~\ref{fig:figure_2}(a)) we can also write:
\be
P(\tau)=\int_{-\infty}^{\infty}dv\, p\left(\tau|v\right)\varphi_{coll}(v).
\label{eq:free_flight_prob_def}
\ee
For low $\eta$, it can be shown that $P\left(\tau|\mathbf{v}\right)=\frac{1}{\langle\tau\left(\mathbf{v}\right)\rangle} e^{-\tau/\langle\tau\left(\mathbf{v}\right)\rangle}$ \cite{Puglisi2006,Visco2008},
a consequence of the molecular chaos assumption, or  {\it strosszahlansatz}, which 
holds in the dilute limit. In this limit the collisional dynamics of a tagged particle is 
exactly a Markov process in the space of velocities, where the hopping probability from velocity $\mathbf{v}'$ to
 velocity $\mathbf{v}$, given by $W(\mathbf{v}|\mathbf{v}')$, can be calculated analytically
 ~\cite{Puglisi2006}. We find that also the single component distribution function 
$p\left(\tau|v\right)$ exhibits an exponential shape $p\left(\tau|v\right)=
\frac{1}{\langle\tau\left(v\right)\rangle} e^{-\tau/\langle\tau\left(v\right)\rangle}$ in the dilute 
limit $\eta\to 0$, as can be appreciated in Fig.~\ref{fig:figure_2}(a). However, though 
both the complete and single component conditional  probabilities $P\left(\tau|\mathbf{v}\right)$ 
and $p\left(\tau|v\right)$ are exponential  in the dilute limit, it must be stressed that 
the characteristic free-flight time given a single component velocity, $\langle\tau(v)\rangle$ shown in 
Fig.~\ref{fig:figure_2}(b), differs from $\langle\tau\left(\mathbf{v}\right)\rangle$.
Deviation of $P\left(\tau|\mathbf{v}\right)$ from a pure exponential
 is expected at large $\eta$ due to the loss of the molecular chaos ansatz, and to the 
ensuing hopping dynamics which is no longer Markovian in velocity space. 
This deviation is clearly observed also in the single component 
$p\left(\tau|v\right)$, as displayed in Fig.~\ref{fig:figure_2}(a). In any case substituting 
Eq.~(\ref{eq:phi_coll}) in Eq.~(\ref{eq:free_flight_prob_def_vect}) means 
that $P(\tau)$ cannot be an  exponential either, as shown in Fig.~\ref{fig:figure_3}(a). The same argument can be specified to 
$p\left(\tau|v\right)$ and $\varphi_{coll}(v)$ in Eq.~(\ref{eq:free_flight_prob_def}). 

\begin{figure}[t]
\includegraphics[width=0.5\textwidth]{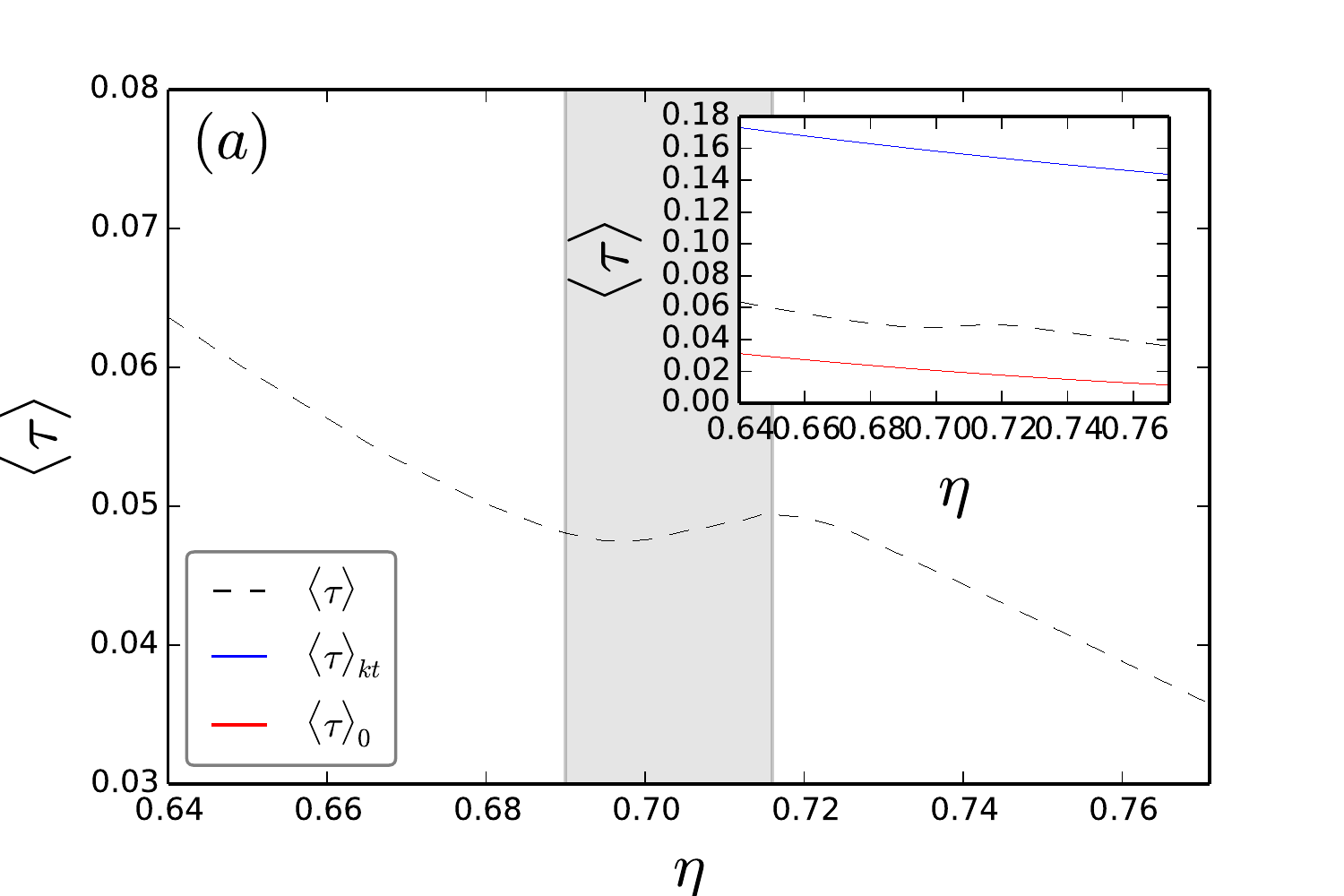}
\includegraphics[width=0.5\textwidth]{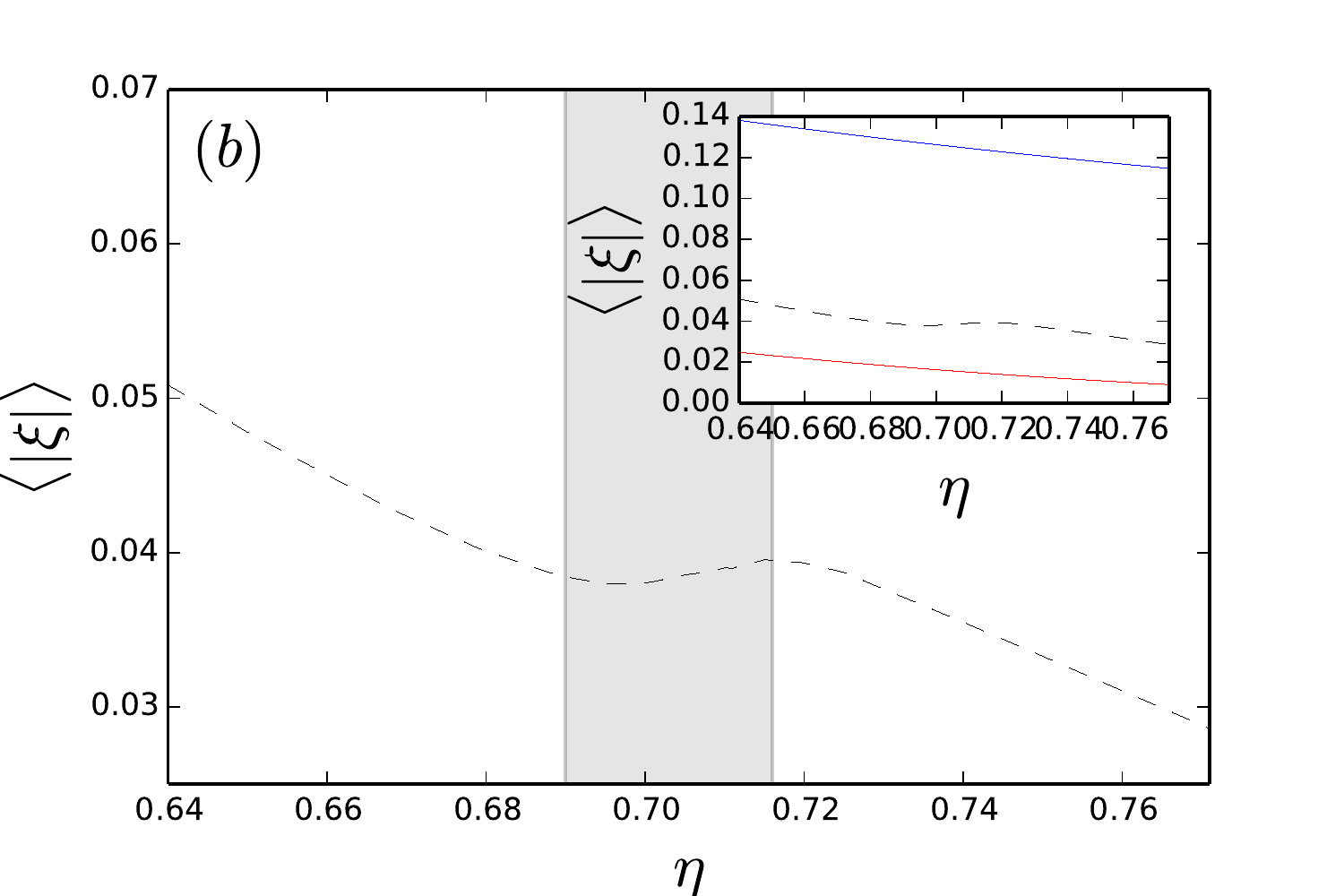}
\caption{(Color online) Average free flight time (a) and average free path  (b) shown as a function of $\eta$. It is seen that they both exhibit a plateau in correspondence of the coexistence phase $0.69\leq \eta\leq 0.716$ \cite{truskette1998} (blurred regions), while they decay monotonically in the fluid and solid phases. Insets: blue curves show the kinetic theory predictions (Eqs.(\ref{app:tau_kinetic_theory}) and (\ref{app:xi_relation_kinetic_theory_new}) respectively), while red curves refer to the Enskog theory Eqs.(\ref{app:tau_enskog_theory}) and (\ref{app:xi_enskog_theory}).}
\label{fig:figure_4}
\end{figure}

 The characteristic free-flight time is defined as
$\langle\tau\rangle=\int_{0}^{\infty}d\tau\, P(\tau)$. Substituting Eq.~(\ref{eq:free_flight_prob_def_vect}) or (\ref{eq:free_flight_prob_def}) yields $\langle\tau\rangle=\int_{-\infty}^{\infty}d\mathbf{v}\,\langle\tau\left(\mathbf{v}\right)\rangle \phi_{coll}(\mathbf{v})$ or $\langle\tau\rangle=\int_{-\infty}^{\infty}dv\,\langle\tau\left(v\right)\rangle \varphi_{coll}(v)$ accordingly. $\langle\tau\rangle$ is displayed in
Fig.~\ref{fig:figure_4}(a) for different $\eta$, spanning from the fluid
to the solid regime. Interestingly, we observe a plateau associated to the coexistence phase consistent to a  scenario where spatial regions with localized particles
(crystallites) can be observed in between regions of mobile particles  \cite{alder1962,lee1992,zollweg1992,Tobochnik1982,Bakker1984,hoover1968,huerta2006,mak2006,mayer1965,mitus1997,ryzhov1995,strandburg1984,weber1995,zollweg1992,binder2002,chui1983,kleinert1983,ramakrishnan1982}. 

The relation between the characteristic free flight time $\langle\tau\rangle$ and $\langle\tau\left(\mathbf{v}\right)\rangle $ follows~\cite{Visco2008}:
\be
\langle\tau\left(\mathbf{v}\right)\rangle\phi_{\text{coll}}(\mathbf{v})=\langle\tau\rangle\phi_{MB}(\mathbf{v}).
\label{eq:time_relation_visco}
\ee
Changing to polar coordinates and integrating over the angle yields~\cite{Lue2005}:
\be
\langle\tau\left(\mathbf{v}\right)\rangle f_{\text{coll}}\left(\left|\mathbf{v}\right|\right)=\langle\tau\rangle f_{MB}\left(\left|\mathbf{v}\right|\right),
\label{eq:time_relation_lue}
\ee
We find that the same equality is valid for the single components, as detailed in Appendix~\ref{app:A}:
\be
\langle\tau\left(v\right)\rangle\varphi_{\text{coll}}(v)=\langle\tau\rangle\varphi_{MB}(v).
\label{eq:time_relation}
\ee
Substituting in Eq.(\ref{eq:time_relation}) the integral expression of $\varphi_{\text{coll}}(v)$, and the numerical value of $\langle\tau\rangle$ (Fig.\ref{fig:figure_4}(a)) leads to a semi-analytic estimate of $\langle\tau\left(v\right)\rangle$, shown in Fig.~\ref{fig:figure_2}(b) to be in excellent agreement
with simulation data.

\begin{figure}[t]
\includegraphics[width=0.5\textwidth]{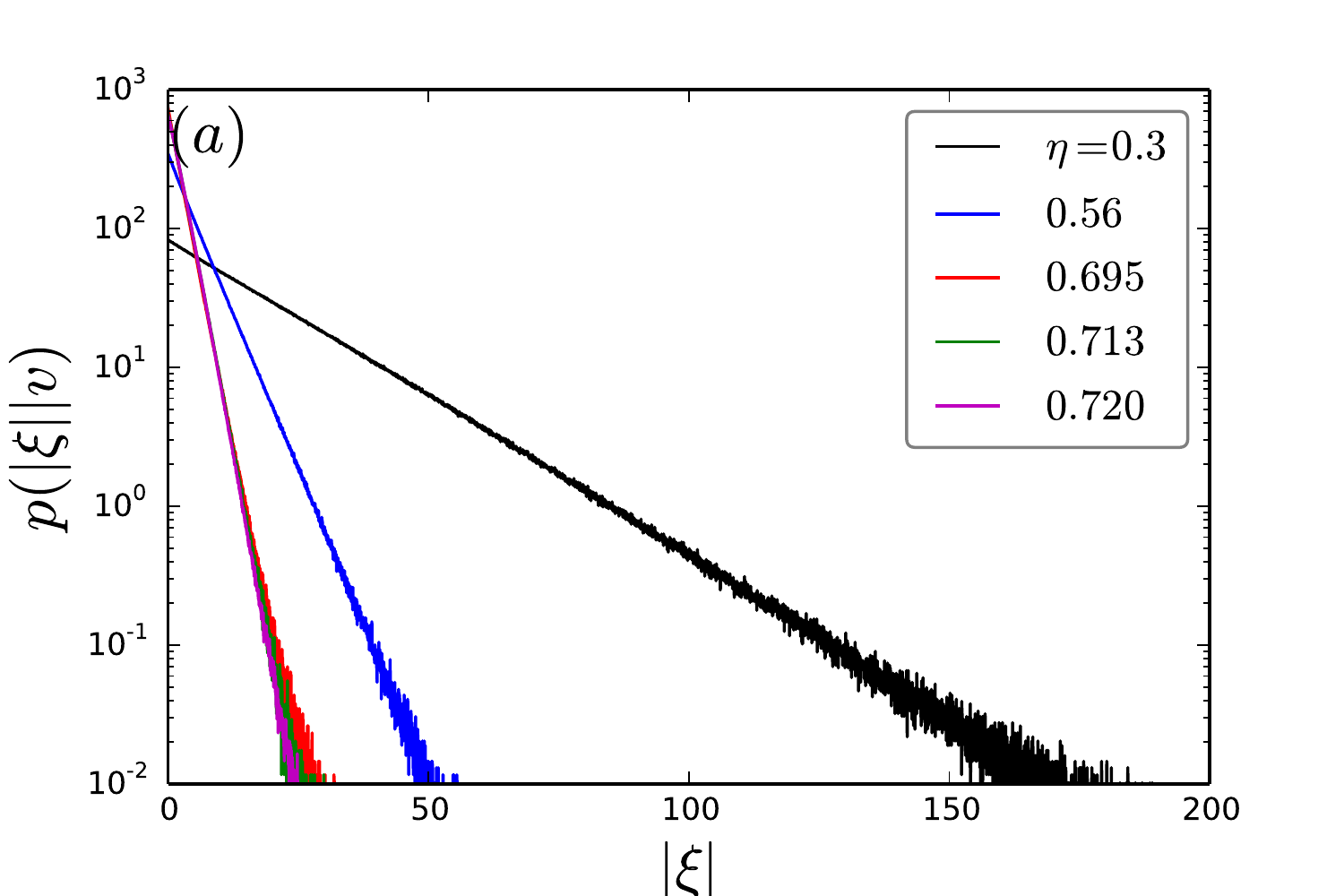}
\includegraphics[width=0.5\textwidth]{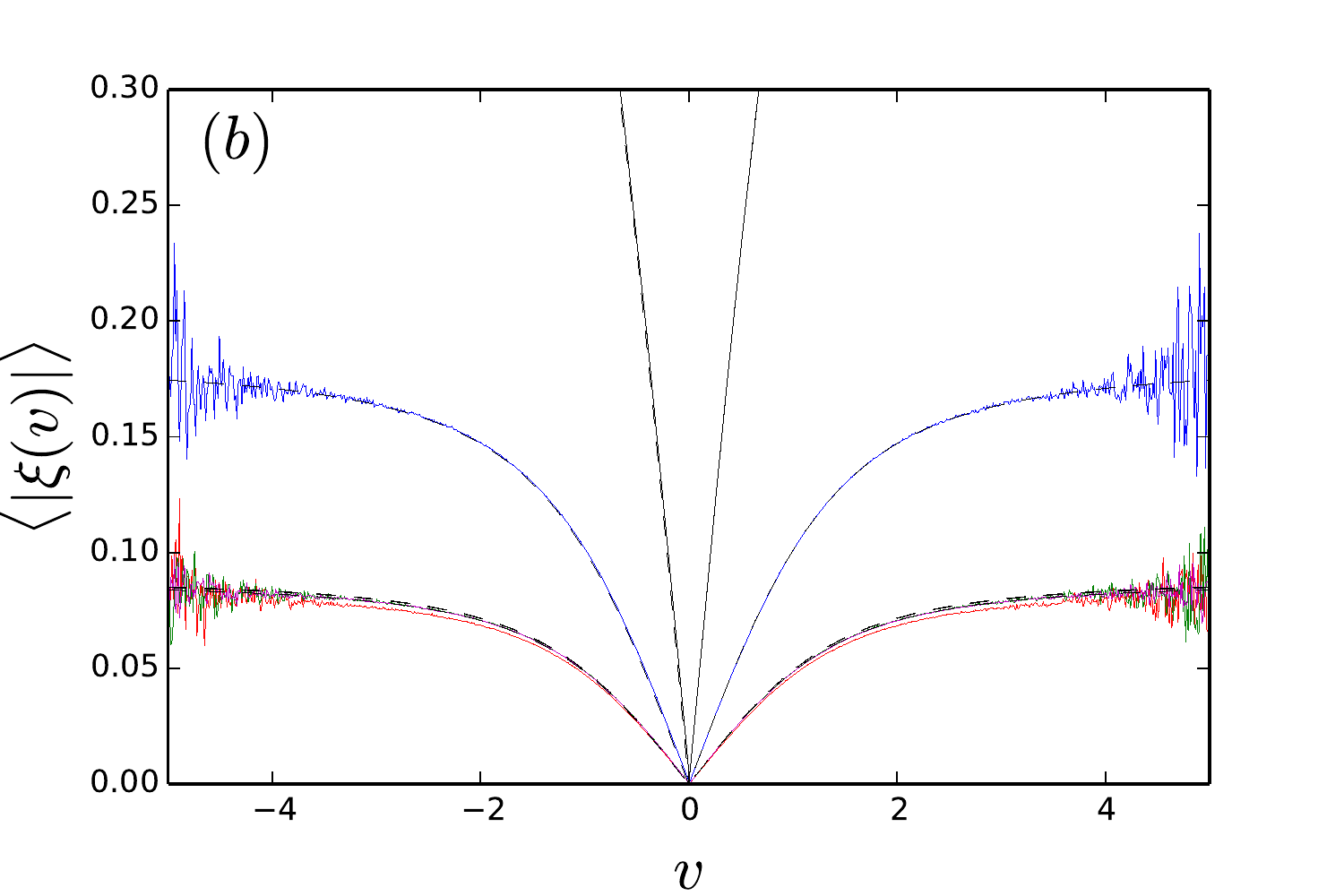}
\caption{(Color online) Dependence of free path length on velocity . (a) The probability distribution $p(|\xi||v)$ for different $\eta$ for $v=0.0\pm 0.05$ (arbitrary units). The exponential expression $\frac{e^{-|\xi(v)|/\langle|\xi(v)|\rangle}}{\langle|\xi(v)|\rangle}$ (dashed lines) describes the numerical outcomes only for the lower value of $\eta$, i.e. $\eta=0.3, 0.56$.   (b) $\langle|\xi|(v)\rangle$ as a function of the velocity for several PFs. The dashed lines are obtained by the relation $\langle|\xi(v)|\rangle=|v|\langle\tau(v)\rangle$.}
\label{fig:figure_5}
\end{figure}

\subsection{Free-path length distribution function}

The  free-path length $r$ is defined as $\left|\boldsymbol{\xi}\right|=\left|\mathbf{v}\right|\tau$, and the single component obeys $\left|\xi\right|=\left|v\right|\tau$. In this sub-section we analyze the single component free-path length distribution function $P(\left|\xi\right|)$,
displayed in Fig.~\ref{fig:figure_3}(b) for several values of $\eta$. In step with the analysis made
 for the free-flight distribution function in the previous sub-section,
 we can define a conditional free-path length distribution function as~\cite{Visco2008}:
\be
P\left(r\left|\mathbf{v}\right.\right)=\int d\tau\, \delta\left(\frac{r}{\left|\mathbf{v}\right|}-\tau\right)\frac{P(\tau|\mathbf{v})}{\left|\mathbf{v}\right|}.
\label{eq:free_path_cond_prob_def_whole}
\ee
and, for the single component:
\be
p(\left|\xi\right||v)=\int d\tau\, \delta\left(\frac{\left|\xi\right|}{|v|}-\tau\right)\frac{p(\tau|v)}{|v|}.
\label{eq:free_path_cond_prob_def}
\ee
Using Eqs.~(\ref{eq:free_path_cond_prob_def_whole}) and (\ref{eq:free_path_cond_prob_def}) together with
the analysis reported in the previous sub-section, it can be shown that in the low limit of $\eta$, $P\left(\left|\boldsymbol{\xi}\right|\left|\mathbf{v}\right.\right)$ and $p(\left|\xi\right||v)$ are both exponentials~\cite{Visco2008}. 
At the same time, a marked deviation from the exponential form is expected for high $\eta$ (as shown
in Fig.~\ref{fig:figure_5}(a)). The conditional mean free-path, given a velocity $\mathbf{v}$, is given by $\langle \left|\boldsymbol{\xi(\mathbf{v})}\right|\rangle = |\mathbf{v}|\langle\tau(\mathbf{v})\rangle$, and $\langle \left|\xi\right(v)|\rangle = |v|\langle\tau(v)\rangle$  for what concerns the single component reported in Fig.~\ref{fig:figure_5}(b). 
Expressing $\langle \left|\xi(v)\right|\rangle$ in terms of $\langle\tau(v)\rangle$ allows the semi-analytical
evaluation of the mean free-path single component, in excellent agreement with the numerics.

The free-path length distribution function is defined as:
\be
P(\left|\boldsymbol{\xi}\right|)=\int_{-\infty}^{\infty}d\mathbf{v} \, P\left(r\left|\mathbf{v}\right.\right)\phi_{coll}(\mathbf{v}),
\label{eq:P_r}
\ee 
and has been the subject of an extensive numerical analysis in Ref.~\cite{Einwohner1968}. In this
early study it was shown that: $(i)$ $P\left(\left|\boldsymbol{\xi}\right|\right)$ is nearly exponential in the fluid and solid limit; $(ii)$ the kinetic theory predicts a very accurate mean free-path at all packing fractions, namely $\langle \left|\boldsymbol{\xi}\right|\rangle_{kt}=\frac{\pi\sigma}{2^{5/2}\eta}$; $(iii)$ the dimensionless free path distribution as a function of the rescaled free-path length, $\langle \left|\boldsymbol{\xi}\right|\rangle_{kt}P\left(\frac{\left|\boldsymbol{\xi}\right|}{\langle \left|\boldsymbol{\xi}\right|\rangle_{kt}}\right)$, exhibits a 
nearly universal behavior, independent of $\eta$; $(iv)$ the rescaled free-path length distribution 
does not agree well with the zero density scaled distribution $\langle  \left|\boldsymbol{\xi}\right|\rangle_{0}P\left(\frac{ \left|\boldsymbol{\xi}\right|}{\langle  \left|\boldsymbol{\xi}\right|\rangle_{0}}\right)$, where $\langle  \left|\boldsymbol{\xi}\right|\rangle_{0}$ is the mean free-path achieved from the Enskog theory
\cite{Enskog1922}. In step with Eq.(\ref{eq:P_r}), the single component path length distribution function is defined as:
\be
P(\left|\xi\right|)=\int_{-\infty}^{\infty}dv \, p(\left|\xi\right||v)\varphi_{coll}(v).
\label{eq:P_xi}
\ee 
Eqs.~(\ref{eq:P_r}) and~(\ref{eq:P_xi}) yield  non-exponential  free path distribution functions for all
$\eta$, once the expressions for $\phi_{coll}(\mathbf{v})$ and $\varphi_{coll}(v)$ are substituted (see Fig.~\ref{fig:figure_3}(b)). In the very dilute case this result coincides with the finding in Ref.\cite{Visco2008}.  
In Fig.~\ref{fig:figure_4}(b) we plot the average free-path with $\eta$ ranging from the fluid to
the solid phase: $\langle \left|\xi\right|\rangle=\int_0^{\infty}d\left|\xi\right|\, \left|\xi\right| P(\left|\xi\right|)$. As for the average free-flight time, the plateau can
be considered as the signature of the coexistence phase \cite{truskette1998}. In Appendix
\ref{app:kinetic_theory} the single component mean
free-path furnished by the kinetic theory is shown to be
$\langle\left|\xi\right|\rangle_{kt}=\frac{\sigma}{2^{7/2}\eta}$ (Eq.(\ref{app:xi_relation_kinetic_theory_new})) and in the inset of  Fig.~\ref{fig:figure_4}(b), one can see that
this estimate  reproduces just qualitatively the numerical
mean free-path $\langle\left|\xi\right|\rangle$, in contradiction to what has been
observed in Ref.~\cite{Einwohner1968}. As a consequence, no universal behavior is expected from the rescaled single component mean free path distribution $\langle\left|\xi\right|\rangle_{kt}P( \frac{\left|\xi\right|}{\langle\left|\xi\right|\rangle_{kt}} )$ (see Fig.\ref{fig:figure_A1}(a)). Moreover, this
rescaling does not provide a correct collapse of the numerical curves, neither when the kinetic theory estimate $\langle\left|\xi\right|\rangle_{kt}$ is replaced by the true $\langle\left|\xi\right|\rangle$, as displayed in  Fig.~\ref{fig:figure_A1}(b). In Appendix \ref{app:kinetic_theory} we also provide the mean free-path length estimate $\langle \left|\xi\right|\rangle_0$ which arises from the Enskog theory (see Eq.(\ref{app:xi_enskog_theory})). Plotting the rescaled free-path probability distribution by $\langle \left|\xi\right|\rangle_0$ corroborates the finding of Ref.~\cite{Einwohner1968}: the agreement of the data with the zero-density limit is indeed unsatisfactory (Fig.~\ref{fig:figure_A1}(c)). In the insets of both panels of Fig.~\ref{fig:figure_4} we report the Enskog theory prediction for the mean free flight time and the mean  free path, $\langle\tau\rangle_0$ and $\langle \left|\xi\right|\rangle_0$ respectively. The expressions furnished in Eqs.~(\ref{app:xi_enskog_theory}) and (\ref{app:tau_enskog_theory}) do not describe the observed behavior also in the limit of very dilute systems.

Finally,   defining  the mean free path as $\langle \left|\boldsymbol{\xi}\right|\rangle=\int_{-\infty}^{\infty}d\mathbf{v}\, \langle \left|\boldsymbol{\xi}(\mathbf{v})\right|\rangle \phi_{coll}(\mathbf{v})$ yields, in view of Eq.(\ref{eq:free_path_cond_prob_def_whole}) and Eq.(\ref{eq:time_relation_visco}), the
following relation:
\be
\langle \left|\boldsymbol{\xi}\right|\rangle=\langle|\mathbf{v}|\rangle_{\text{cont}}\langle\tau\rangle,
\label{eq:mean_free_path_avg}
\ee
or
\be
\langle \left|\xi\right|\rangle=\langle|v|\rangle_{\text{cont}}\langle\tau\rangle,
\label{eq:xi_avg}
\ee
for the single component. The same relation in Eq.~(\ref{eq:mean_free_path_avg}) is imposed 
within the framework of the kinetic theory to define the mean free-path (see Eq.~(\ref{app:xi_relation_kinetic_theory}) and Ref.\cite{Chapman1939}). Moreover it has been rigorously 
validated also in one-dimensional elastic rod systems, also known as Jepsen's gases \cite{marchesoni2007}. Indeed, as in 1D  the velocity distribution function is invariant upon elastic collisions \cite{lebowitz1967},  the Eq.~(\ref{eq:xi_avg})   has been shown to hold for any type of distributions $\varphi\left( v\right)$,  not only for the Maxwell-Boltzmann distribution $\varphi_{MB}\left(v\right)$.

%
%

 \section{Collisional representation: single particle dynamics}
Our analysis now turns to the dynamical properties of the single particle, or tracer. In particular we reinterpret the stochastic motion of a tagged particle in 
positional space $x_n$ and velocity space $v_n$ as a global collisional process, where the continuous time $t$ is now replaced by the collisional index $n$. Correspondingly, our analysis will extend to the study of the properties of  $\tau_n$. 

In what follows we will first consider the velocity autocorrelation function,  discuss the mean squared displacement (MSD) and the connection among them through the Green-Kubo relation. We then
continue to investigate autocorrelation functions of free flight path and free flight time. Lastly we complete the picture by discussing the intermediate scattering function and the self-part of the van Hove function, or the propagator.

\subsection{Velocity autocorrelation function \label{sec_ACF_vel}}

 \begin{figure}
 \begin{center}
 \includegraphics[width=0.5\textwidth]{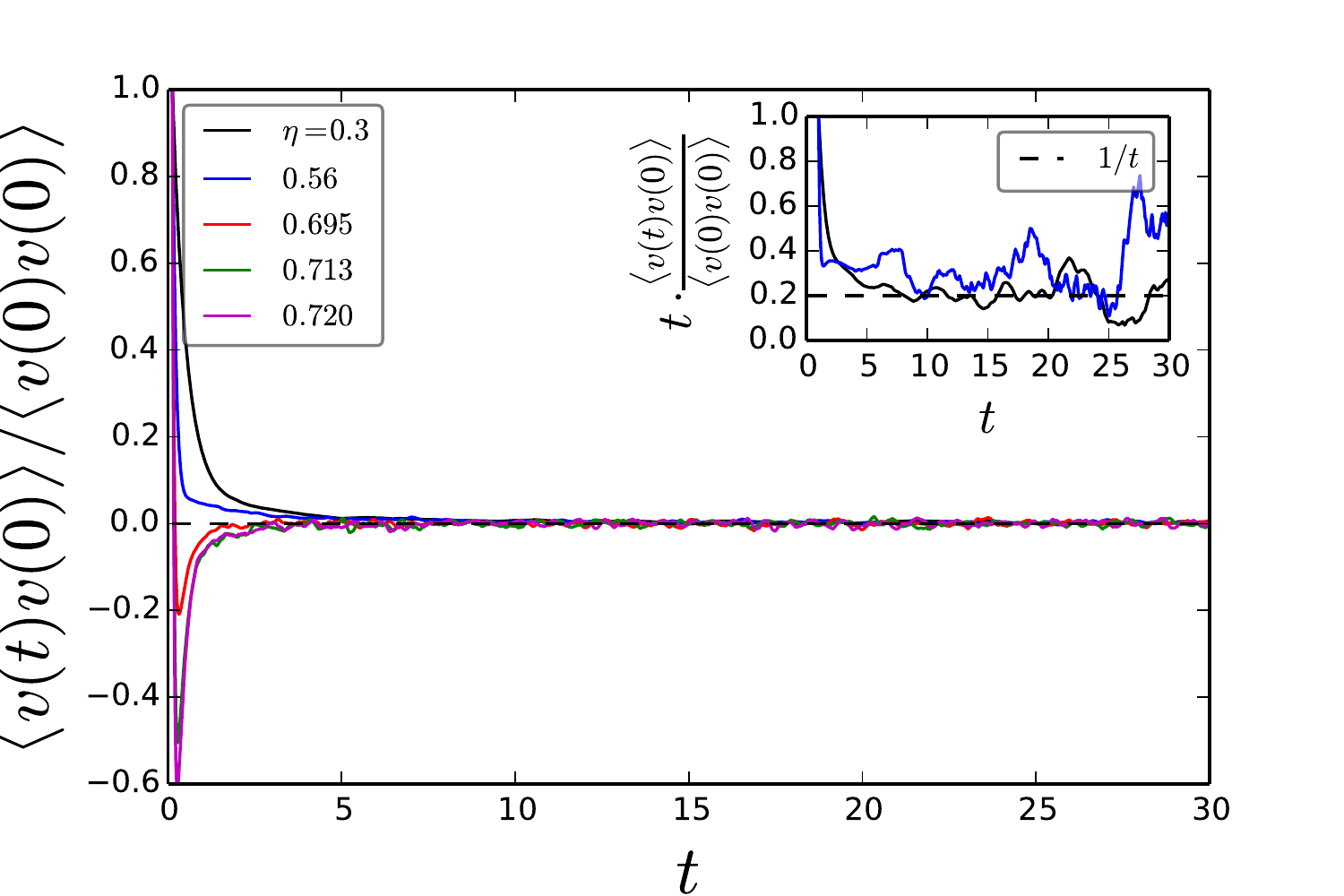}
 \includegraphics[width=0.5\textwidth]{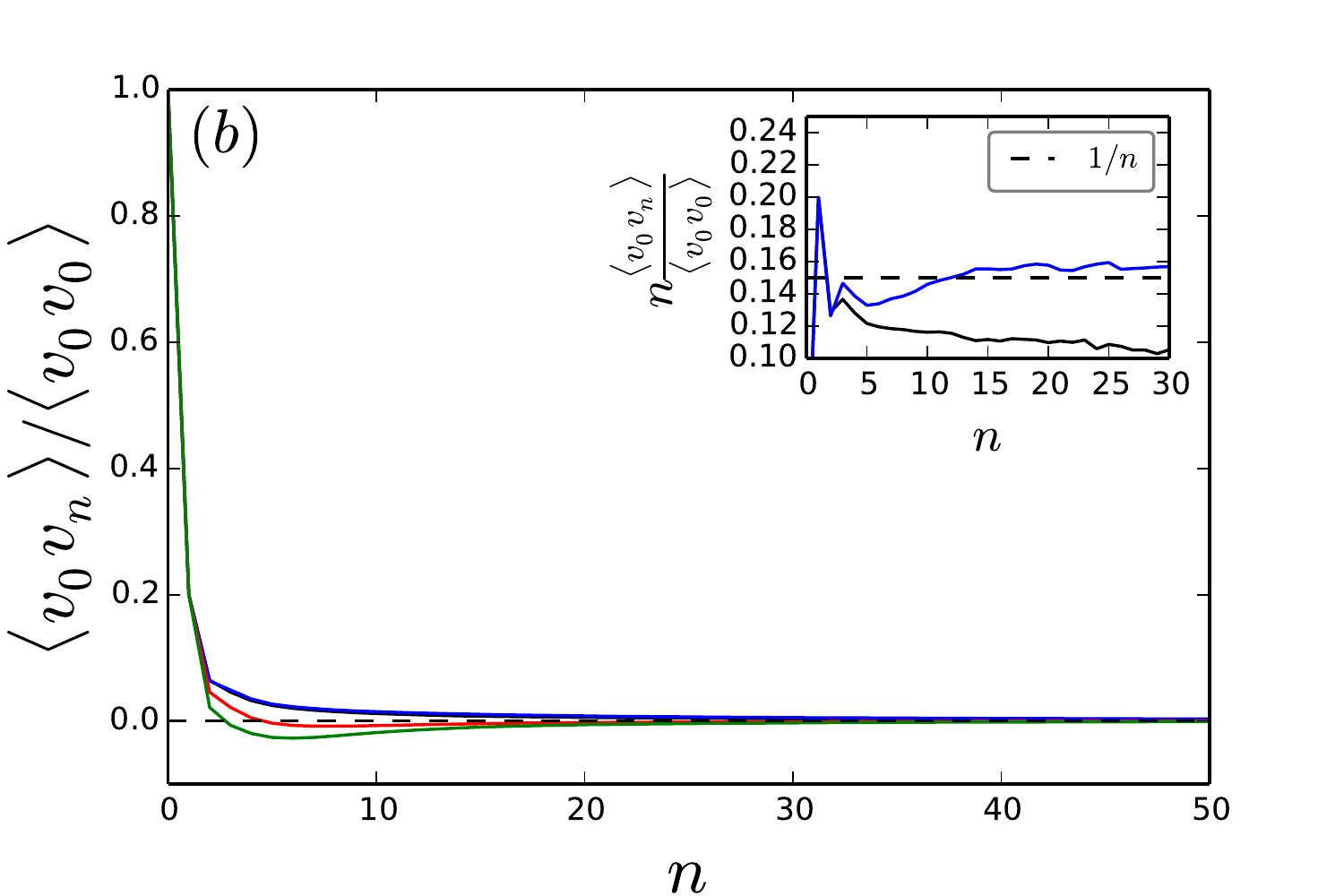}
  \end{center}
 \caption{(Color online) Normalized velocity ACF. (a) Continuous time. The velocity ACF exhibits persistent memory effects in the low PF limit and in the high PF region. The predicted $1/t$ behavior is evident in the inset, while the antipersistent tails characterize the caging effect for high $\eta$ (b) Collisional  representation. The velocity appears to be non-Markovian also in the collisional representation, showing a surprisingly analogy with the continuous time counterpart. In the inset the $1/n$ behavior appears to define the low $\eta$ systems, while the antipersistent memory effects dominate the high $\eta$ regime.}
 \label{fig:figure_6}
 \end{figure}

We wish to follow the argument of Markovianity and its intrinsic connection  to the  ansatz of molecular chaos in dilute systems. This issue is particularly important and compelling since it constitutes the fundamental hypothesis from which  the on-collision velocity distribution function in Eq.~(\ref{eq:phi_coll}) is derived, and which leads to the exponential shape of the free-flight time conditional probability distribution $P(\tau|\mathbf{v})$ in the low density limit. Thus, let us briefly recall what is the theoretical framework in Refs.~\cite{Puglisi2006,ViscoJCP2008,Visco2008} that hinges on this assumption. The diluteness of the system guarantees the molecular chaos ansatz, meaning that the evolution of the tracer velocity distribution function $\phi(\mathbf{v})$ is governed by a linear Boltzmann equation. This, in turn, leads to the Maxwell-Boltzmann distribution $\phi_{MB}(\mathbf{v})$ in the stationary regime. The analysis of the Boltzmann equation provided in Ref.\cite{Puglisi2006} clarifies that the tracer undergoes a sequence of collisions, performing a Markov process in velocity space, whose transition rates $W(\mathbf{v}|\mathbf{v}')$ are analytically calculated from the linear Boltzmann equation. Hence, the conditional characteristic free flight time $\langle\tau\left(\mathbf{v}\right)\rangle$ is obtained through the relation $\langle\tau\left(\mathbf{v}\right)\rangle^{-1}= \int d\mathbf{v}'W(\mathbf{v}|\mathbf{v}')$. Finally, substituting this in Eq.~(\ref{eq:time_relation_visco}) yields the full expression of $\phi_{coll}(\mathbf{v})$ in Eq.~(\ref{eq:phi_coll}). It is important to note that Eqs.~(\ref{eq:time_relation_visco}) and (\ref{eq:phi_coll}) are generally valid, since they must hold for any $\eta$, not only in the dilute limit $\eta\to 0$. This means that the ratio $\frac{\langle\tau\left(\mathbf{v}\right)\rangle}{\langle\tau\rangle}$ is a universal function of the speed $\Omega(\left|\mathbf{v}\right|)$, independent of the packing fraction of the system. In view of these simple considerations, it is critical to verify whether the molecular chaos and/or Markovian assumptions are necessary and verified conditions in the dilute limit. However, since we restrict our analysis to the dynamical properties of single particles, we will focus on testing the Markovian assumption by studying the behavior of the velocity autocorrelation function (ACF) in collisional representation: $\langle v_0 v_n \rangle$. Furthermore, the Markov property leads to the exponential shape of $P(\tau|\mathbf{v})$ in the low $\eta$ limit, as is reported in Refs.~\cite{ViscoJCP2008,Visco2008} and  shown in Fig.~\ref{fig:figure_2}(a).

It is instructive to compare the dependence of the velocity ACF $\langle v_0 v_n \rangle$ on the collision index $n$ to the dependence of its continuous time counterpart $\langle v(0) v(t) \rangle$ on $t$.
In continuous time, velocity is known to be a non-Markovian process for any value of $\eta$. The pioneering numerical work of Alder and Wainwright~\cite{alder1967, alder1970} in the dilute limit has shown that $\langle v(0) v(t) \rangle\sim t^{-1}$ for long times. The discovery of these persistent memory effects greatly influenced the further development of nonequilibrium statistical physics of liquid states, and it is often referred to as the "2D long-time-tail problem''~\cite{dorfman1977}. Several theories based on mode-coupling theory~\cite{kawasaki1970}, nonequilibrium statistical mechanics~\cite{ernst1970} and a kinetic approach~\cite{dorfman1970} have been devised to explain this surprising numerical finding,  marking, in the latter case,  the birth of the ``Modern Era'' of kinetic theory (see also the partial review in~\cite{pomeau1975}). However, the exact form of the long-time tail is not fully understood~\cite{petrosky1999}: while an earlier result~\cite{deschepper1977} estimated the first correction to the $t^{-1}$ tail as $\ln(t/t_0)$, mode coupling theories~\cite{kawasaki1971, wainwright1971} and the renormalization group approach~\cite{forster1977} have led to the prediction that the $t^{-1}$ decay corresponds to intermediate times, while at longer times the asymptotic decay takes the form of $(t\sqrt{\ln t})^{-1}$. Extensive numerical simulations carried out using a direct approach~\cite{erpenbeck1982},  a cellular automaton lattice gas model~\cite{frenkel1989,naitoh1990,van1991,van1990,naitoh1991} and an event-driven direct hard-disk simulation scheme~\cite{isobe2008} provided the evidence in favor of the $(t\sqrt{\ln t})^{-1}$ long time behavior. 

In Fig.~\ref{fig:figure_6}(a) we report the single component velocity ACF $\langle v(0)v(t)\rangle$ for different $\eta$. In the inset one can appreciate the asymptotic $t^{-1}$ behavior for the lower $\eta$, while the accuracy of our event-driven simulations cannot clearly distinguish the predicted $(t\sqrt{\ln t})^{-1}$ for larger times.  For higher $\eta$, well inside the coexistence region and the solid phase, it is possible to see how the velocity ACF exhibits negative antipersistent  tails. This antipersistence is due to the backscattering resulting from the ``caging effect'', where particles
are confined to transient cages, resulting in temporal anticorrelations in particle displacements. 
The cage effect has been observed in colloidal systems close to the glass transition~\cite{weeks2000,weeks2002,nagamanasa2011,donati1999,hurley1995}, supercooled liquids~\cite{berthier2004,Gotze1992,chaudhuri2007,donati2002,glotzer2000,kob1993,lavcevic2003,larini2007,gotze1998,toninelli2005,niss2010} or granular systems close to the jamming transition~\cite{Keys2007,dauchot2005,reis2007}, and mainly connected to a reduction of the displacements of the particles trapped in the cages ~\cite{huerta2006,huerta2014}.
Here we claim that velocity anticorrelations might reveal the presence of solid-like structures, characterizing the solid-fluid coexistence phase in hard-disks systems.  
We note that the focus here is not on the continuous time behavior summarized in Fig.~\ref{fig:figure_6}(a), which is meant only to provide a confirmation of the overall behavior in continuous time, but rather on the behavior of the velocity ACF in collisional representation. In Fig.~\ref{fig:figure_6}(b) we report $\langle v_0 v_n \rangle$ for several $\eta$. The inset shows that in the dilute limit $\langle v_0 v_n \rangle\sim n^{-1}$ for large times. The main panel displays the progressive approach from the persistent to the antipersistent regime  as $\eta$ increases. These numerical  results highlight the complete analogy between the  asymptotic behavior exhibited by the velocity ACF in 
 the continuous time and the collisional representations
for any $\eta$. Most importantly, it demonstrates that the Markovian assumption for velocity in collisional representation is never fulfilled. In particular, surprisingly, it does not hold in the limit $\eta\to 0$, revealing that the Markov property is not a necessary condition  for  Eq.~(\ref{eq:phi_coll}) to be valid. Nonetheless this is puzzling, since  the exact expression provided by the  Eq.~(\ref{eq:phi_coll}) is derived by assuming the single particle collisional velocity process to be Markovian in the dilute limit.

 \begin{figure}
 \begin{center}
 \includegraphics[width=0.5\textwidth]{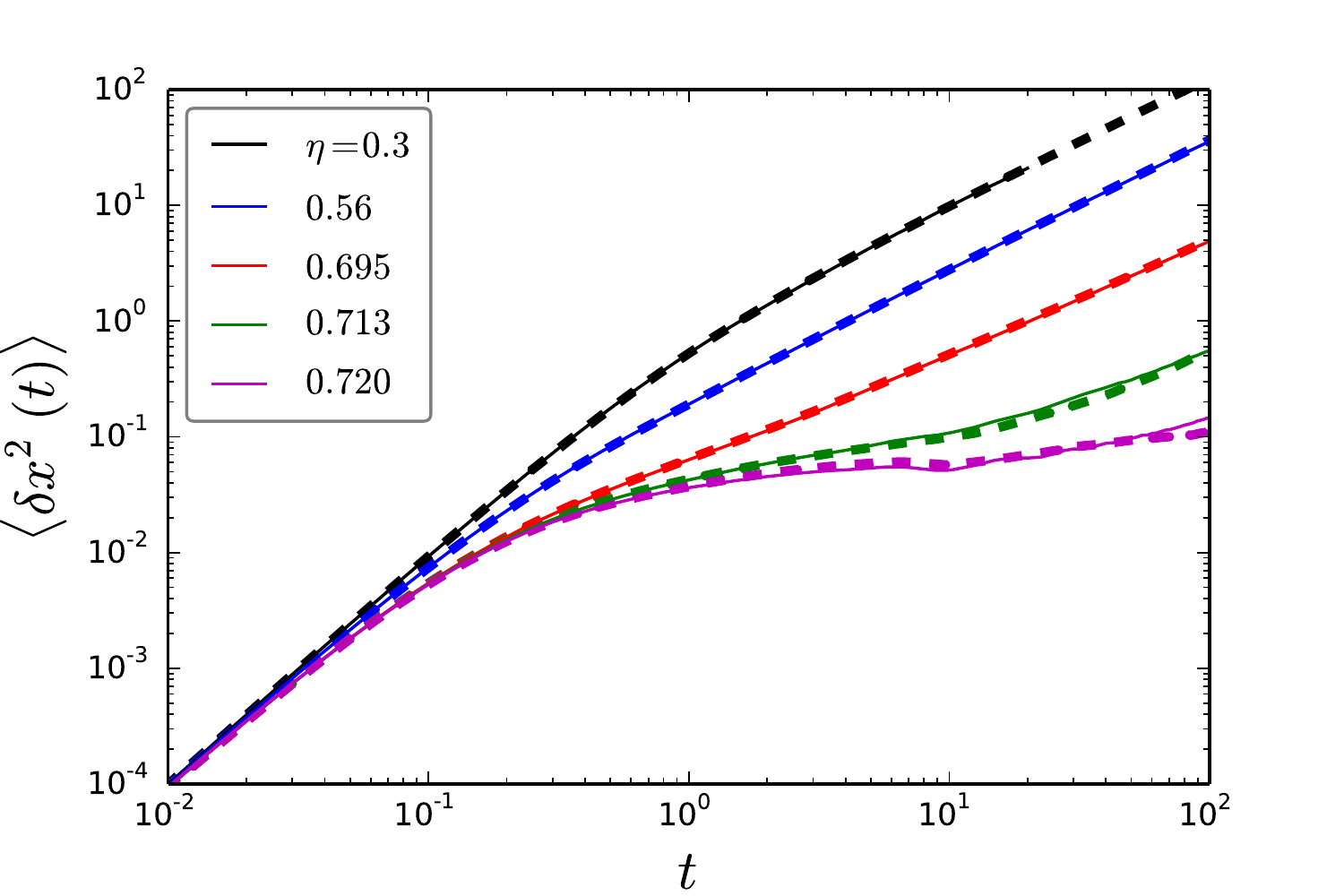}
 \includegraphics[width=0.5\textwidth]{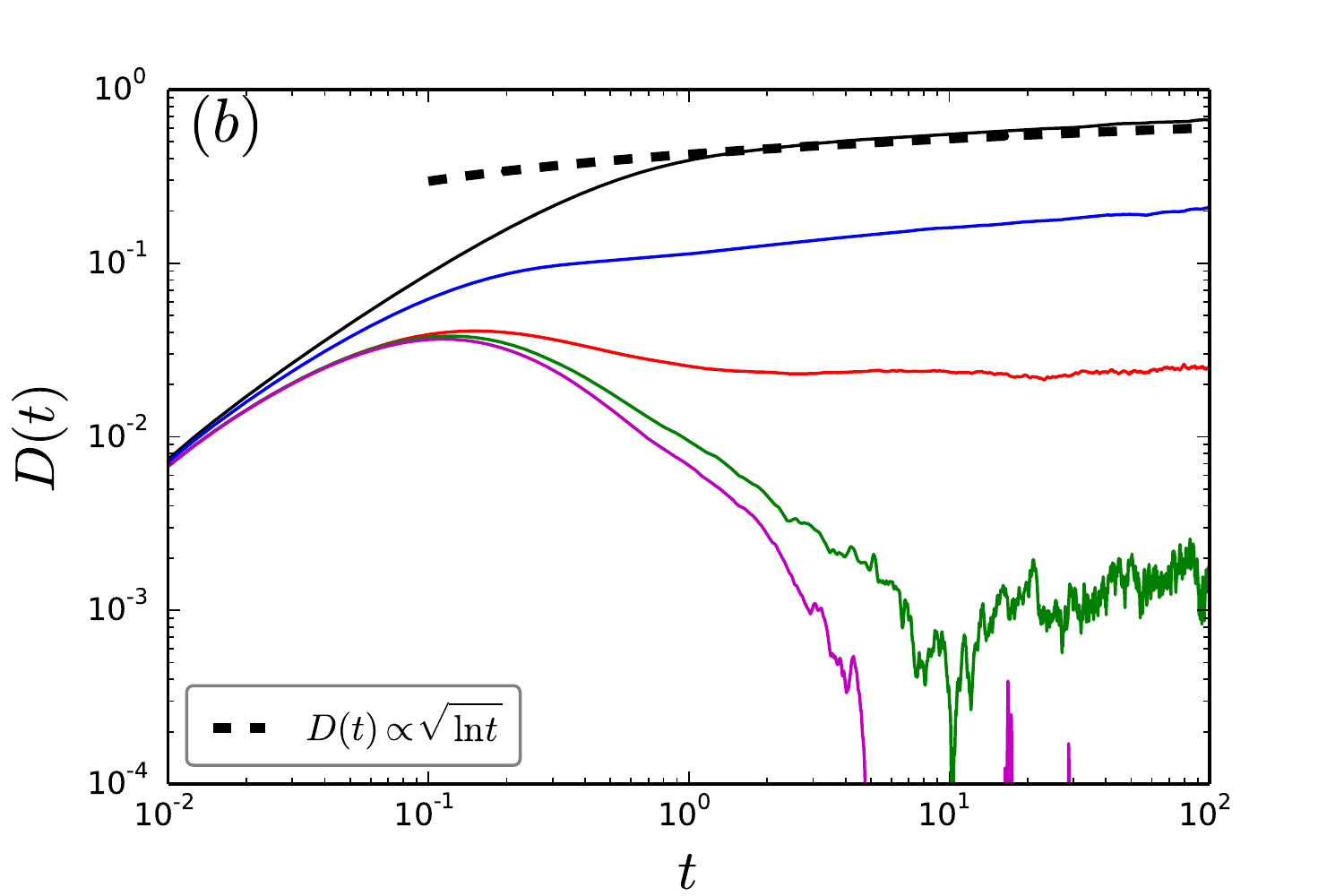}
  \end{center}
 \caption{(Color online)  (a) MSD in continuous time for several $\eta$. The solid lines refer to the Einstein-Helfand formula (\ref{eq:MSD_ein_ct}) while the dashed ones are obtained by the Green-Kubo expression (\ref{eq:MSD_kubo_ct}). They both coincide yielding, after an initial ballistic regime,  normal diffusion for low PF and an intermediate slowdown (plateau) for PF in the coexistence phase and beyond. (b) Time-dependent transport coefficient defined as $D(t)=\int_0^tds\langle v(0)v(s)\rangle$. For the lowest analyzed $\eta$, $\eta=0.3$, $D(t)$ increases as $\sim\sqrt{\ln t}$ consistent with the predicted $\langle v(0)v(t)\rangle\sim (t\sqrt{\ln t})^{-1}
$ \cite{kawasaki1971, wainwright1971,forster1977}.  For the higher PF  shown, $D(t)$ displays a ``subdiffusive'' decay  in correspondence of the caging regime in panel (a) (see Ref.\cite{marchesoni2006}).}
 \label{fig:figure_7}
 \end{figure}

Finally let us compare the  results obtained in 2D with the properties of Jepsen gas, mentioned earlier. In 1D, the velocity process in collisional representation (i.e. $\langle v_0v_n\rangle$) is Markovian only in systems with bimodal (or dichotomic) velocity distributions, i.e.  $\varphi(v)=\frac{1}{2}\left[\delta(v-c)+\delta(v+c)\right]$ where $\delta(x)$ stands for the Dirac's delta function. The on-collision velocity process is non-Markovian for systems with any other distribution function, including $\varphi_{MB}(v)$~\cite{marchesoni2007}. On the other hand, in continuous time, the behavior of $\langle v(0)v(t)\rangle$ in 1D considerably differs from the collisional representation, in contrast to what we have shown in 2D. For bimodal velocity distribution functions the continuous time velocity ACF is strictly exponential~\cite{lebowitz1967,balakrishnan2002,marchesoni2007}, while 
weak memory effects of the form $\langle v(0)v(t)\rangle\sim -t^{-3}$ result from
velocity distribution functions such as Maxwell-Boltzmann ~\cite{jepsen1965,lebowitz1967,balakrishnan2002,marchesoni2007}, uniform $\varphi(v)=\frac{1}{2c}\theta(v+c)\theta(v-c)$~\cite{balakrishnan2002} or $\varphi(v)=\frac{c^2}{2}(c^2+v^2)^{3/2}$~\cite{lebowitz1967}, where $\theta(x)$ and $c$ represent the Heaviside's function and a constant respectively. For velocity distribution functions like the three-modal  $\varphi(v)=\mu\delta(v)+\frac{1-\mu}{2}\left[\delta(v-c)+\delta(v+c)\right]$~\cite{balakrishnan2002,marchesoni2007} or four-modal $\varphi(v)=\frac{\mu}{2}\left[\delta(v-c_1)+\delta(v+c_1)\right]+\frac{1-\mu}{2}\left[\delta(v-c_2)+\delta(v+c_2)\right]$~\cite{marchesoni2007} ($0<\mu<1$, $c_1$ and $c_2$ constants), the memory effects are stronger: $\langle v(0)v(t)\rangle\sim -t^{-3/2}$. Based on the numerical evidence, it was argued in Ref.~\cite{marchesoni2007} that the non-exponential behavior  observed in the velocity ACF in continuous time are ascribable only to the non-Markovian nature of the underlying collisional mechanism. This conjecture seems to be fulfilled also in 2D.

\subsection{Mean squared displacement and free path autocorrelation function \label{sec:MSD}}

 \begin{figure}
 \begin{center}
 \includegraphics[width=0.5\textwidth]{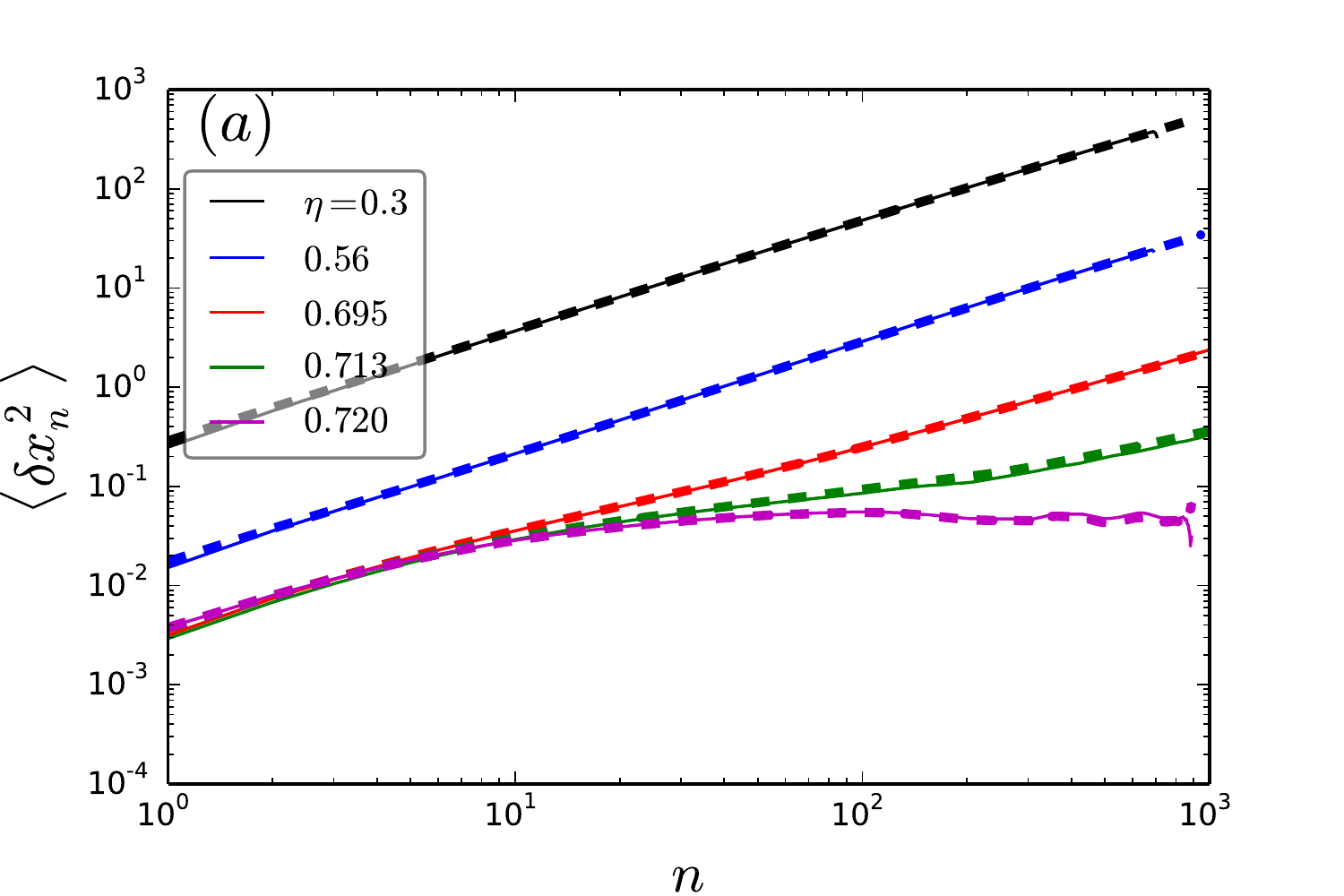}
 \includegraphics[width=0.5\textwidth]{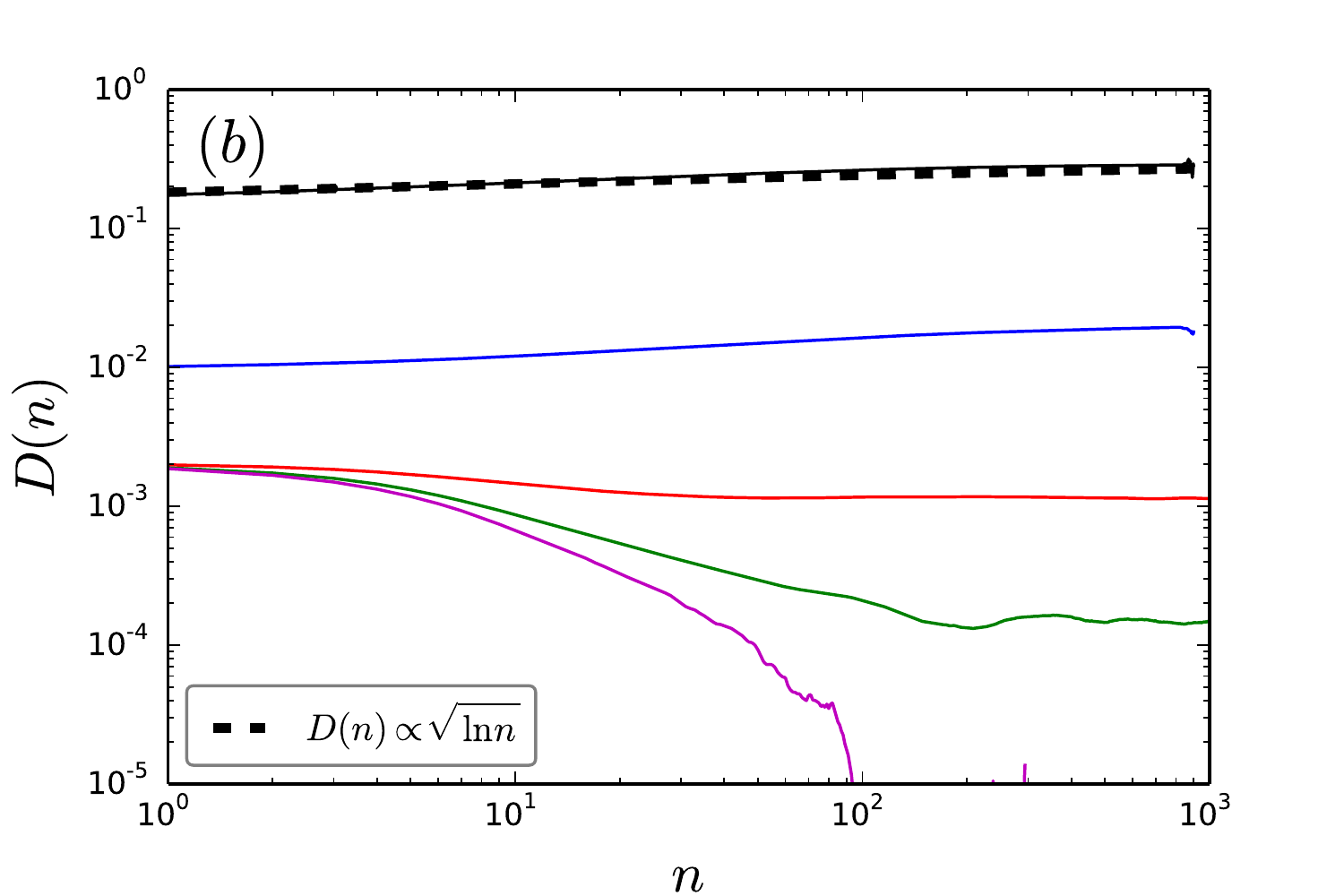}
  \end{center}
 \caption{(Color online)  (a) MSD in collisional representation for several $\eta$. The solid lines refer to the Einstein-Hellfand formula (\ref{eq:MSD_ein_coll}) while the dashed ones are obtained by the Green-Kubo expression (\ref{eq:MSD_kubo_coll}). They both coincide, displaying normal diffusion for low PF and an intermediate slowdown (plateau) for PF in the coexistence phase and beyond, in analogy with the results shown in Fig.\ref{fig:figure_7}(a). (b) Transport coefficient defined as $D_n=\sum_{m=0}^{n-1}C_{\xi\xi}(m)$. $D_n$ exhibits a persistent increasing behavior consistent with $\sim \sqrt{n\ln n}$ for $\eta=0.3$. This is in agreement to the continuous time behavior displayed in Fig.\ref{fig:figure_7}(b) and with the power-law tails in Fig.\ref{fig:figure_9}. For the higher PF, $D_n$ shows the same  ``subdiffusive'' decay of $D(t)$ (Fig.\ref{fig:figure_7}(b)) in correspondence of the caging regime in panel (b) (see the detailed analysis in appendix \ref{app:theorem}).}
 \label{fig:figure_8}
 \end{figure}

We analyze the single particle mean squared displacement (MSD) in collisional representation and compare it to its counterpart in continuous time. For the sake of clarity, let us start by discussing the MSD in the common continuous time representation. The MSD is defined as
\be
\langle \delta x^2(t)\rangle=\langle \left[x(t)-x(0)\right]^2\rangle.
\label{eq:MSD_ein_ct}
\ee
We will be referring to Eq.~(\ref{eq:MSD_ein_ct}) as the Einstein-Helfand expression, in clear connection to the well-established Einstein-Helfand formula for the diffusion coefficient $D=\frac{1}{2}\lim_{t\to\infty}\frac{d\langle \delta x^2(t)\rangle}{dt}$~\cite{helfand1960,gass1969,resibois1977,allen1987,mclennan1989}. In Fig.~\ref{fig:figure_7}(a) we report the outcome of our numerical simulations for several $\eta$. After a transient  ballistic regime, it is clearly seen that systems characterized by a lower $\eta$ exhibit normal diffusion behavior, whereas, for higher $\eta$, the MSD reaches a plateau which eventually turns into a linear regime. This  plateau can be considered as further evidence of the caging  undergone by a tracer during the coexistence phase, in analogy with  colloids~\cite{donati1998,nagamanasa2011,weeks2000,weeks2002}, supercooled liquids~\cite{donati1998,kob1993,lavcevic2003,larini2007,niss2010}, and granular systems~\cite{Keys2007,reis2007}. Due to the stationarity of the velocity process, the MSD can be also written as 
\be
\langle \delta x^2(t)\rangle=2\int_0^t ds \langle v(0)v(s)\rangle (t-s),
\label{eq:MSD_kubo_ct}
\ee
referred to as the Green-Kubo MSD expression, in relation to the definition of the diffusion coefficient according to the Green-Kubo relation, $D=\int_0^\infty dt \langle v(0)v(t)\rangle$~\cite{zwanzig1965}. Fig.~\ref{fig:figure_7}(a) shows the excellent agreement between the MSD calculated according to the Einstein-Helfand expression from Eq.~(\ref{eq:MSD_ein_ct}) (solid lines) and that resulting from the Green-Kubo formula in Eq.~(\ref{eq:MSD_kubo_ct}) (dashed lines). 
The  logarithmic tail of the velocity ACF shall lead to the weak divergence  of the diffusion coefficient $D$ in the long-time limit ($D\sim[\ln(\infty)+const]^{1/2}$)~\cite{kawasaki1971, wainwright1971}. However, as stressed in Ref.\cite{rojo2006}, in spite of the enormous numerical effort devoted to the identification of the $ \langle v(0)v(t)\rangle$ logarithmic asymptotic regime, very little work has been dedicated to the study of its influence on the actual value of the diffusion coefficient~\cite{alderJCP1970,rojo2006}. Indeed, the  numerical simulations shown in Fig.~\ref{fig:figure_7}(a) seem to point out the absence of any correction to the linear behavior exhibited by the  MSD, in agreement with the results reported in Refs.~\cite{rojo2006,viscardy2003}. However, plotting the time-dependent transport coefficient $D(t)=\frac{1}{2}\frac{d\langle \delta x^2(t)\rangle}{dt}=\int_0^t ds \langle v(0)v(s)\rangle$, as  shown in Fig.~\ref{fig:figure_7}(b), shows a small but persistent increase within the timescale  corresponding to the linear region exhibited by the MSD, for which one would have expected a constant value.   For $\eta=0.3$ we observe $D(t)\sim \sqrt{\ln(t)}$, consistent with the velocity ACF form $\langle v(0)v(t)\rangle\sim \left(t\sqrt{\ln t}\right)^{-1} $, and in contradiction to the numerical finding already present in one of the earliest numerical works on this subject~\cite{alderJCP1970}.  For larger $\eta$, $D(t)$ exhibits a ``subdiffusive'' decay consistent with the appearance of negative tails of $\langle v(0)v(t)\rangle$ (shown in Fig.~\ref{fig:figure_6}(a)) and with the corresponding plateau observed in the MSD (shown in Fig.~\ref{fig:figure_7}(a)) for intermediate times; asymptotically it approaches a constant value, in line with the asymptotic diffusive regime.
The rigorous connection between anomalous diffusion and power-law tails exhibited by $\langle v(0)v(t)\rangle$ is provided by the generalization of Kubo's theorem in Ref.~\cite{marchesoni2006}.

 \begin{figure}
 \begin{center}
 \includegraphics[width=0.5\textwidth]{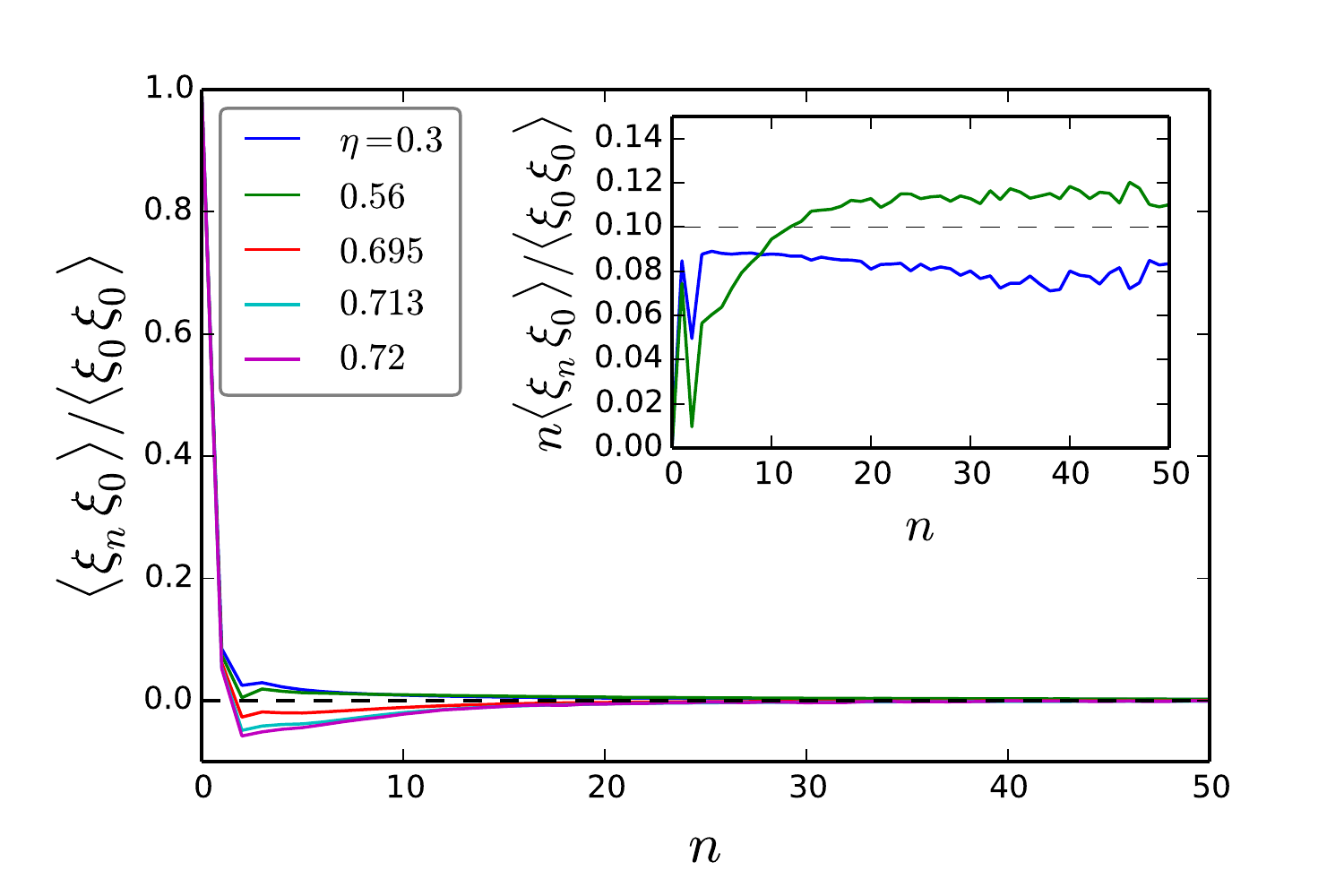}
  \end{center}
 \caption{(Color online)  Normalized free path length ACF. The non-Markovianity of the free path length is apparent both at low and high $\eta$. For low $\eta$ (inset) the $1/n$ behaviour is clearly shown, consistent with the velocity ACF in Fig.~\ref{fig:figure_6}(b). The systems with high $\eta$ are displayed in the main panel, exhibiting a negative part, i.e. antipersistence due to the caging phenomenon.}
 \label{fig:figure_9}
 \end{figure}

In collisional representation the analog of the Einstein-Hellfand MSD can be expressed as
\be
\langle  \delta x^2_n\rangle=\langle \left[x_n-x_0\right]^2\rangle
\label{eq:MSD_ein_coll}
\ee
and is reported in Fig.\ref{fig:figure_8}(a) (solid lines). For low $\eta$ the MSD undergoes normal diffusion in the collision index $n$. We note that in collisional representation the initial ballistic regime is suppressed by definition.  Increasing $\eta$, the same intermediate plateau as in continuous time  is observed. Since the single-component trajectory is expressed as 
\be
x_n-x_0=\sum_{l=0}^{n-1}\xi_l,
\label{eq:traj_coll}
\ee
and assuming the stationarity of the free-path process, one arrives at the Green-Kubo expression in collisional representation:
\be
\langle  \delta x^2_n\rangle= n\langle\xi^2\rangle+2\sum_{m=1}^{n-1}\langle \xi_0\xi_m\rangle(n-m),
\label{eq:MSD_kubo_coll}
\ee 
where $\langle\xi^2\rangle=\int_{-\infty}^\infty d\xi \xi^2 P(\xi). $ Hence, the numerical evaluation of Eq.~(\ref{eq:MSD_kubo_coll}) requires the calculation of the free-path ACF $\langle \xi_0\xi_n\rangle$, which is reported in Fig.~\ref{fig:figure_9}. The observed behavior resembles that of the velocity ACF. Indeed, for any $\eta$  the free-path process $\xi_n$ is non-Markovian, with persistent asymptotic tails $\propto n^{-1}$ for small $\eta$ (see the inset of Fig.~\ref{fig:figure_9}), and antipersistent  as $\eta$ increases. When calculated according to the Green-Kubo expression in Eq.~(\ref{eq:MSD_kubo_coll}), the MSD  (dashed lines in Fig.~\ref{fig:figure_8}(a)) is shown to coincide with the Einstein-Hellfand expression  in Eq.~(\ref{eq:MSD_ein_coll}). In analogy with the continuous time analysis, the asymptotic $n^{-1}$ regime should lead to the divergence of the MSD in Eq.~(\ref{eq:MSD_kubo_coll}). However, no appreciable deviation from the linear MSD trend  is detected, albeit in the range for which $\langle \xi_0\xi_n\rangle$  displays the $n^{-1}$ behavior. We now recast the Green-Kubo MSD expression in Eq.~(\ref{eq:MSD_kubo_coll}) by introducing the ``symmetrized'' free path autocorrelation function as $\tilde{C}_{\xi\xi}(n)=\langle \xi_0\xi_n\rangle-\frac{1}{2}\langle \xi^2\rangle\delta_{n,0}$~\cite{Taloni2006,marchesoni2006}:
\be
\langle  \delta x^2_n\rangle= 2\sum_{m=0}^{n-1}\tilde{C}_{\xi\xi}(m)(n-m).
\label{eq:MSD_kubo_coll_sym}
\ee 
We now identify the on-collision transport coefficient $D_n= \sum_{m=0}^{n-1}\tilde{C}_{\xi\xi}(m)$, and report it in Fig.\ref{fig:figure_8}(b). This quantity dictates the $n$-dependence of the MSD in collisional representation as $D(t)$ does for $\langle \delta x^2(t)\rangle$, detailed in the theorem in Appendix~\ref{app:theorem}. This theorem  can be viewed as the extension to impact dynamics of the generalized Kubo theorem valid in continuous time~\cite{marchesoni2006}. In particular when $D_n$ attains a constant value, one expects normal diffusion $\langle \delta x^2_n\rangle\sim n$  as displayed in Fig.\ref{fig:figure_8}(a) for low $\eta$. However, $D_n$ behaves as $\sim \sqrt{n\ln n}$ for the lowest packing fraction $\eta=0.3$, in complete analogy with the continuous time analysis reported in  Fig.\ref{fig:figure_7}(b), and in partial agreement with the observed persistent tails exhibited by the free path ACF $\langle \xi_0\xi_n\rangle$ (Fig.\ref{fig:figure_9}).
When $D_n$ exhibits a decay in the form of a power-law,  $\langle \delta x^2_n\rangle$  is sublinear in the collision index $n$. This is indeed the case for  intermediate $n$ with large  $\eta$, ultimately ascribable to the caging effect. 
 
\subsection{Free flight time autocorrelation function and free flight and free path cross correlation function \label{sec:FFT_ACF}}

 \begin{figure}
 \begin{center}
 \includegraphics[width=0.5\textwidth]{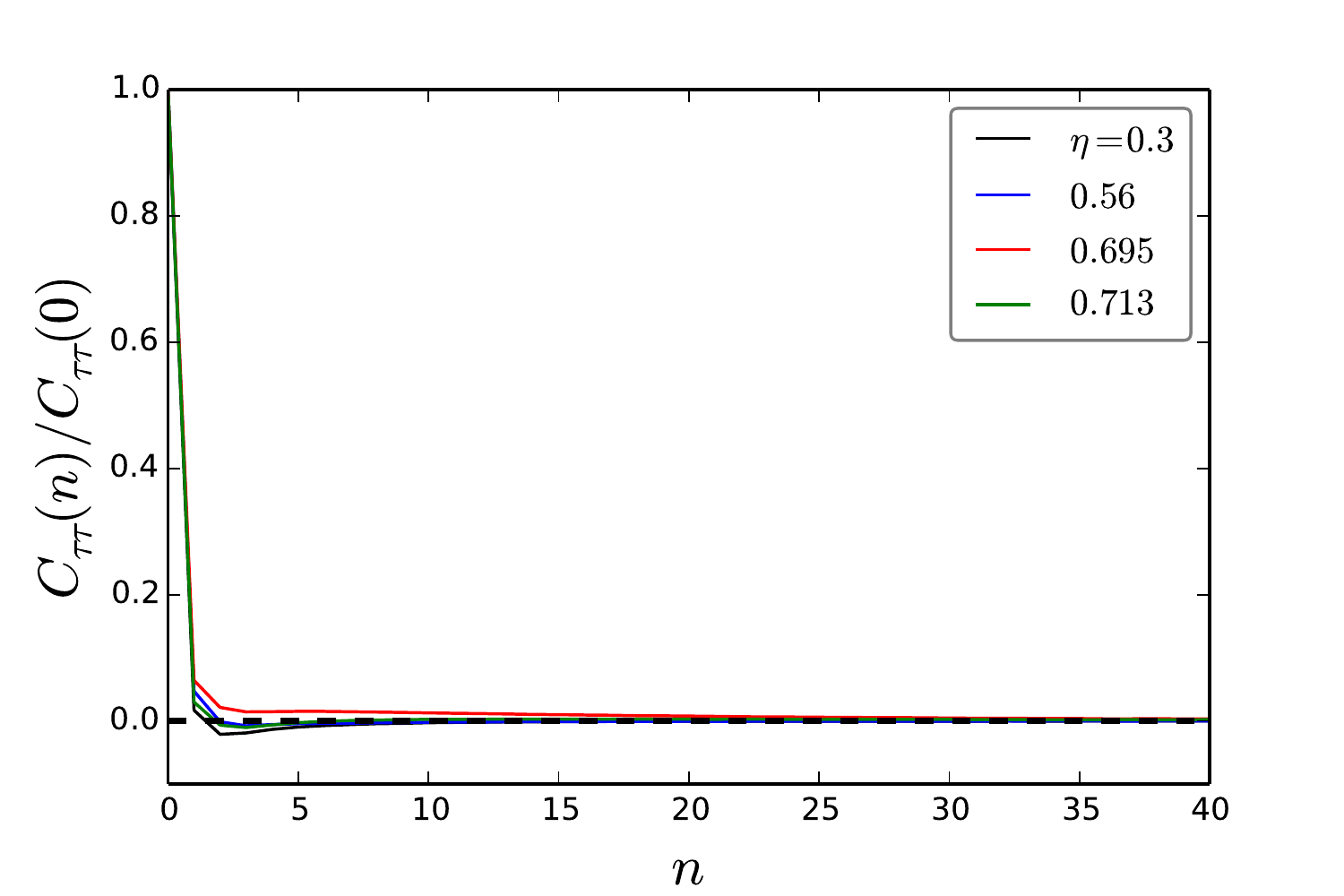}
  \end{center}
 \caption{(Color online)  Normalized free flight time ACF, $C_{\tau\tau}(n)=\langle \tau_0\tau_n\rangle -\langle\tau\rangle^2$. The non-Markovianity of the free flight time process $\tau_n$ characterizes both low and high $\eta$.}
 \label{fig:figure_10}
 \end{figure}

 Since in collisional representation space and time can be studied separately, we now focus on the free flight time autocorrelation function defined as $C_{\tau\tau}(n) = \langle\tau_0\tau_n\rangle -\langle\tau\rangle^2$, reported in Fig.~\ref{fig:figure_10} for different $\eta$. As for the free path $\xi_n$, the free flight time $\tau_n$ is a stationary process. 
For an uncorrelated  process one would expect $\langle \tau_0\tau_n\rangle -\langle\tau\rangle^2=\left[\langle \tau^2\rangle -\langle\tau\rangle^2\right]\delta_{n,0}$. However, the analysis in Sec.~\ref{sec:free_flight} demonstrates that the time elapsed between subsequent collisions is velocity-dependent;  this velocity, on the other hand, is a correlated process both at low and high packing fractions, as discussed in Sec.~\ref{sec_ACF_vel}. Therefore  correlations should also characterize the discrete dynamics of free flight times of single particle~\cite{Visco2008}, as indeed confirmed by the numerical outcomes  in Fig.~\ref{fig:figure_10}. The expression for the time ACF $C_{\tau\tau}(n)$ is formally defined through the correlation
\be
\begin{split}
\langle \tau_0\tau_n\rangle=\int d\tau_0\cdots d\tau_n\,dv_0\cdots dv_n\,\tau_0\tau_n \\
p(\tau_0,\cdots\tau_n|v_0,\cdots v_n)\phi_{coll}(v_0,\cdots v_n).
\end{split}
\label{eq:FFT_ACF_def}
\ee
Hence as long as times depend on velocities, the uncorrelation of the velocity process implies the uncorrelation of the free flight times collisional dynamics and vice versa. Indeed, if we assume the on-collision velocities to be uncorrelated, i.e. $\phi_{coll}(v_0,\cdots v_n)=\phi_{coll}(v)^n$, then we have $p(\tau_0,\cdots\tau_n|v_0,\cdots v_n)=p(\tau|v)^n$, yielding $\langle \tau_0\tau_n\rangle=\langle \tau^2\rangle$ and therefore $C_{\tau\tau}(n)=\left[\langle \tau^2\rangle -\langle\tau\rangle^2\right]\delta_{n,0}$.

Now, let us make the hypothesis that free flight times are independent of the velocities, i.e. $p(\tau_0,\cdots\tau_n|v_0,\cdots v_n)\equiv p(\tau_0,\cdots\tau_n)$. We want to discuss the implication of this simplification on the free path ACF. In general, since $\xi_n=v_n\tau_n$, the  free path  ACF is expressed as $\langle \xi_0\xi_n\rangle=\langle\tau_0\tau_n v_0v_n\rangle$, where
\be
\begin{split}
\langle \tau_0\tau_nv_0v_n\rangle=\int d\tau_0\cdots d\tau_n\,dv_0\cdots dv_n\,\tau_0\tau_n v_0v_n\\
p(\tau_0,\cdots\tau_n|v_0,\cdots v_n)\phi_{coll}(v_0,\cdots v_n).
\end{split}
\label{eq:FFT_vel_ACF_def}
\ee

Consistent to our assumption, we obtain  $\langle \xi_0\xi_n\rangle=\langle\tau_0\tau_n\rangle \langle v_0v_n\rangle$. Moreover, if we make the additional hypothesis that $\tau_n$ is an uncorrelated process, the free path ACF transforms to $\langle \xi_0\xi_n\rangle=\langle\tau\rangle^2 \langle v_0v_n\rangle$. Substituting this into Eq.~(\ref{eq:MSD_kubo_coll_sym}) we then obtain:

 \begin{figure}
 \begin{center}
 \includegraphics[width=0.5\textwidth]{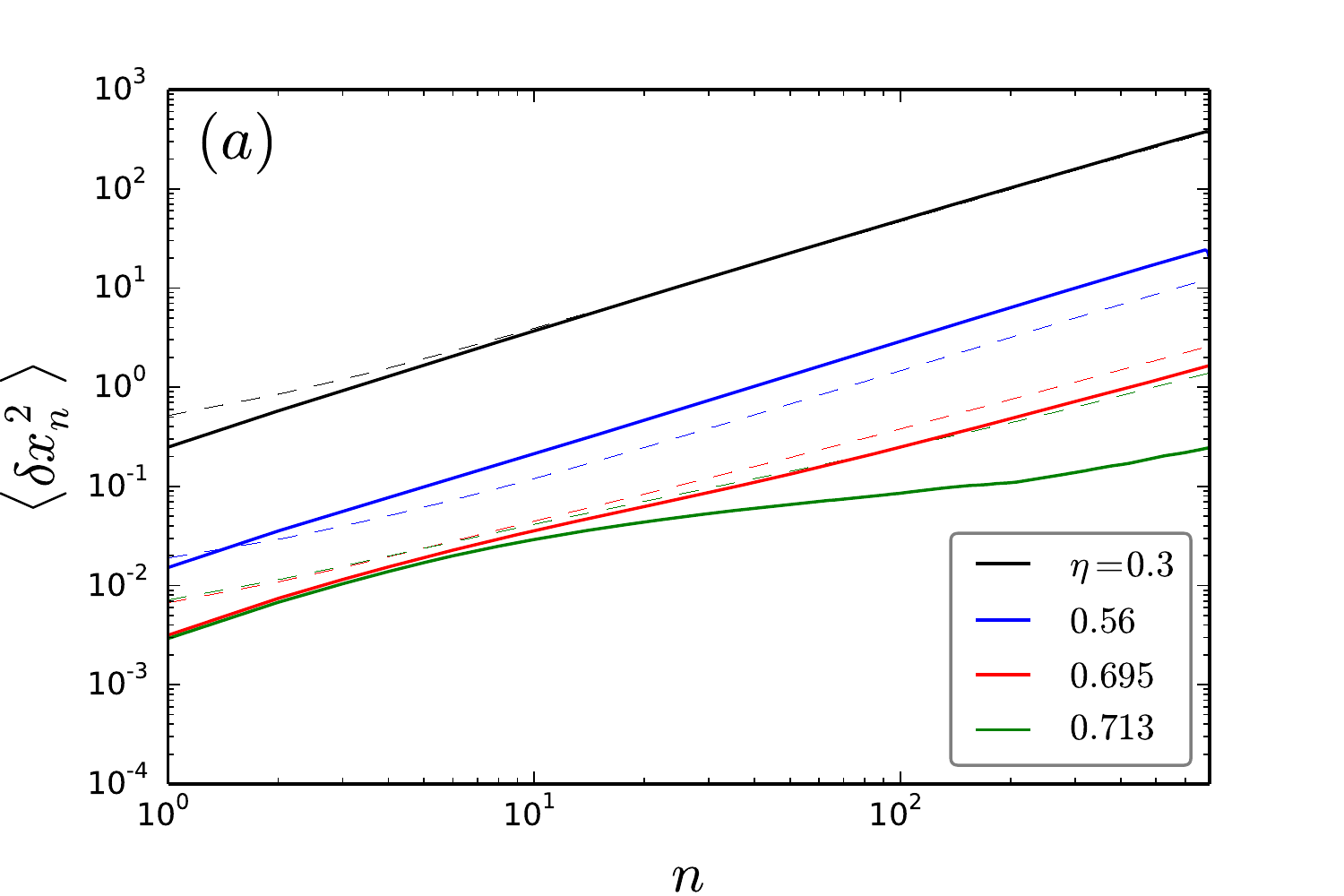}
 \includegraphics[width=0.5\textwidth]{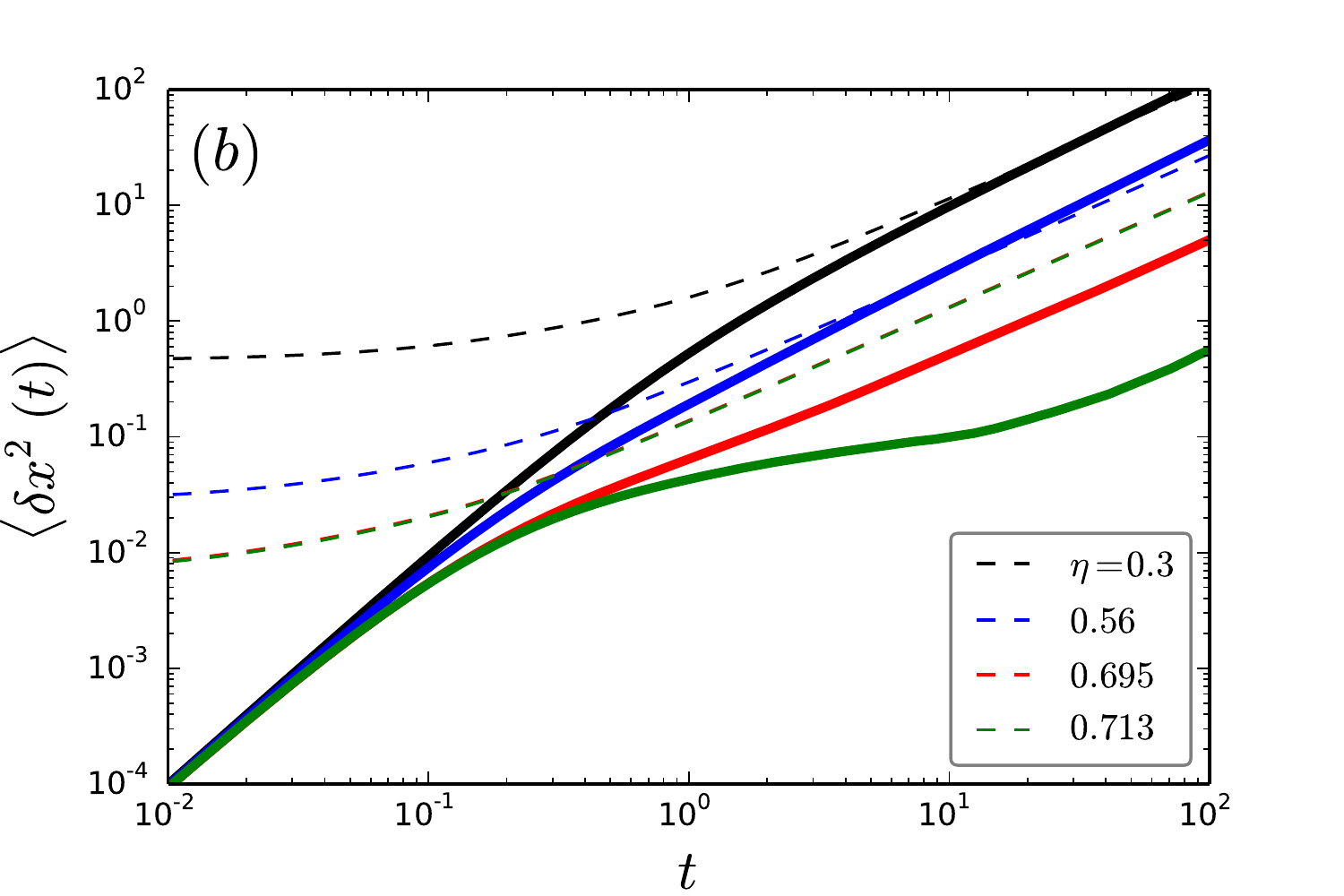}
  \end{center}
 \caption{(Color online)  Poisson approximated expression for the MSD. (a) MSD in collisional representation for several $\eta$ calculated according to the Poisson approximation in Eq.~(\ref{eq:MSD_kubo_coll_approx}) (dashed lines). The solid lines refer to the  Green-Kubo expression from Eq.~(\ref{eq:MSD_kubo_coll}),
coinciding with the Poissonian approximation only in the low $\eta$ limit.
(b) MSD in continuous time representation using the first two terms of the Poisson formula  appearing in Eq.~(\ref{eq:MSD_coll->ct_poiss_approx}) (dashed lines), 
and the Green-Kubo expression in Eq.~(\ref{eq:MSD_kubo_ct}) (solid lines). Here too the low $\eta$ systems appear to be well captured by the approximated formula. The high $\eta$ limit requires a complete determination of the correlations appearing in Eq.~(\ref{eq:MSD_coll->ct}).}
 \label{fig:figure_10_emmezzo}
 \end{figure}

\be
\langle  \delta x^2_n\rangle\simeq\langle \tau\rangle^2 \left[n\langle v^2\rangle+2\sum_{m=1}^{n-1}\langle v_0v_m\rangle(n-m)\right],
\label{eq:MSD_kubo_coll_approx}
\ee 
with $\langle\tau\rangle=\int_{0}^{\infty}d\tau P(\tau)\tau$ and $\langle v^2\rangle=\int_{-\infty}^{\infty}dv\phi_{coll}(v)v^2$. The former approximation corresponds to a Poisson process, as we will discuss in Sec.~\ref{sec:poisson}. We highlight that the only two assumptions made to define a Poisson process are i) independence between free flight times and velocities in collisional representation, ii) no memory effects characterizing the free flight collisional process. The numerical evaluation of the approximated MSD in Eq.~(\ref{eq:MSD_kubo_coll_approx}) is reported in Fig.~\ref{fig:figure_10_emmezzo}(a) and compared to the expression in Eq.~(\ref{eq:MSD_ein_coll}), showing a satisfactory agreement only in the low packing fraction limit. For more dense fluids,  the intermediate caging regime appears to be poorly captured by Eq.~(\ref{eq:MSD_kubo_coll_approx}). The negative antipersistent part of the free path ACF  (Fig.~\ref{fig:figure_9}), is indeed less pronounced in the approximation $\langle \xi_0\xi_n\rangle\simeq\langle\tau\rangle^2\langle v_0v_n\rangle$.

 \begin{figure}
 \begin{center}
 \includegraphics[width=0.5\textwidth]{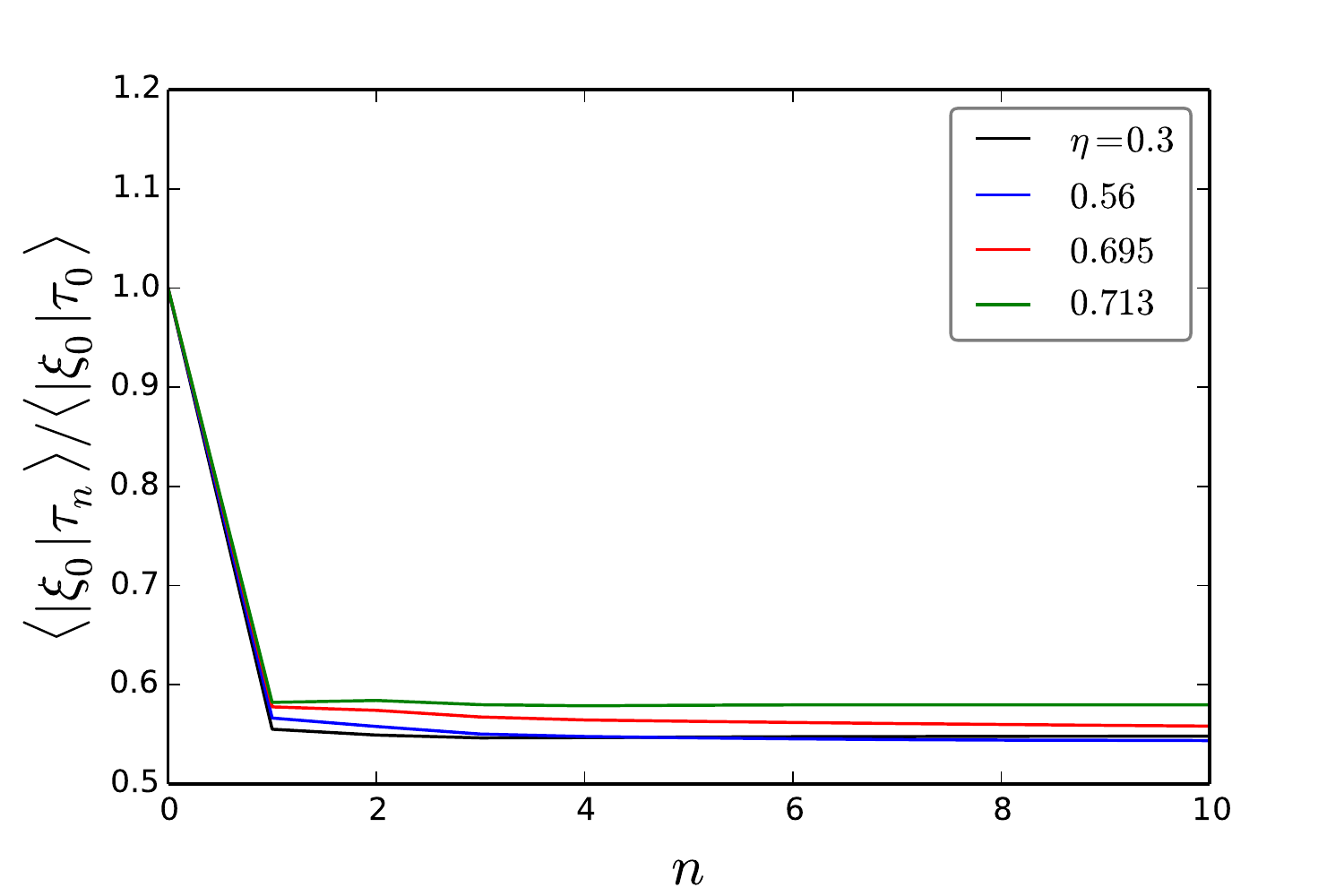}
  \end{center}
 \caption{Color online)  Normalized cross correlation of  free flight time and free path length.}
 \label{fig:figure_11}
 \end{figure}

We finally study the cross-correlation functions $\langle \xi_0\tau_n\rangle$ and $\langle \tau_0\xi_n\rangle$. Both vanish since $\langle\xi\rangle=0$, we therefore consider the absolute values, i.e. $\langle |\xi_0|\tau_n\rangle$ and $\langle \tau_0|\xi_n| \rangle$.  Fig.\ref{fig:figure_11}
shows $\langle |\xi_0|\tau_n\rangle$.  The first term can be analyzed in the low packing fraction limit, reading:
\be
\langle|\xi_0|\tau_0\rangle= \int_{-\infty}^{\infty} dv\, |v|\phi_{coll}(v)\int_0^{\infty}d\tau\, \tau^2 p(\tau|v)
\label{eq:cross_ACF_first}
\ee
where we dropped the index $0$ on the RHS, to simplify the notation.
The conditional probability $p(\tau|v)$ is exponential for low $\eta$ (see Fig.\ref{fig:figure_2}(a)), and substituting Eq.~(\ref{eq:time_relation}) we have:
\be
\langle|\xi_0|\tau_0\rangle= 2\langle\tau\rangle^2\int_{-\infty}^{\infty} dv\, |v|\frac{\varphi_{MB}^2(v)}{\varphi_{coll}(v)}.
\label{eq:cross_ACF_first_formula}
\ee
In 1D (Jepsen gas), where $\varphi_{coll}(v)\equiv\varphi_{MB}(v)$, one obtains  $\langle|\xi_0|\tau_0\rangle=2^{3/2}\langle\tau\rangle^2\sqrt{\frac{k_BT}{\pi}}$~\cite{marchesoni2007}.

\subsection{Intermediate scattering function}

 \begin{figure}
 \begin{center}
 \includegraphics[width=0.5\textwidth]{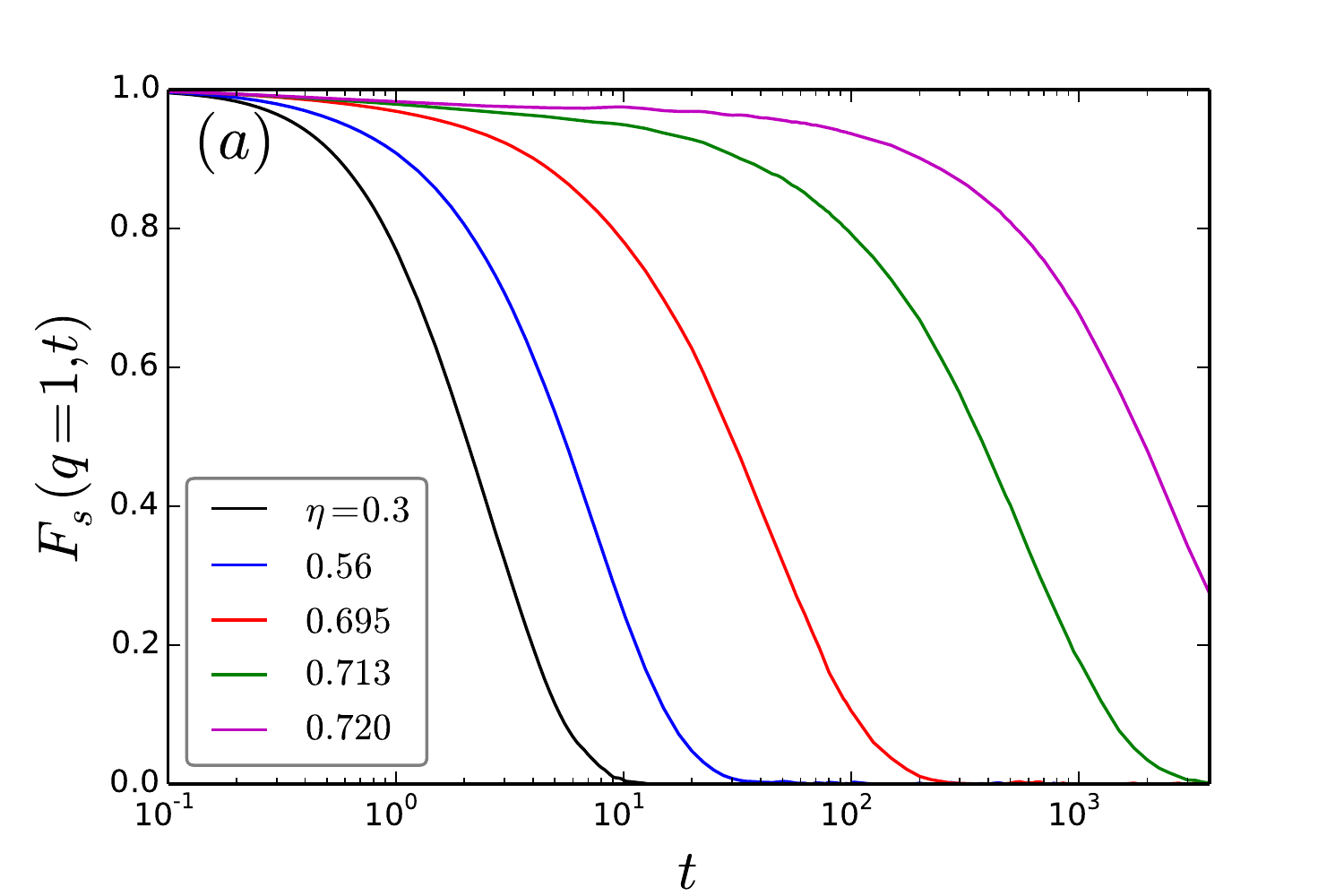}
 \includegraphics[width=0.5\textwidth]{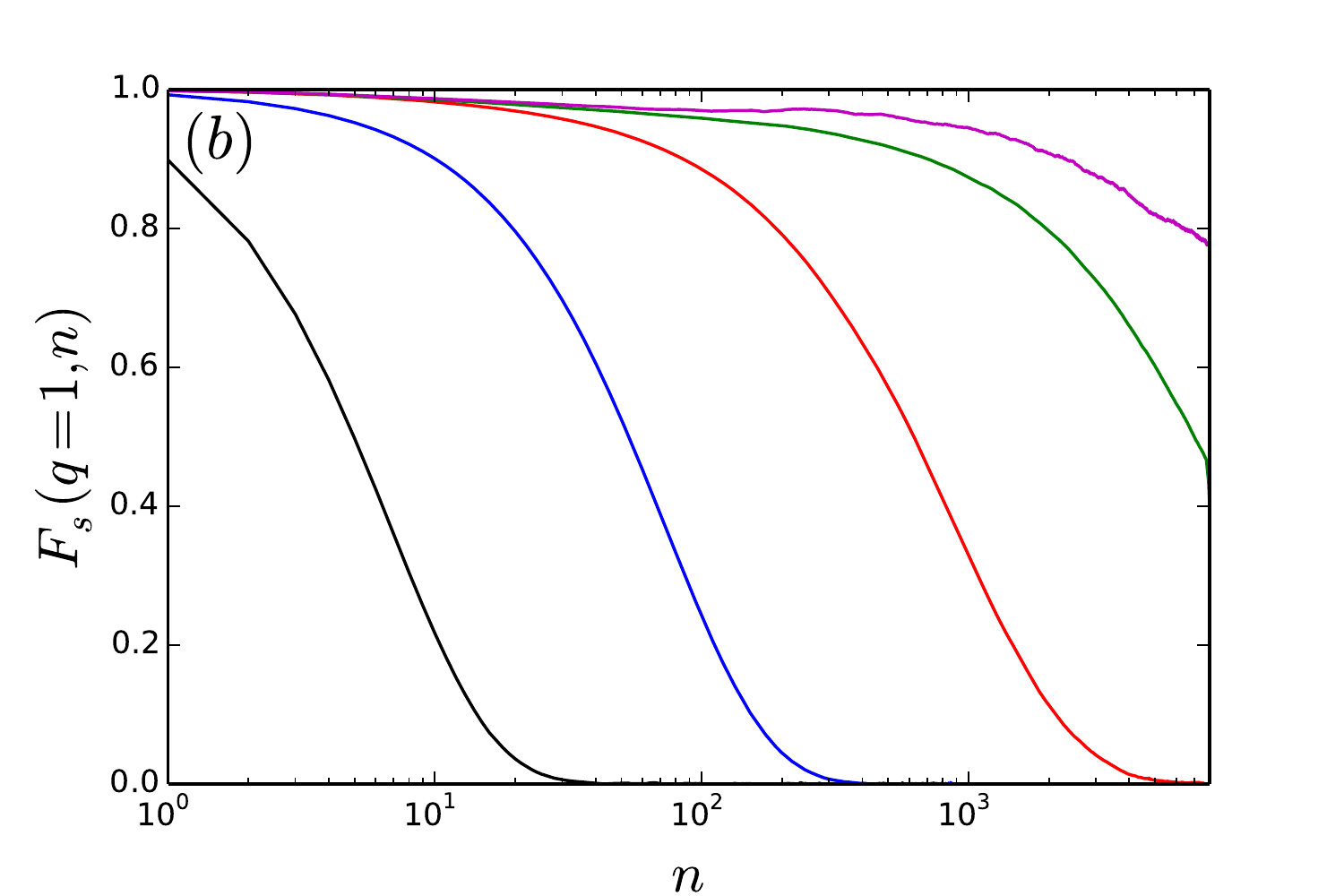}
  \end{center}
 \caption{(Color online)  Intermediate scattering function. (a) Continuous time $F_S(q,t)$ for several $\eta$ calculated according to Eq.(\ref{eq:FS_cont}). The passage from the exponential decay to the stretched exponential regime, particular to caging in granular and colloidal systems is evident. (b) Collisional representation $F_S(q,n)$, Eq.~(\ref{eq:FS_coll}). The exhibited trend traces that of panel (a).}
 \label{fig:figure_12}
 \end{figure}

We now study a different dynamical function at the single particle level,  the single component intermediate scattering function. In continuous time it is defined as:
\be
F_S(q,t)=\langle \cos\left[q\left(x(t)-x(0)\right)\right]\rangle.
\label{eq:FS_cont}
\ee
This function is the real part of the $x$ component of the single particle (incoherent) dynamic structure factor with a wave vector $\mathbf{q}$, $F_S(\mathbf{q},t)=\langle e^{i\mathbf{q}\cdot\left(\mathbf{r}(t)-\mathbf{r}(0)\right)}\rangle$, and has been the subject of  an extremely large amount of numerical and theoretical work in recent years.  While a single exponential decay is expected for a normal fluid state, the emergence of a $q$-dependent 2-step decay at high packing fractions has been recognized to characterize  the behavior of systems in different areas of physics, ranging from supercooled liquids and granular materials to glasses and colloidal systems \cite{sillescu1999, berthier2011, schweizer2007,reis2007, ediger2000,larini2007,kawasaki2010,pham2002,zaccarelli2013}.  Intuitively, this non-exponential regime (often referred to as a stretched exponential decay) should be related to the cage effect, which has been discussed as the main reason for the plateau-like region observed in the MSD (Fig.~\ref{fig:figure_7}(a)) and the antipersistent back-scattering phenomenon exhibited by the velocity ACF (Fig.~\ref{fig:figure_6}(a)). The 2 steps of the $F_S$ relaxation dynamics consist in a fast and local ``caged'' $\beta$-process, and a slow ``cage escape'' $\alpha$-process, related to the cage restructuring,  occurring in a more and more cooperative manner as the molecular packing increases~\cite{donati1998}. The apparent stretching of the relaxation functions can also be explained by spatial heterogeneity, where relaxation occurs exponentially on different timescales in each spatial domain~\cite{palmer1984}. In Fig.~\ref{fig:figure_12}(a) we report the numerical study of the intermediate scattering function Eq.~(\ref{eq:FS_cont}) for several $\eta$. The progressive approach from the single exponential toward the 2-steps stretched exponential relaxation is clearly shown, and, to our knowledge, this is the first time that such evidence is  presented for a hard-disks system. This result, together with the plots in Figs.~\ref{fig:figure_6}(a) and~\ref{fig:figure_7}(a), demonstrates  2D hard-core systems present several important analogies with complex colloids, glasses and supercooled liquids.

Inspired by the previous analogy between continuous time and collisional dynamical quantities, we study the intermediate scattering function in collisional representation, which takes the form:
\be
F_S(q,n)=\langle \cos\left[q\left(x_n-x_0\right)\right]\rangle.
\label{eq:FS_coll}
\ee
Once again, if one neglects the early stage regime characterizing the ballistic behavior in the continuous time analysis, the decay exhibited by  $F_S(q,n)$ in Fig.\ref{fig:figure_12}(b) closely resembles that of  $F_S(q,t)$. In particular the exponential relaxation expected for the fluid state is recovered for low $\eta$, while the stretched exponential decay appears to be more marked as as $\eta$ increases.

\subsection{Self part of the van Hove function}

 \begin{figure}
 \begin{center}
 \includegraphics[width=0.5\textwidth]{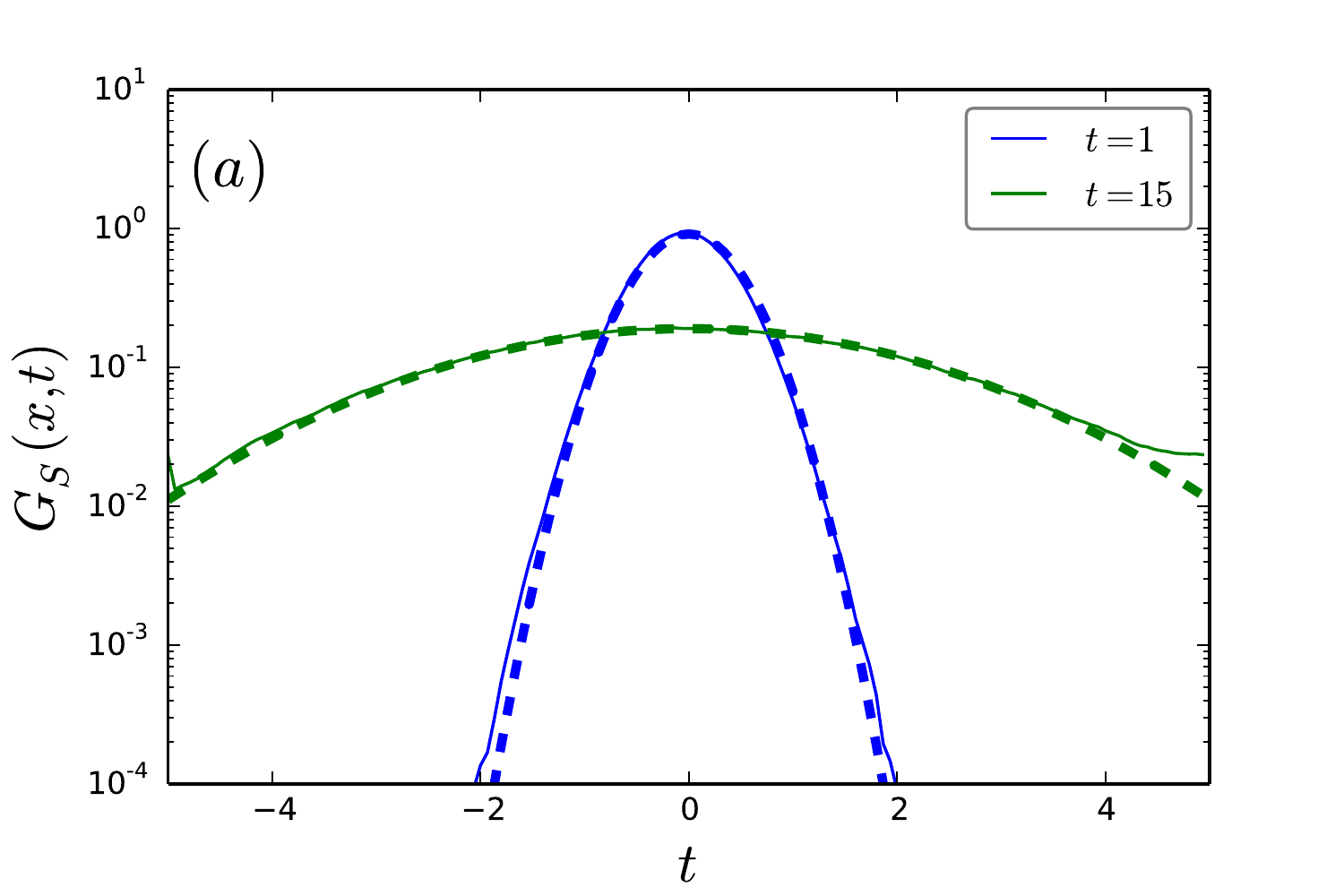}
 \includegraphics[width=0.5\textwidth]{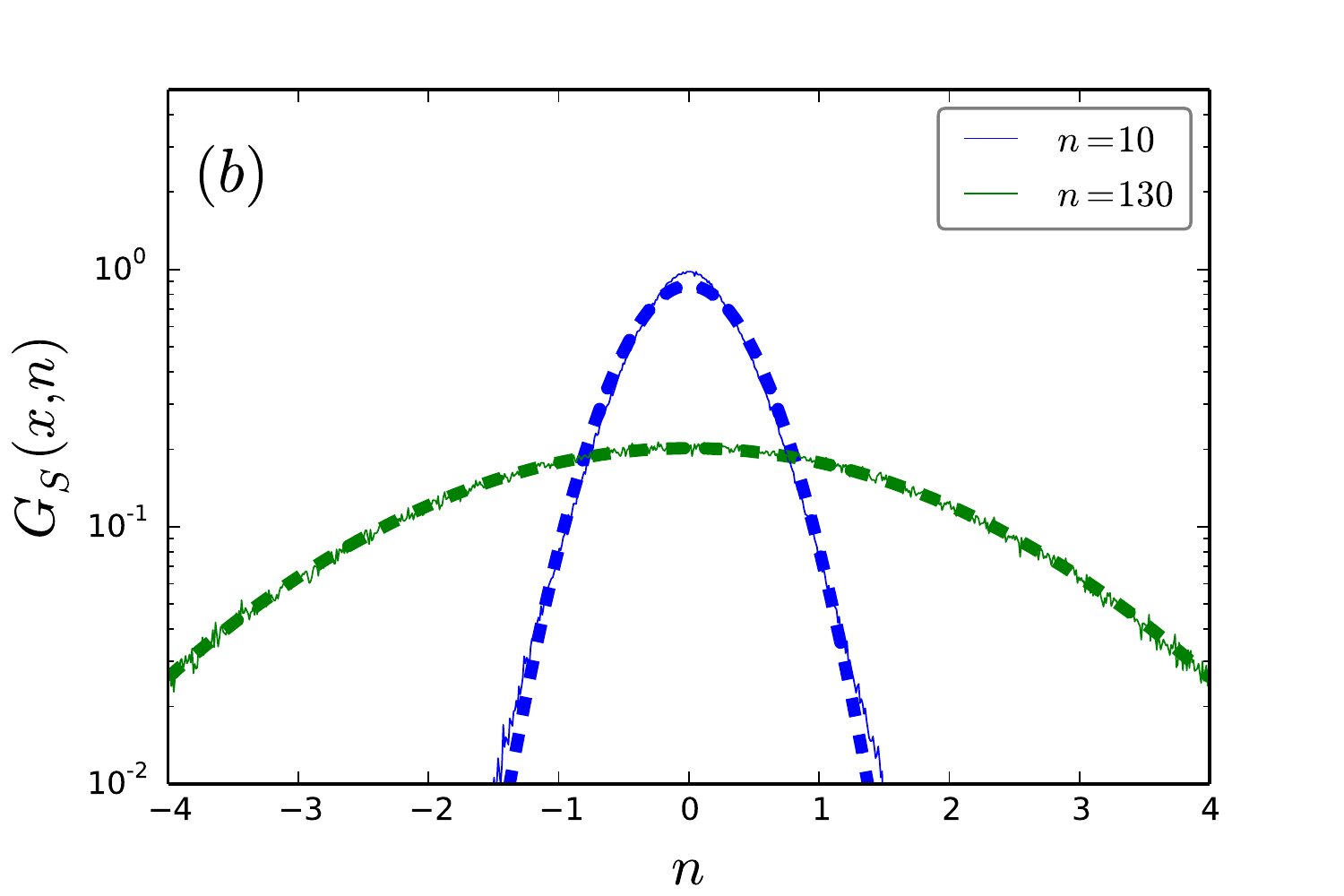}
  \end{center}
 \caption{(Color online)  Self part of the van Hove function. (a) Continuous time $G_S(x,t)$  for $\eta=0.3$, calculated according to Eq.(\ref{eq:vanhove_ct}), for two different times $t$ (solid lines). The dashed lines represent the Gaussian form $\frac{1}{\sqrt{2\pi\langle\delta x^2(t)\rangle}}\exp {(-x^2/2\langle\delta x^2(t)\rangle)}$, where the values of $\langle\delta x^2(t)\rangle$ have been drawn from Fig.{\ref{fig:figure_7}}(a). (b) Collisional representation $G_S(x,n)$  for $\eta=0.3$, calculated according to Eq.(\ref{eq:vanhove_coll}), for two different values of $n$ (solid lines). The dashed lines represent the Gaussian form $\frac{1}{\sqrt{2\pi\langle\delta x^2_n\rangle}} \exp{(-x^2/2\langle\delta x^2_n\rangle)}$, where the values of $\langle\delta x^2_n\rangle$ have been drawn from Fig.~{\ref{fig:figure_8}}(a).}
 \label{fig:figure_13}
 \end{figure}

 \begin{figure}
 \begin{center}
 \includegraphics[width=0.5\textwidth]{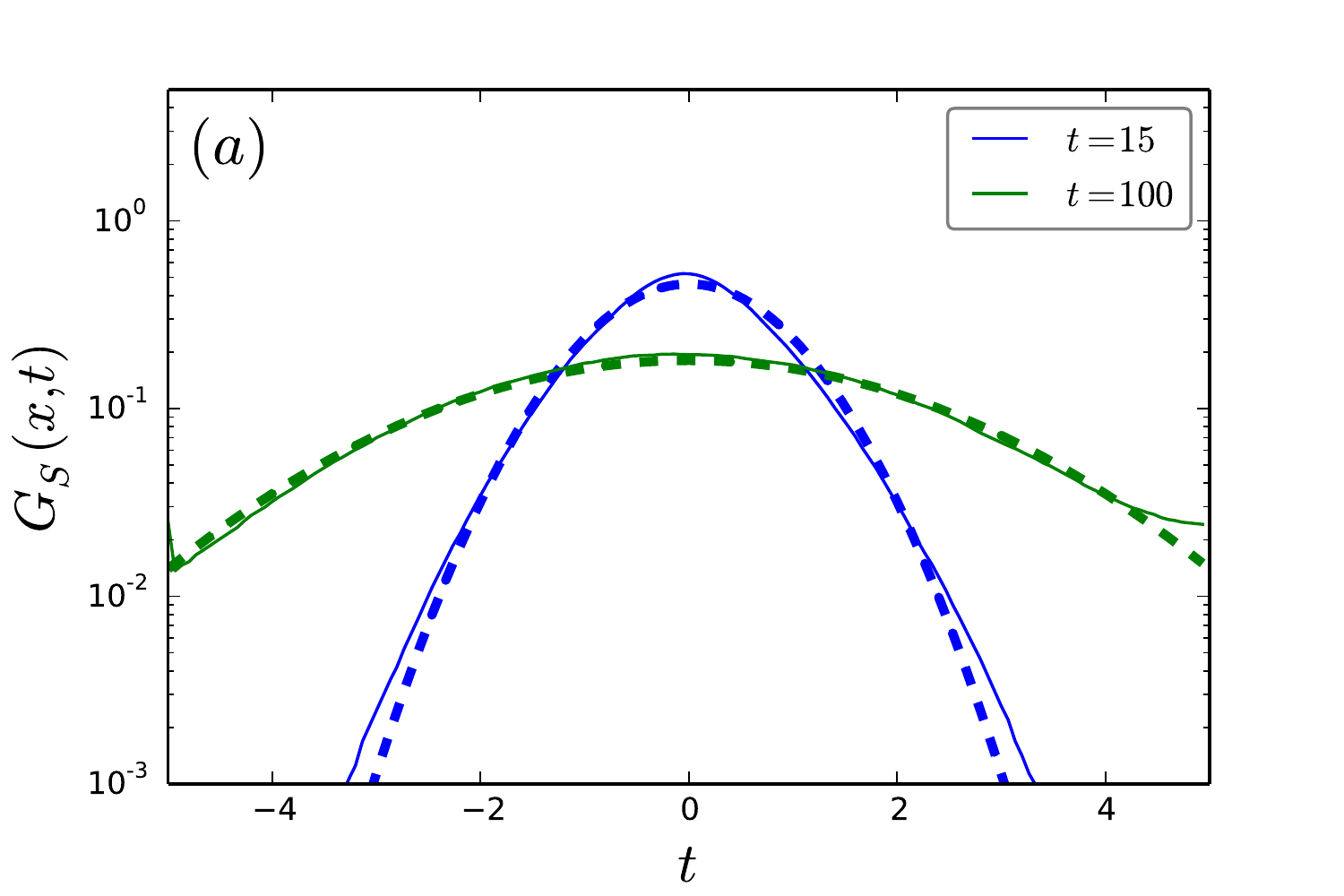}
 \includegraphics[width=0.5\textwidth]{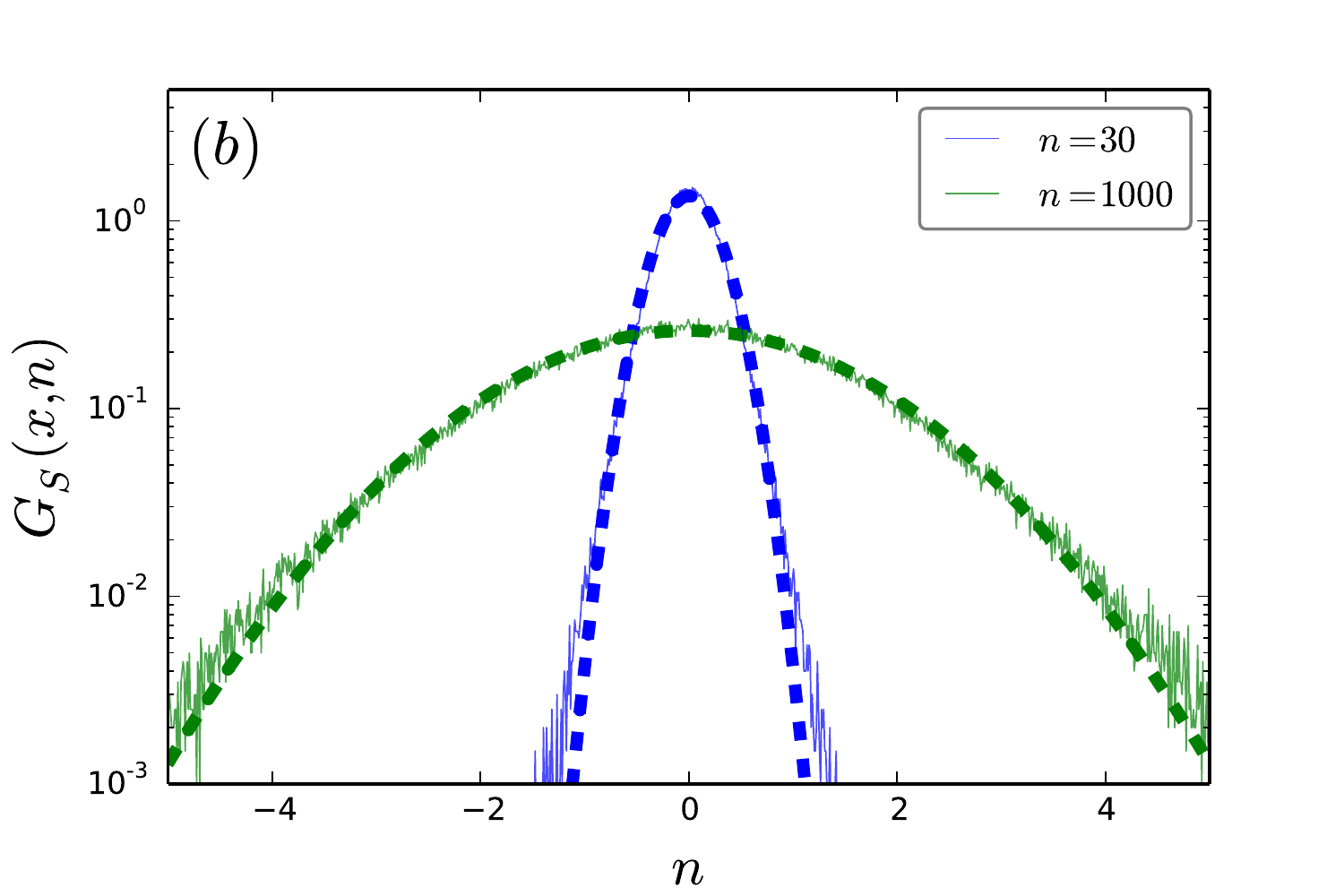}
  \end{center}
 \caption{(Color online)  Self part of the van Hove function. (a) Continuous time $G_S(x,t)$  for $\eta=0.695$, calculated according to Eq.~(\ref{eq:vanhove_ct}), for two different times $t$ (solid lines). The dashed lines represent the Gaussian form $\frac{1}{\sqrt{2\pi\langle\delta x^2(t)\rangle}}\exp {(-x^2/2\langle\delta x^2(t)\rangle)}$, where the values of $\langle\delta x^2(t)\rangle$ have been drawn from Fig.{\ref{fig:figure_7}}(a). Small but persistent deviations from the Gaussian expression in the tails of the distributions are displayed at both times.  (b) Collisional representation $G_S(x,n)$  for $\eta=0.3$, calculated according to Eq.(\ref{eq:vanhove_coll}), for two different values of $n$ (solid lines). The dashed lines represent the Gaussian form $\frac{1}{\sqrt{2\pi\langle\delta x^2_n\rangle}}\exp {(-x^2/2\langle\delta x^2_n\rangle)}$, where the values of $\langle\delta x^2_n\rangle$ have been drawn from Fig.{\ref{fig:figure_8}}(a). Deviations similar to the ontinuous time case appear.}
 \label{fig:figure_14}
 \end{figure}

 \begin{figure}
 \begin{center}
 \includegraphics[width=0.5\textwidth]{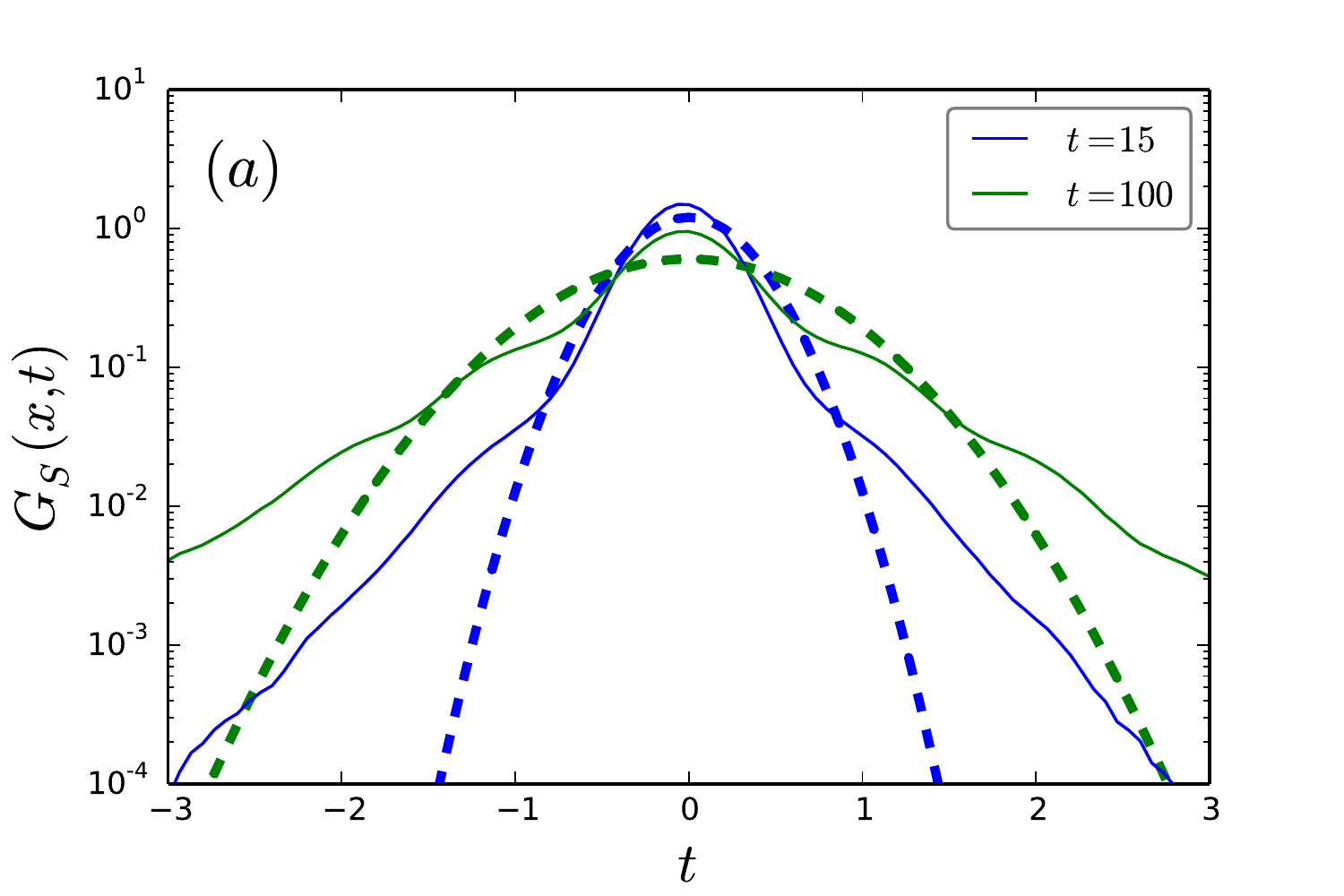}
 \includegraphics[width=0.5\textwidth]{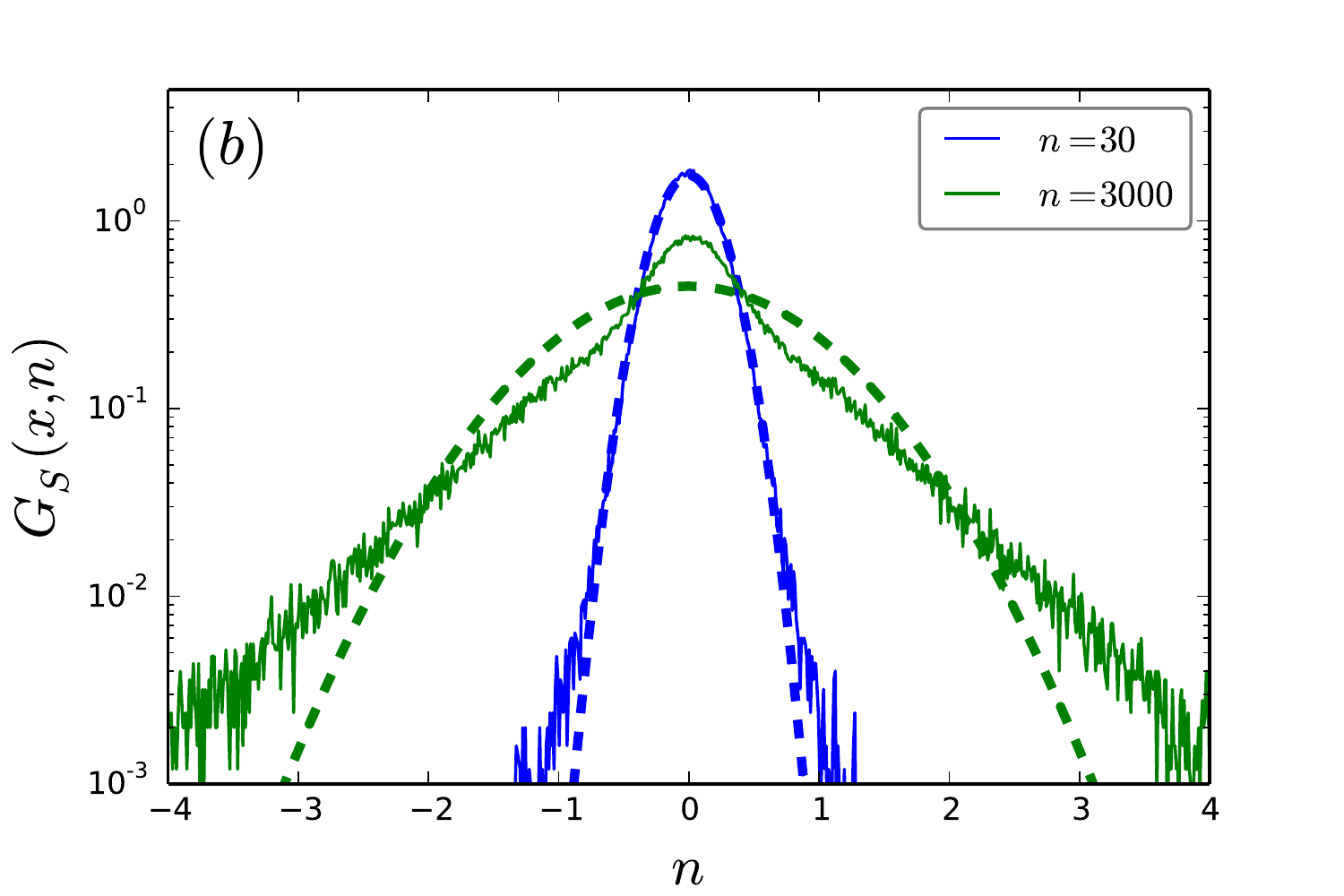}
  \end{center}
 \caption{(Color online)  Self part of the van Hove function. (a) Continuous time $G_S(x,t)$  for $\eta=0.713$, calculated according to Eq.(\ref{eq:vanhove_ct}), for two different times $t$ (solid lines). The dashed lines represent the Gaussian form $\frac{1}{\sqrt{2\pi\langle\delta x^2(t)\rangle}}  \exp{(-x^2/2\langle\delta x^2(t)\rangle)} $, where the values of $\langle\delta x^2(t)\rangle$ have been drawn from Fig.{\ref{fig:figure_7}}(a). Deviations  from the Gaussian form are clearly displayed: in particular, for the longer time corresponding to the asymptotic linear regime  in Fig.{\ref{fig:figure_7}}(a) non-Gaussian tails are reminiscent of what is found in granular gases and colloids.  (b) Collisional representation $G_S(x,n)$  for $\eta=0.3$, calculated according to Eq.~(\ref{eq:vanhove_coll}), for two different values of $n$ (solid lines). The dashed lines represent the Gaussian form $\frac{1}{\sqrt{2\pi\langle\delta x^2_n\rangle}} \exp{(-x^2/2\langle\delta x^2_n\rangle)}$, where the values of $\langle\delta x^2_n\rangle$ have been drawn from Fig.~{\ref{fig:figure_8}}(a). Deviations from the Gaussian diffusive form appear. Moreover non-Gaussian tails similar to those in panel (a) characterize the free path dynamics,  for $n$ well beyond the plateau in Fig.~{\ref{fig:figure_8}}(a). We note that oscillations present in the continuous, representing spatial order due to high packing fractions, are not apparent in the collisional representation, due to the decoupling of space and time (detailed in main text).}
 \label{fig:figure_15}
 \end{figure}

The last dynamical quantity that we study at the single particle level is the self part of the van Hove function~\cite{van1954}. This function is the equilibrium, canonically averaged,
probability distribution of the particle displacements (the propagator),  that for technical reasons is used in experiments in its one-dimensional form:
\be
G_S(x,t)=\langle \delta\left(x-\left[x(t)-x(0)\right]\right)\rangle.
\label{eq:vanhove_ct}
\ee
Generally, in materials close to the glass or jamming transition and colloidal gelation, the non-Fickian character of the single particle displacements is indicated by three factors: $(i)$ the relaxation functions decay non-exponentially, $(ii)$ the MSD exhibits a subdiffusive plateau at intermediate time scales, $(iii)$ the intermediate scattering function $F_S$ exhibits stretched-exponential decay, $(iv)$ the self-part of the van Hove distribution function is non-Gaussian. In the previous sections we have provided the numerical evidence of the first three phenomena in both continuous time and collisional representation, and we  now tackle the last one. Moreover, the deviation from the Gaussian form is generally interpreted as the key signature of the so-called dynamical heterogeneity \cite{ediger2000,berthier2011}, i.e. when slow particles tend to cluster together forming cages, as well as faster particles forming fluid-like regions~\cite{chaudhuri2007,gao2009,chaudhuri2008,kegel2000,nagamanasa2011,weeks2002,weeks2000}.
In Fig.~\ref{fig:figure_13}(a) we report the van Hove function from Eq.~(\ref{eq:vanhove_ct}) calculated for $\eta=0.56$, in the dilute regime (solid line). The dashed lines represent the Gaussian expression $\frac{1}{\sqrt{2\pi\langle\delta x^2(t)\rangle}}e^{-\frac{x^2}{2\langle\delta x^2(t)\rangle}}$ where the value of the MSD has been obtained from data in  Fig.~\ref{fig:figure_7}(a). It is apparent that in this case the Fickian behavior is respected and the Gaussian form reproduces with remarkable accuracy the numerics at two different times.  In Fig.~\ref{fig:figure_14}(a) the van Hove distribution function is displayed for $\eta=0.695$,  within the coexistence phase. $G_S(x,t)$ has been calculated at two different times according to the different stages attained by the MSD in Fig.~\ref{fig:figure_7}(a):  within the subdiffusive plateau and the subsequent diffusive phase, once the tracer has escaped from the cage.
In this case one can clearly see that the dashed lines accounting for the Gaussian Fickian behavior do not capture the tails of the van Hove function for times corresponding to the plateau region in Fig.~\ref{fig:figure_7}(a). When the dynamics is diffusive, the Gaussian behavior is instead restored. The appearance of non-Gaussian tails is even more apparent for $\eta=0.713$ (Fig.~\ref{fig:figure_15}(a)) where at short and  larger times the exponential tails characterize the behavior of the van Hove function, suggesting that the  particle dynamics can be represented by a number subsequent caging events. This is in line with the observation of plateau-like regions  detected in the MSD (Fig.~\ref{fig:figure_7}(a)), and with the velocity ACF antipersistent tails observed in Fig.~\ref{fig:figure_6}(a).

Studying the single component van Hove function in collisional representation through the definition 
\be
G_S(x,n)=\langle \delta\left(x-\left[x_n-x_0\right]\right)\rangle
\label{eq:vanhove_coll}
\ee
traces the behavior exhibited in continuous time. In Figs.~\ref{fig:figure_13}(b), \ref{fig:figure_14}(b) and \ref{fig:figure_15}(b) the solid lines represent the numerical data, whereas the dashed curves account for the Gaussian form $\frac{1}{\sqrt{2\pi\langle\delta x^2_n\rangle}}e^{-\frac{x^2}{2\langle\delta x^2_n\rangle}}$, where $\langle\delta x^2_n\rangle$ is drawn from the results in Fig.~\ref{fig:figure_9}(a). For the lowest packing fraction the agreement observed in continuous time is respected. For $\eta$=0.695 shown in Fig.~\ref{fig:figure_14}(b), the agreement between numerical curves and Gaussian expressions seems to be better for $n=1000$ than  $n=30$. For $\eta=0.713$ (Fig.~\ref{fig:figure_15}(b)) the exponential non-Gaussian tails characterize the subdiffusive regime as well as the ensuing diffusive phase. We also note the continuous-time propagator exhibits oscillations related to the disk packing, which seem to disappear in the collisional representation. This can be attributed to the fact that solid packing is a collective process, orchestrated with external time $t$.

%
%

\section{From collisional to continuous time representation\label{sec:poisson}}

We now wish to study the relation between continuous time and collisional representation. In Sec.\ref{sec:FFT_ACF} we have provided evidence that the free flight time $\tau_n$ evolves according to a non-Markovian dynamics, and this property is fulfilled by gases at any $\eta$. The distribution of times $t$ required for a particle to undergo  $n$ collisions, i.e. the probability that a particle makes \emph{exactly} $n$ collisions up to a time $t$,  is given by:
\be
\begin{split}
P(n,t)=\int d\tau_0\cdots d\tau_{n-1}\,dv_0\cdots dv_{n-1}\,\delta\left(t-\sum_{i=0}^{n-1}\tau_i\right)\\
p(\tau_0,\cdots\tau_{n-1}|v_0,\cdots v_{n-1})\phi_{coll}(v_0,\cdots v_{n-1}).
\end{split}
\label{eq:times_after_n} 
\ee
Hence, the exact expression of $P(n,t)$ can be derived uniquely upon knowing the joint probability distribution of the velocities, and the ensuing conditional probability of the free flight times.  Following Ref.~\cite{Visco2008}, we take the Laplace transform of both sides of Eq.(\ref{eq:times_after_n}):
\be
\begin{split}
P(n,s)=\int d\tau_0\cdots d\tau_{n-1}\,dv_0\cdots dv_{n-1}e^{-s\sum_{i=0}^{n-1}\tau_i}\\
p(\tau_0,\cdots\tau_{n-1}|v_0,\cdots v_{n-1})\phi_{coll}(v_0,\cdots v_{n-1}).
\end{split}
\label{eq:times_after_n_lt} 
\ee
Assuming the velocity process to be uncorrelated, the former Laplace transform then reads:
\be
P(n,s)=\left[\int dv\, p(s|v)\phi_{coll}(v)\right]^{n},
\label{eq:times_after_n_lt_markov} 
\ee
where $P(s)=\int dv\, p(s|v)\phi_{coll}(v)$ is the Laplace transform of the free flight time distribution function $P(\tau)$ in Eq.~(\ref{eq:free_flight_prob_def}). In the dilute limit, the exponential form of $p(\tau|v)$ leads to: 
\be
P(n,s)=\left[\int dv\, \frac{1}{\langle\tau\rangle s\phi_{coll}(v)+\phi_{MB}(v)}
\right]^{n},
\label{eq:times_after_n_lt_markov_lowPF} 
\ee
where we made use of Eq.~(\ref{eq:time_relation_visco}). The same expression has been derived in Ref.~\cite{Visco2008} assuming the velocity collisional process to be Markovian, where the numerical Laplace inversion of Eq.~(\ref{eq:times_after_n_lt_markov_lowPF}), or analogously of $P(s)^n$, has been shown to  closely agree with $P(n,t)$ extracted from simulations for $\eta$ in the fluid phase.
This tells us that the  uncorrelation assumption of the velocity collisional process hold in the dilute limit, if one considers $P(n,t)$ or $P(\tau)$ \cite{Visco2008}. In the large packing fraction limit, however, this assumption fails and the inverse Laplace transform of Eq.~(\ref{eq:times_after_n_lt_markov_lowPF}) does not capture the probability $P(n,t)$. 

Let us consider the case in which the free flight time distribution takes the exponential form $P(\tau)=\frac{1}{\langle\tau\rangle}e^{-t/\langle\tau\rangle}$. In view of Eq.~(\ref{eq:free_flight_prob_def}), we are implicitly assuming $p(\tau|v)=\frac{1}{\langle\tau\rangle}e^{-t/\langle\tau\rangle}$, i.e. a unique time scale characterizes the collisional process irrespective of the undergone velocity. This corresponds, \emph{de facto}, to taking  free flight times independent of velocities. We can thus write the former Eq.~(\ref{eq:free_flight_prob_def}) as:
\be
P(n,t)=\int d\tau_0\cdots d\tau_{n-1}\,\delta\left(t-\sum_{i=0}^{n-1}\tau_i\right)
P(\tau_0,\cdots\tau_{n-1}),
\label{eq:times_after_n_poiss_approx1} 
\ee
and, if we make the further assumption of considering the free flight times to be uncorrelated, one finally obtains for the  Laplace transform (\ref{eq:times_after_n_lt}) :
\be
P(n,s)= \frac{1}{(\langle\tau\rangle s+1)^{n}}.
\label{eq:times_after_n_poiss_approx_lt} 
\ee
Inverting back in time domain, yields the Poisson formula ~\cite{Lue2005}:
\be
P(n,t)= \frac{1}{\langle\tau\rangle}\left(\frac{t}{\langle\tau\rangle}\right)^{n-1}\frac{e^{-\frac{t}{\langle\tau\rangle}}}{(n-1)!}.
\label{eq:times_after_n_poiss_approx_final} 
\ee
which has shown a large discrepancy to Eq.(\ref{eq:times_after_n}) even in the dilute limit \cite{Lue2005,Visco2008}. However, the collisional MSD expression in Eq.~(\ref{eq:MSD_kubo_coll_approx}) derived in this framework seems to work fairly well in the  low $\eta$ limit (see the black curve corresponding to $\eta=0.3$ in Fig.\ref{fig:figure_10_emmezzo}(a)).

\begin{figure}
 \begin{center}
 \includegraphics[width=0.5\textwidth]{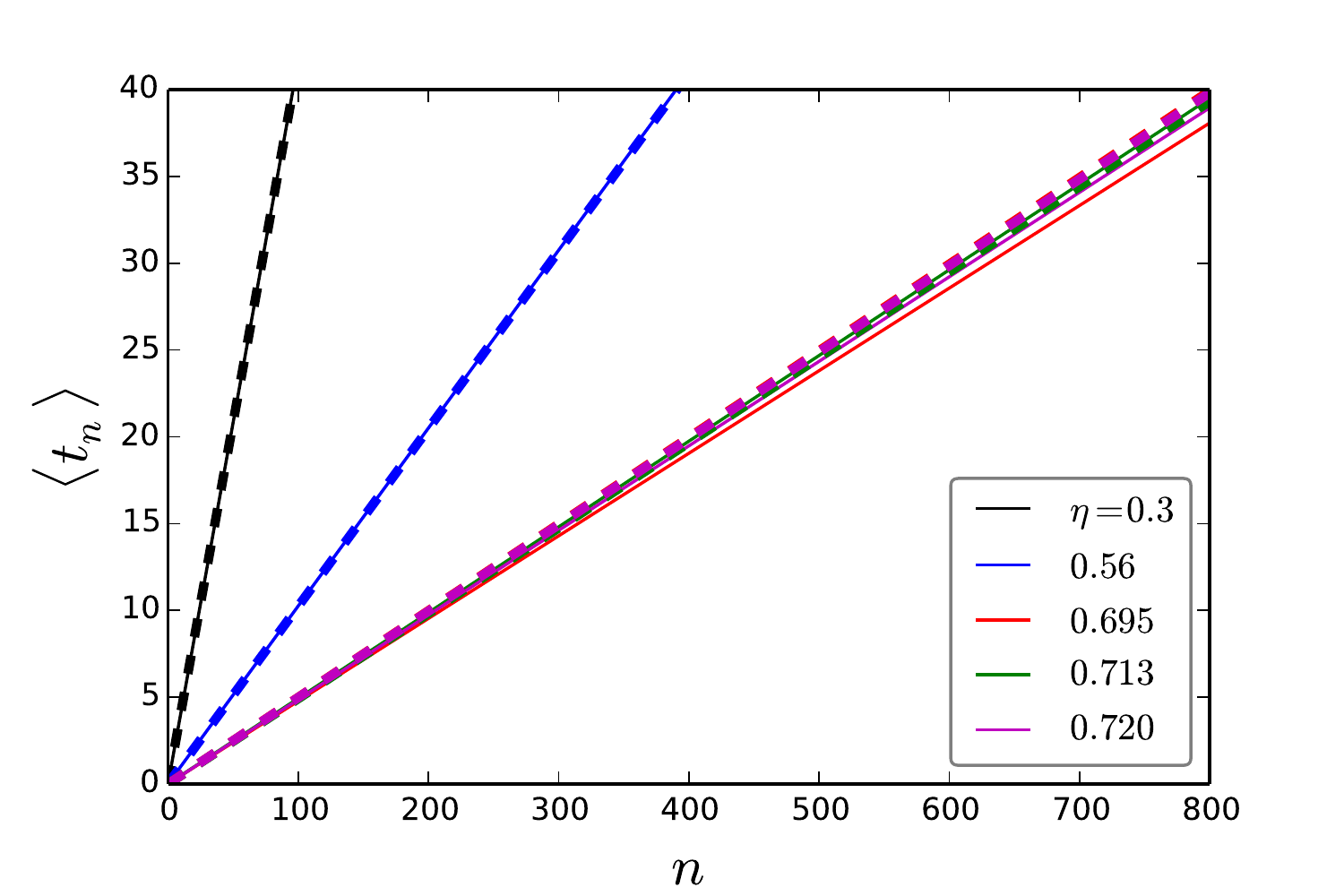}
  \end{center}
 \caption{(Color online)  Average time after $n$ collisions, $\langle t_n \rangle$,  for several $\eta$ $\eta$ (solid lines). Dashed lines correspond to the Poissonian approximation in Eq.~(\ref{eq:average_time_poiss_approx}).}
 \label{fig:figure_16}
 \end{figure}

We now turn to the average time between $n$ collisions, i.e. the first moment of $P(n,t)$, within the context of the Poisson approximation, i.e.  using the Poisson approximated expression in Eq.~(\ref{eq:times_after_n_poiss_approx_final}). A straightforward calculation yields:
\be
\langle t_n\rangle=n\langle\tau\rangle,
\label{eq:average_time_poiss_approx} 
\ee
and, surprisingly, it reproduces very well the numerical values reported in  Fig.\ref{fig:figure_16} not only for low $\eta$ but also for $\eta$ in the solid regime.

To conclude this section, let us turn to the single particle propagator. To obtain the self part of the van Hove function in continuous time of Eq.~ (\ref{eq:vanhove_ct}) from the corresponding function in collisional representation of Eq.~(\ref{eq:vanhove_coll}), we have to recur to the well-known subordination relation~\cite{feller}:
\be
G(x,t)=\sum_{n=1}^{\infty}G(x,n)\Pi(n,t),
\label{eq:van_hove_coll->ct}
\ee
where $\Pi(n,t)$ is the single particle probability of having $n$ collisions at time $t$. The connection between the formerly introduced probability $P(n,t)$ in Eq.~(\ref{eq:times_after_n}) and $\Pi(n,t)$ is given, for large $n$, by $\Pi(n,t)=\langle\tau\rangle P(n,t)$. Thus, the MSD can be easily cast ed in the following form:
\be
\langle\delta x^2(t)\rangle=\langle\tau\rangle\sum_{n=1}^{\infty}\langle\delta x^2_n\rangle P(n,t).
\label{eq:MSD_coll->ct}
\ee
This expression constitutes the relation between the MSD in collisional representation and that in continuous time, but the rigorous evaluation at any $\eta$ requires the knowledge of the conditional probability distributions in Eq.~(\ref{eq:times_after_n}). However, a semianalytical expression of the MSD in continuous time can be achieved if we use the  Poissonian approximation. Thanks to Eq.~(\ref{eq:MSD_kubo_coll_approx}) and Eq.~(\ref{eq:times_after_n_poiss_approx_final}) we have, after elementary algebraic passages:
\be
\begin{split}
\langle\delta x^2(t)\rangle=\langle\tau\rangle^2\langle v^2\rangle+
\langle\tau\rangle\langle v^2\rangle t+\\
2\sum_{n=2}^{\infty}\sum_{m=1}^{n-1}\langle v_0v_m\rangle(n-m)\left(\frac{t}{\langle\tau\rangle}\right)^{n-1}\frac{e^{-\frac{t}{\langle\tau\rangle}}}{(n-1)!}.
\end{split}
\label{eq:MSD_coll->ct_poiss_approx}
\ee
This approximated expression can be considered valid only  in the low $\eta$ regime and, as such, the three terms in it can easily be interpreted. The first term accounts for the ballistic regime, the second is the leading diffusive term which dominates the single particle dynamics after the first collision. The third term represents the logarithmic corrections expected for an hard-disks gas. It is interesting to notice that the second moment of the velocity $\langle v^2\rangle$ is obtained using the collisional stationary distribution of velocity $\varphi_{coll}(v)$, instead of the regular Maxwell-Boltzmann $\varphi_{MB}(v)$. As a first approximation we neglect the logarithmic corrections in Eq.~(\ref{eq:MSD_coll->ct_poiss_approx}) which are believed to play a significant role for large times (see the discussion in Sec.\ref{sec:MSD}).
In Fig.~\ref{fig:figure_10_emmezzo}(b) we observe that the approximated Poisson formula (\ref{eq:MSD_coll->ct_poiss_approx}) accurately reproduces the MSD in continuous time for low $\eta$, whereas it clearly fails to account for the high packing regime, as indeed is expected.

\section{Discussion}

We have studied statistical and dynamical properties of a 2D system of hard-disks in collisional representation, with packing fractions ranging from the fluid to the solid phase. Throughout this paper we tried to place our results withing a broader context, resulting in a comprehensive study of 2D hard disk systems. The first part of our analysis focused on the statistics of velocity, free flight time and the path undergone by a disk between subsequent collisions. We have provided the numerical and theoretical evidence that the velocity $x$ and $y$ components are correlated in collisional representation, together with the fact that free flight times are strongly correlated to the on-collision velocities. This result has previously been investigated~\cite{Puglisi2006,Visco2008,ViscoJCP2008} only in the low packing fraction limit, and we have extended it to the coexistence and solid phase. We have shown that kinetic  and Enskog theory do not furnish correct expressions for  the average free flight time $\langle\tau\rangle$ and path $\langle\left|\xi\right|\rangle$ as a function of the packing fraction $\eta$, not even in very dilute systems. Furthermore our numerical analysis spanning packing fractions up to the high $\eta$ limit has unveiled plateau-like regions of $\langle\tau\rangle$ and  $\langle\left|\xi\right|\rangle$ corresponding to the fluid-solid coexistence region, which to the best of our knowledge have not been reported before.

The second part of this study has been devoted to the dynamical single particle properties in collisional representation, and the comparison to their continuous time equivalents. We have shown a full analogy between observables calculated in continuous time (as a function of $t$) and in collisional representation (as a function of the number of collisions, $n$), in spite of the fact that the process is not truly Poissonian~\cite{Visco2008}. We have shown that the velocity is never a Markovian process both in continuous time and  collisional representation. This is attributed to the persistent memory effects characterizing the velocity autocorrelation function at low packing fractions, that eventually turn into antipersistent tails at high packing fractions. In particular, this finding contradicts the common assumption of Markovianity of velocity in the collisional representation at low packing fractions~\cite{Puglisi2006}.
Furthermore we have analyzed the mean squared displacement, intermediate scattering function and self-part of the van Hove function, or propagator, both in continuous time and collisional representation, showing remarkable similarities with colloidal systems, glassy systems and supercooled liquids close to the glass or jamming transition. These analogies provide a strong indication that within the coexistence phase the simplistic hard-disk model captures properties of glassy behavior.

Finally our analysis shows that, according to the observable of interest, different approximations yield better descriptions of the data.
Indeed we have shown that, although the system is generally non-Poissonian, at low packing fractions it can be fairly well described as a Poissonian model, i.e. considering uncorrelated free flight times independent of velocities. This approximation works well if one considers the mean squared displacements $\langle \delta^2x(t)\rangle$ and $\langle \delta^2x_n\rangle$ or the average time as a function of the number of collisions $\langle t_n\rangle$. On the other hand we have  pointed out that a different approximation, considering the on-collision  velocity process uncorrelated or Markovian but still correlated to the free flight times, provides excellent predictions in describing the probability $P(\tau)$ at low $\eta$, the probability $\Pi(n,t)$ of having $n$ collisions up to a time $t$ in the dilute limit, or when one attempts to derive the on-collision stationary velocity distribution function $\varphi_{coll}(v)$ \cite{Visco2008}. However,  none of the former  assumptions is entirely correct, since velocities are correlated in collisional representation at any $\eta$, and they are also correlated with free  flight times: therefore none of them can furnish an overall satisfactory theoretical framework, not even in the very dilute regime.

\section{Acknowledgments}
This work is supported by CONACYT under project 152431 and Promep of M\'exico.
Y. M. received support from the Weizmann Institute of
Science, National Postdoctoral Award Program for
Advancing Women in Science. A. T. acknowledges CNR through the ERANET ComplexityNet pilot project LOCAT and  ERC AdG-2011 SIZEFFECTS. A. H. acknowledges discussions with Andrij Trokhymchuk and ``Red Tem\'atica de la Materia Condensada Blanda" of Conacyt M\'exico.

%
%
\appendix

\section{Single component mean free-flight times relation }
\label{app:A}
Let us assume the sequence of velocities a particle acquires during $N$ collisions
 is $\left\{v_i\right\}$ where $i\in [1,N]$. The number of occurrences of some velocity
$v$ is denoted:
\be
n_v = \sum_i^{N}{\int_{-\infty}^{\infty}{dv \delta(v_i - v)}},
\ee
and the total time spent on the velocity v is then given by $T_v = \sum_i^{N}{\tau_i\int{dv \delta(v_i - v)}}$.
The velocity distribution function in collisional representation follows:
\be \label{app:phi_col}
\varphi_{coll}(v)=\frac{n_v}{N},
\ee
and the velocity distribution function in continuous time can be approximated as:
\be \label{app:phi_cont}
\varphi_{MB}(v)=\frac{T_v}{\langle \tau \rangle}.
\ee
Multiplying and dividing the RHS by N, substituting Eq.(\ref{app:phi_col}) and
realizing that $T_V$ can be approximated as $T_v = n_v\langle\tau(v) \rangle$
yields the relation:
\be
\varphi_{MB}(v)=\frac{\langle\tau(v)\rangle}{\langle\tau\rangle}\varphi_{coll}(v).
\label{app:phi_cont_final}
\ee

\section{Mean free-path and mean free flight time from kinetic and Enskog theory}
\label{app:kinetic_theory}

The familiar expression for the average free-flight time arising from kinetic theory for a system of elastic hard disks is~\cite{Chapman1939}:
\be
\langle\tau\rangle_{kt}=\frac{1}{4\sqrt{\pi k_BT} n \sigma}
\label{app:tau_kinetic_theory},
\ee
where $n$ is the density number, i.e. $n=\frac{N}{L^2}$. Recalling that the packing fraction is $\eta=n\pi\sigma^2/4$, we obtain $\langle\tau\rangle_{kt}=\frac{\sigma}{16\eta}\sqrt{\frac{\pi}{k_BT}}$. Fig.~\ref{fig:figure_4}(a) shows that
Eq.~(\ref{app:tau_kinetic_theory}) agrees with the numerical data only
qualitatively (blue line). The mean free-path is defined as:
\be
\langle\left|\boldsymbol{\xi}\right|\rangle_{kt}=\langle\left|\mathbf{v}\right|\rangle_{cont}\langle\tau\rangle_{kt}=\frac{\pi\sigma}{2^{9/2}\eta}.
\label{app:xi_relation_kinetic_theory}
\ee
The corresponding expression for the single component reads:
\be
\langle\xi\rangle_{kt}=\langle\left| v\right|\rangle_{cont}\langle\tau\rangle_{kt}=\frac{\sigma}{2^{7/2}\eta}.
\label{app:xi_relation_kinetic_theory_new}
\ee
The mean free path value furnished by the Enskog theory for a system of 2D hard disks is given by:
\be
\langle \left|\boldsymbol{\xi}\right|\rangle_{0}=\frac{\pi\sigma}{2^{9/2}\eta Y[\eta]},
\label{app:r_enskog_theory}
\ee
where $Y[x]=\left(1-\frac{7}{16}x\right)/(1-x)^2$ is the Enskog factor. Notice that the expression furnished in Eq.~(\ref{app:r_enskog_theory}) differs from the formula provided in Refs.~\cite{isobe2008,Gaspard2004} by a factor $1/2$ to match the kinetic theory expression in Eq.~(\ref{app:xi_relation_kinetic_theory}), as should be the case. The single component is then:
\be
\langle \xi\rangle_{0}=\frac{\sigma}{2^{7/2}\eta Y[\eta]}.
\label{app:xi_enskog_theory}
\ee
The expressions in Eqs.~(\ref{app:xi_enskog_theory}) and (\ref{app:xi_relation_kinetic_theory_new}) are displayed in the inset of Fig.~\ref{fig:figure_4}(b), showing a non-satisfactory agreement with the average free path obtained in the numerical simulations (red and blue lines respectively).  The mean free flight time can be obtained through:
\be
\langle\tau\rangle_{0}=\frac{\langle \left|\boldsymbol{\xi}\right|\rangle_{0}}{\langle\left|\mathbf{v}\right|\rangle_{cont}}=\frac{\sigma}{16\eta Y[\eta]}\sqrt{\frac{\pi}{k_BT}}.
\label{app:tau_enskog_theory}
\ee
and is plotted in the inset of Fig.~\ref{fig:figure_4}(a) with the expression in Eq.~(\ref{app:xi_enskog_theory}) for the mean free path (red line).

\begin{figure}[t]
\includegraphics[width=0.5\textwidth]{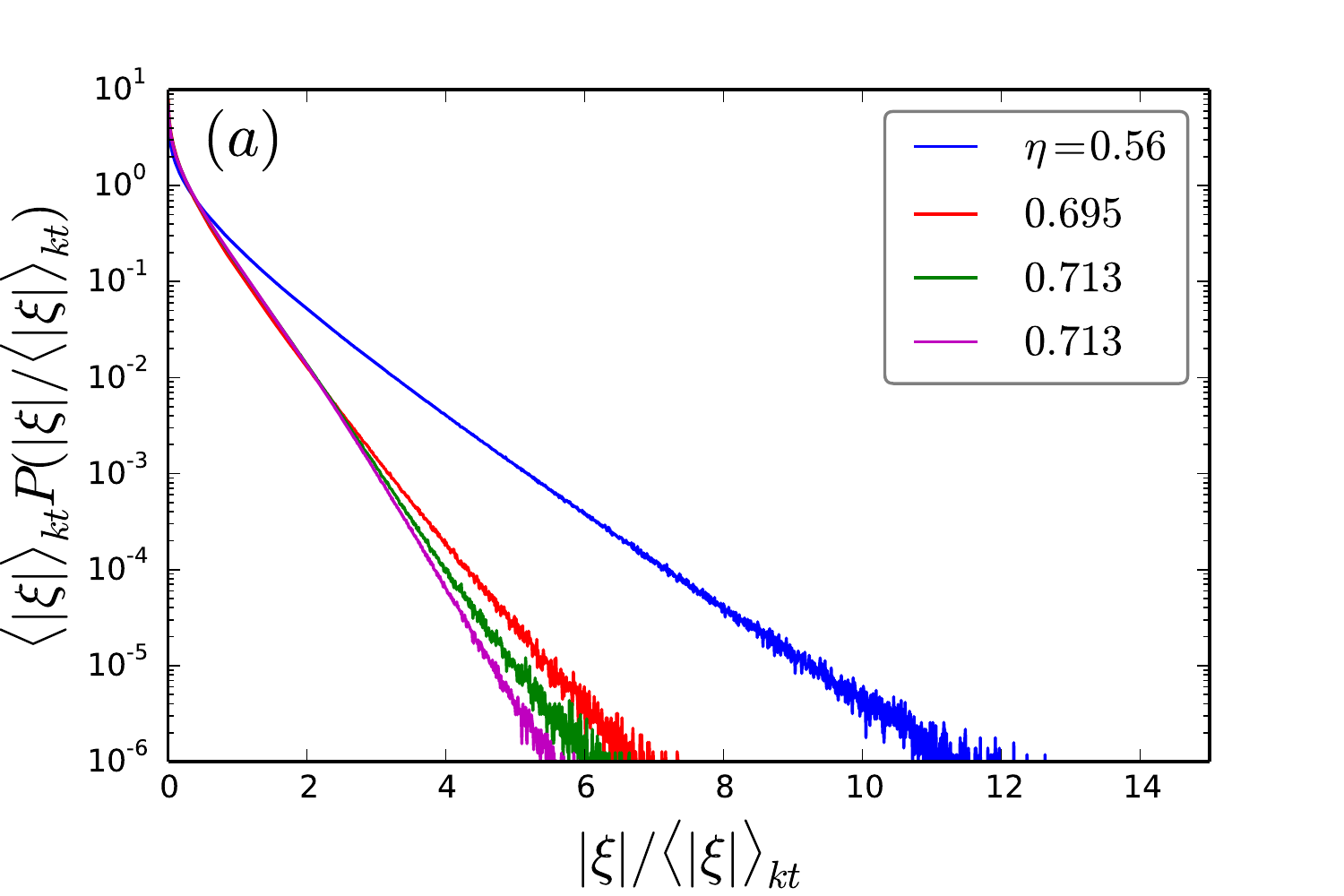}
\includegraphics[width=0.5\textwidth]{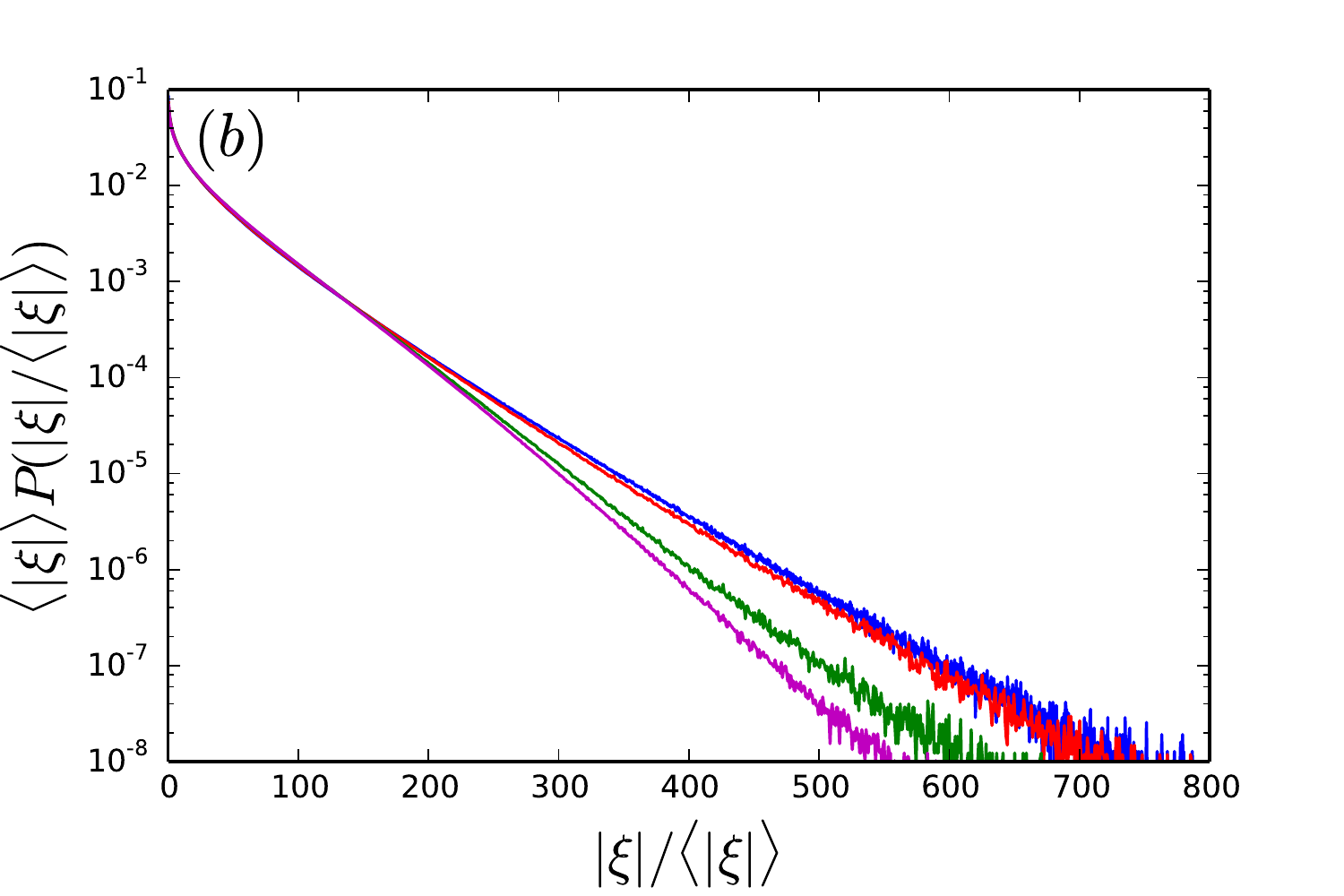}
\includegraphics[width=0.5\textwidth]{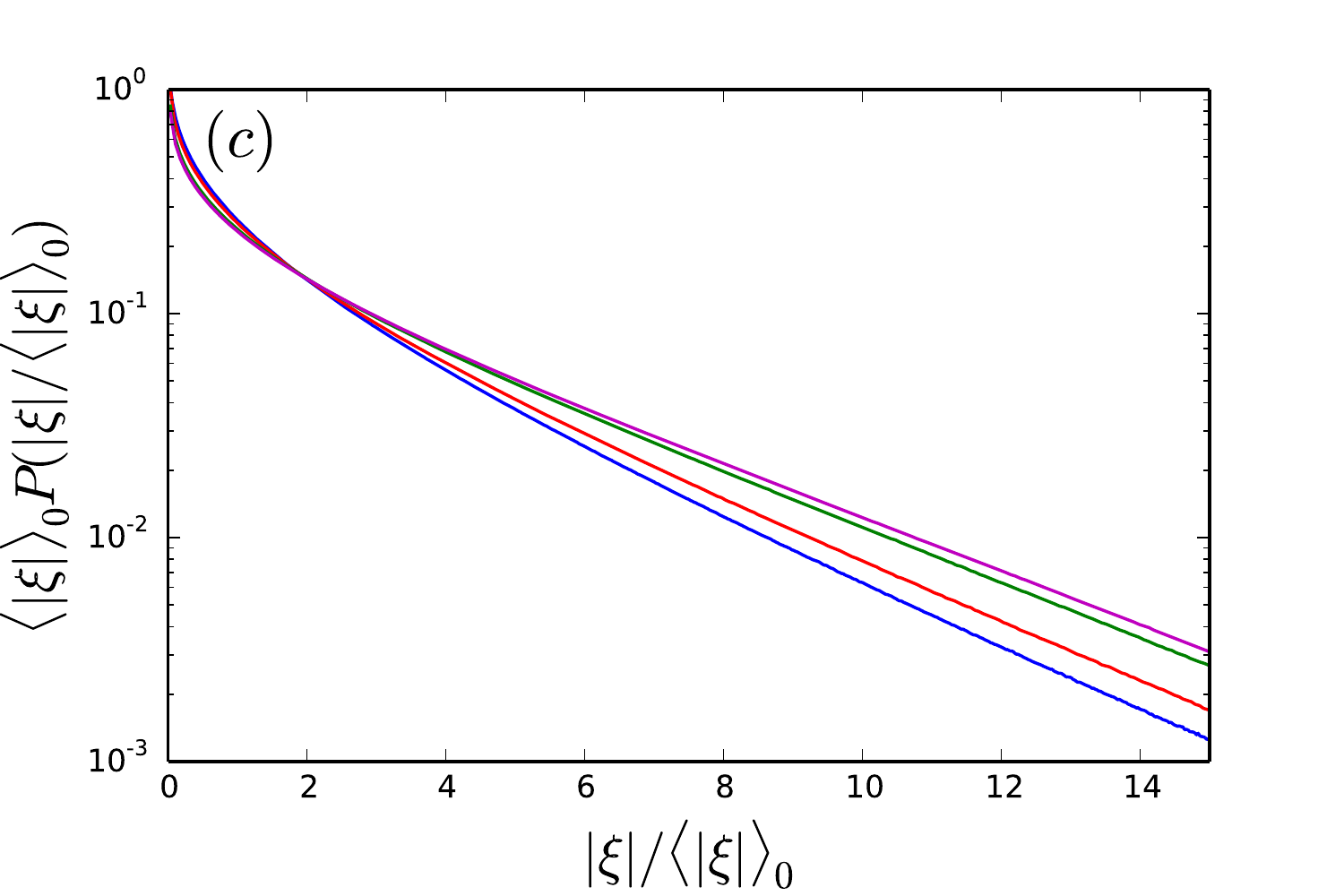}
\caption{(Color online). Free path length probability distribution (appearing in Fig.~\ref{fig:figure_3}(b)) rescaled by: (a) the kinetic theory expression (Eq.(\ref{app:xi_relation_kinetic_theory_new})),  (b) the average free path (Fig.\ref{fig:figure_4}(b))  and (c) the free path furnished by the Enskog theory (Eq.(\ref{app:xi_enskog_theory})). None of them shows good collapse of the curves.}
\label{fig:figure_A1}
\end{figure}

\section{Generalized Kubo theorem in collisional representation}
\label{app:theorem}

Let us consider a discrete stochastic process $\xi_n$ such that $\langle\xi_n\rangle = 0$. The relation between the MSD and the symmetrized ACF is provided by Eq.~(\ref{eq:MSD_kubo_coll_sym}), and brought again here:
\be
\langle  \delta x^2_n\rangle= 2\sum_{m=0}^{n-1}\tilde{C}_{\xi\xi}(m)(n-m).
\label{app:MSD_kubo_coll_sym}
\ee 
The on-collision transport coefficient $D_n$ is defined as:
\be
D_n=\sum_{m=0}^{n-1}\tilde{C}_{\xi\xi}(m),
\label{app:Dn}
\ee 
and it is possible to  study its behavior once the dependence of $\tilde{C}_{\xi\xi}$ on $n$ is defined. Two main situations can arise: $(i)$ $D_n \sim const$, and $(ii)$ $D_n\sim n^{1-\beta}$ ($\beta > 0$).

$(i)$ Suppose that $\tilde{C}_{\xi\xi}$ is such that $D_n=\sum_{m=0}^{n-1}\tilde{C}_{\xi\xi}(m)\sim D$  with $D=const$. Hence from  Eq.~(\ref{app:MSD_kubo_coll_sym}) we have:
\be
\langle  \delta x^2_n\rangle= 2D_n-2\sum_{m=0}^{n-1}\tilde{C}_{\xi\xi}(m)m.
\label{app:MSD_kubo_coll_sym_const}
\ee 
The second term on the RHS of Eq.~(\ref{app:MSD_kubo_coll_sym_const}) is $\sum_{m=0}^{n-1}\tilde{C}_{\xi\xi}(m)m\leq Dn$, yielding:
\be
\langle  \delta x^2_n\rangle\sim 2D_n
\label{app:MSD_kubo_coll_sym_const_final}
\ee 
for a large enough $n$.

$(ii)$ Consider now a process for which $\tilde{C}_{\xi\xi}(n)\sim c_\beta n^{-\beta}$. The corresponding discrete transport coefficient is then given by~\cite{prudnikov1986,abramowitz1972}:
\be
D_n=c_\beta\left[\zeta(\beta)+\frac{(n-1)^{1-\beta}}{1-\beta}+\beta\int_{n-1}^{\infty}dx \frac{x-[x]}{x^{\beta+1}}\right],
\label{app:transport_coeff_pl}
\ee 
where $\zeta(\beta)$ is the Riemann zeta function and $[\cdot\, ]$ represents the integer part. Hence for large $n$ we can write $D_n\sim D+c_\beta\frac{n^{1-\beta}}{1-\beta}$ where $D$ is a constant. From Eq.~\ref{app:MSD_kubo_coll_sym} it then follows:
\be
\langle  \delta x^2_n\rangle\sim 2Dn + 2c_\beta \frac{n^{2-\beta}}{(1-\beta)(2-\beta)}.
\label{app:MSD_kubo_coll_sym_pl}
\ee 
If $0<\beta<1$ the process is superdiffusive. Indeed $c_\beta$ is positive
meaning that persistent memory effects characterize the free path $\xi_n$. Moreover $D_n$ diverges as $\sim n^{1-\beta}$ according to Eq.~(\ref{app:transport_coeff_pl}).
If $1<\beta<2$ and $D=0$ the process is subdiffusive. In this case $c_\beta<0$, which defines the antipersistent tails of the symmetrized ACF $\tilde{C}_{\xi\xi}(n)$. $D_n$ is positive and decays to zero as $n^{1-\beta}$. If $D\neq 0$ the  process is diffusive and $D_n \sim D$.
If $\beta\geq 2$, $D\neq 0$, one has normal diffusion and $D_n \sim D$.

\bibliography{biblio}

\end {document}